\documentclass[twocolumn,aps,superscriptaddress,prx,showpacs]{revtex4-2}
\usepackage{graphicx}
\usepackage{epsfig}
\usepackage{float}
\usepackage{amssymb,amsmath,amsfonts,hyperref}
\usepackage{wasysym}
\usepackage{latexsym}
\usepackage{verbatim}
\usepackage{float}
\usepackage[normalem]{ulem}
\usepackage{color}
\usepackage{todonotes}
\definecolor{labelkey}{RGB}{192,0,0}

\newcommand{\be}{\begin{eqnarray}}
	\newcommand{\ee}{\end{eqnarray}}

\newcommand{\calR}{{\mathcal R}}
\newcommand{\calP}{{\mathcal P}}
\newcommand{\calB}{{\mathcal B}}
\newcommand{\calI}{{\mathcal I}}
\newcommand{\calG}{{\mathcal G}}
\newcommand{\mtotR}{{m_{\rm tot}^{\mathcal R}}}
\newcommand{\mtotP}{{m_{\rm tot}^{\mathcal P}}}
\newcommand{\mtoto}{{m_{\rm tot}^{\rm o}}}
\newcommand{\mtote}{{m_{\rm tot}^{\rm e}}}
\newcommand{\ntotB}{n_{\rm tot}^{{\mathcal B}}}

\newcommand{\mmax}{m_{\rm max}}
\newcommand{\mmaxR}{{m_{\rm max}^{\mathcal R}}}
\newcommand{\mmaxP}{{m_{\rm max}^{\mathcal P}}}
\newcommand{\Rmax}{R_{\rm max}}
\newcommand{\RmaxR}{{R_{\rm max}^{\mathcal R}}}
\newcommand{\RmaxP}{{R_{\rm max}^{\mathcal P}}}
\newcommand{\mmaxB}{{m_{\rm max}^{\mathcal B}}}
\newcommand{\nR}{{n_{\mathcal R}}}
\newcommand{\nP}{{n_{\mathcal P}}}

\begin{document}
	\title{Chaotic percolation in the random geometry of maximum-density dimer packings}
	\author{Ritesh Bhola}
	\affiliation{\small{Department of Theoretical Physics, Tata Institute of Fundamental Research, Mumbai 400005, India}}
	\author{Kedar Damle}
	\affiliation{\small{Department of Theoretical Physics, Tata Institute of Fundamental Research, Mumbai 400005, India}}
	\begin{abstract}

Maximum-density dimer packings (maximum matchings) of non-bipartite site-diluted lattices, such as the triangular and Shastry-Sutherland lattices in $d=2$ dimensions and the stacked-triangular and corner-sharing octahedral lattices in $d=3$,  generically exhibit a nonzero density of monomers (unmatched vertices). 
Following a construction in the recent literature, we use the structure theory of Gallai and
Edmonds to decompose the disordered lattice into ``${\mathcal R}$-type'' regions which host the monomers of any maximum matching, and  perfectly matched ``${\mathcal P}$-type'' regions from 
which such monomers are excluded. When the density $n_v$ of quenched vacancies lies well within the low-$n_v$ geometrically percolated phase of the disordered lattice, we find that the random geometry of these regions 
exhibits unusual {\em Gallai-Edmonds percolation} phenomena.
In $d=2$, we find two phases separated by a critical point, namely a phase in which all $\calR$-type and $\calP$-type regions are small, and a percolated phase that displays a striking lack of 
self-averaging in the thermodynamic limit: Each sample has a single percolating region which is of type 
${\mathcal R}$  with probability $f_{\calR}$ and type $\calP$ with probability $1-f_{\calR}$, where $f_{\calR} \approx 0.50(2)$ is independent of $n_v$ (away from the critical region). In this regime, microscopic changes in the vacancy configuration lead to chaotic changes in the large-scale structure of $\calR$-type and 
$\calP$-type regions.
In $d=3$, apart from a phase with small ${\calR}$-type and ${\calP}$-type regions, the thermodynamic limit exhibits {\em four} distinct percolated phases separated by critical points at successively lower $n_v$.  In the first such phase, each sample has one percolating ${\mathcal R}$-type region and no such ${\mathcal P}$-type region. In the second, each sample simultaneously has one percolating region of each type (${\mathcal R}$ and ${\mathcal P}$). 
The third phase exhibits a different violation of self-averaging in the thermodynamic limit: $\calR$-type regions percolate with unit probability, while $\calP$-type regions percolate with probability $1-f_{\calR} \approx 0.50(3)$ away from the phase boundaries. 
Finally, the lowest-$n_v$ phase is identical in character to the unusual percolated phase found in $d=2$, again with $f_{\calR} \approx 0.50(2)$ away from the transition region.  We argue that this unusual behavior is expected to have an observable effect on the Majorana fermion contribution to the heat conductance in the disordered vortex lattice state of a class of two-dimensional topological superconductors. In addition, it is expected to lead to observable effects in the susceptibility of disordered short-range resonating valence bond spin liquid states of frustrated magnets, and disordered Majorana spin liquids.

	\end{abstract}

	\maketitle
	
	\section{Introduction and Overview}
	\label{sec:IntroductionOverview}
	Geometric percolation phenomena~\cite{Broadbent_Hammersley_1957,Stauffer_Aharony_1992,DuminilCopin_2019review}, associated with sharply defined thresholds in the end-to-end connectivity of a random medium, have been and remain of interest in multiple disciplines. For instance, they are of interest in the physical sciences and in engineering as they represent a relatively simple and striking example of the effects of quenched micro-scale disorder  on large-scale properties, such as the conductance through a medium or the elasticity of a material~\cite{Isichenko_1992,Sahimi_Hunt_2021,Sahimi_2023}. Viewed from the vantage point of statistical mechanics and phase transition theory, such percolation transitions provide perhaps the simplest and most intuitively appealing {\em geometric} examples of critical phenomena and associated universal scaling behavior~\cite{Cardy_1996,Stauffer_Aharony_1992,Stauffer_1979,Langlands_1994,Cardy_2001_sci,
Fortuin_Kasteleyn_1969,Fortuin_Kasteleyn_1972,Wu_1978,Swendsen_Wang_1987,Chayes_1997,
Bouabci_2000,Graham_Grimmett_2006,Grimmett_book2006}. Lattice models of 
such percolation transitions and generalizations such as first-passage percolation have also been of sustained interest in probability theory~\cite{Grimmett_book2018,Bollobas_Riordan_book,
Kesten_CommMathPhys1987,Kesten_1987,DuminilCopin_2020}.
				\begin{figure*}
			\begin{tabular}{cc}
		\includegraphics[width=\columnwidth]{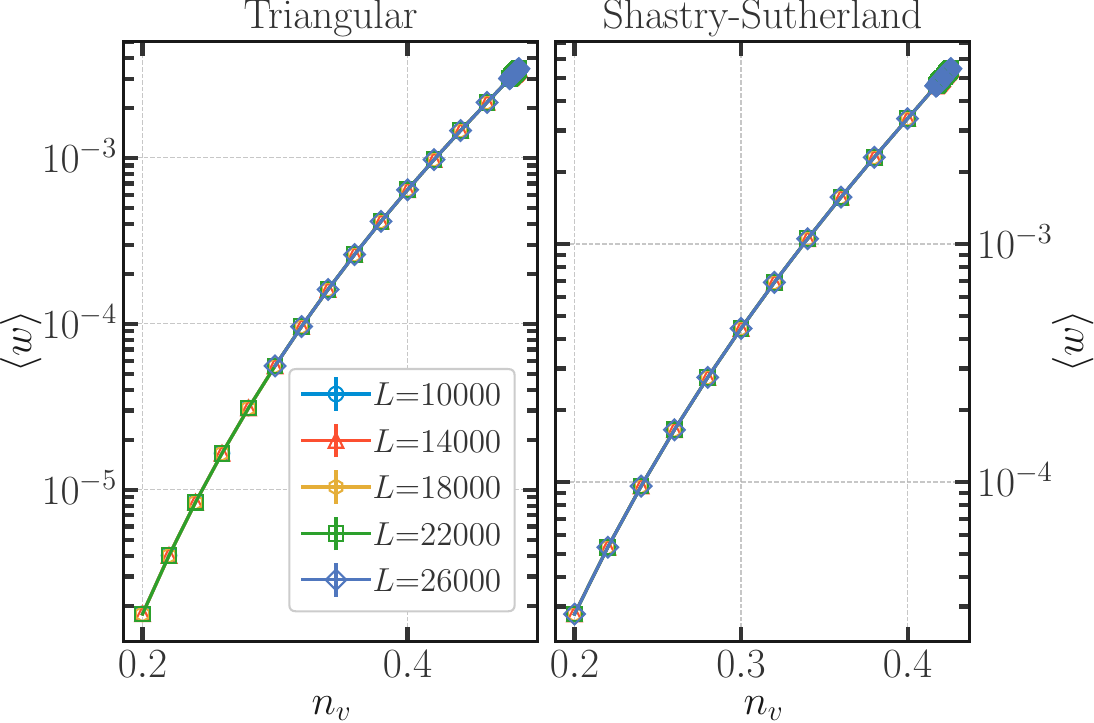}	&
			\includegraphics[width=\columnwidth]{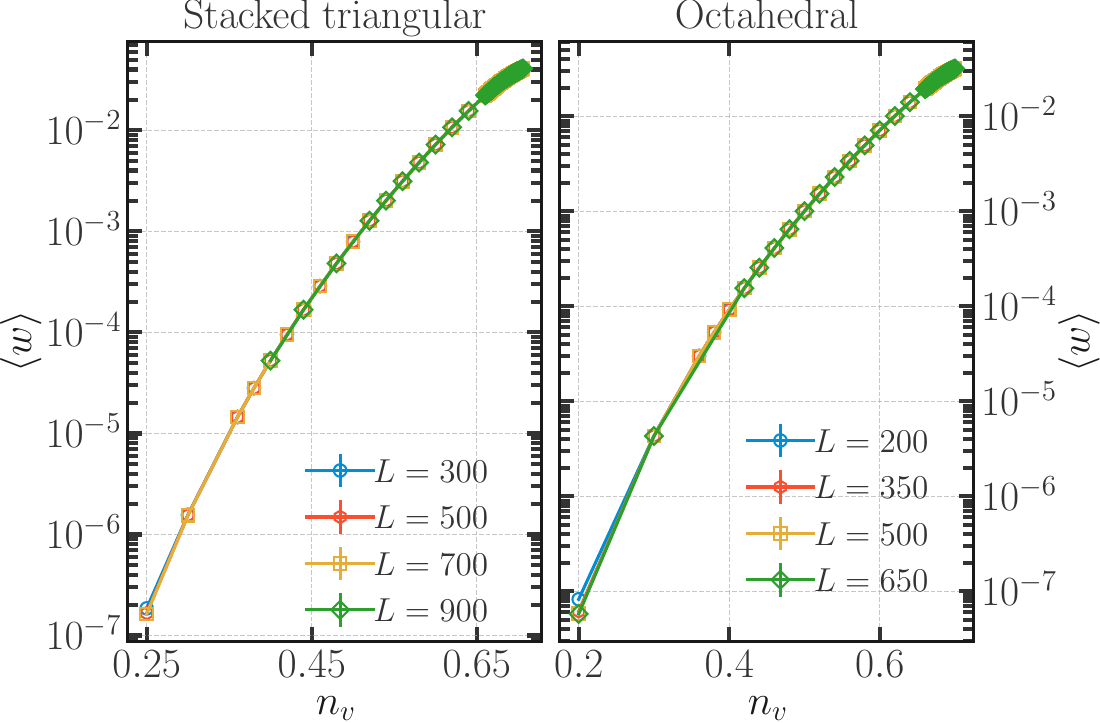} \\	
			\end{tabular}
		\caption{The sample-averaged density of monomers $\langle w \rangle$ in maximum matchings of the largest geometric cluster $\calG$ of site-diluted triangular, Shastry-Sutherland, stacked triangular and corner-sharing octahedral lattices is nonzero in the thermodynamic limit. It decreases monotonically and smoothly with decreasing site-dilution $n_v$. We have checked that this density is self-averaging, in the sense that its histogram has a single narrow peak which sharpens with increasing $L$. See Sec.~\ref{sec:ComputationalMethodsObservables} and Sec.~\ref{sec:ComputationalResults} for details.\label{fig:w}}
	\end{figure*} 

Bernoulli percolation on Euclidean lattices, that is, threshold behavior of the end-to-end connectivity of a randomly site-diluted (or bond-diluted) Euclidean lattice such as the triangular or square lattice in two dimensions, and hypercubic lattices in higher dimensions~\cite{Stauffer_Aharony_1992,Bollobas_Riordan_book}, is perhaps the simplest and most studied model for such a percolation transition. In such Bernoulli percolation models, the dilution (deletion of sites or bonds) is completely uncorrelated, so that each site (or bond) is either present with probability $1-n_{v}$ or deleted with probability $n_{v}$ independent of the other sites (or bonds). 
In two and higher dimensions, there is a transition at a nonzero $n_{v}^{\rm crit} < 1$ which separates two phases, a high-dilution phase in which the probability to have a percolating cluster in the thermodynamic limit is zero, and a low-dilution phase in which this probability is unity~\cite{Stauffer_Aharony_1992,Bollobas_Riordan_book,
Aizenman_Kesten_Newman_1987,Burton_Keane_1989,Grimmett_review2006}. 

In this well-studied Euclidean lattice setting, the percolating cluster responsible for maintaining end-to-end connectivity is a ``giant component'' of the disordered lattice, which contains (in the thermodynamic limit) a nonzero fraction of the sites of the undiluted lattice~\cite{Stauffer_Aharony_1992,Bollobas_Riordan_book}; one may therefore legitimately refer to this giant component as an ``infinite cluster''. More generally, in settings where there is no natural notion of end-to-end connectivity, such as various random graph ensembles~\cite{Bollobas_book2001,Erdos_Renyi_1959,Erdos_Renyi_1960,Erdos_Renyi_1961a,Erdos_Renyi_1961b}, there is still a sharply defined percolation transition having to do with the appearance of such a giant component or infinite cluster.

In Bernoulli percolation on Euclidean lattices, there is only one percolated phase, and there is exactly one such giant component of the lattice in this percolated phase~\cite{Stauffer_Aharony_1992,Bollobas_Riordan_book,
Aizenman_Kesten_Newman_1987,Burton_Keane_1989,Grimmett_review2006}.  However, on hyperbolic lattices such as hyperbolic tilings of the unit circle, even Bernoulli percolation has a more intricate phase diagram~\cite{Benjamini_Schramm_1996,Benjamini_Schramm_2000,
Baek_Minnhagen_Kim_2009,Gu_Ziff_2012,Mertens_Moore_2017,Grimmett_Li_2022}. Indeed, apart from the high-dilution phase, there are two percolated phases: an intermediate dilution phase with multiple giant components, and another percolated phase at lower dilution in which there is a single giant component. 

Variants of percolation with some correlations in the disorder ensemble have also been of interest both in statistical mechanics and at the interface of probability theory and geometry. These share many but not all properties of Bernoulli percolation.
A prominent example is Fortuin-Kastelyn percolation, which provides a geometric view of the spontaneous symmetry breaking phase transitions of spin systems in statistical mechanics~\cite{Fortuin_Kasteleyn_1969,Fortuin_Kasteleyn_1972,Wu_1978,Swendsen_Wang_1987,Chayes_1997,
Bouabci_2000,Graham_Grimmett_2006,
Grimmett_book2006,Grimmett_book2018}. The key idea is to map the phase transition to a percolation transition in  certain cluster representations of the underlying spin system. This is achieved by using an ensemble with bond dilution, in which the probability distribution of bond configurations incorporates correlations inherited from the equilibrium correlations of the corresponding spin system. Restated in algorithmic terms, this is nothing but the percolation transition (as a function of temperature) of the clusters constructed by the Swendsen-Wang Monte Carlo simulation algorithm for the Ising model, and its generalizations to other problems~\cite{Swendsen_Wang_1987}. 

Another such variant, motivated by the morphology of growing interfaces as well as the statistical mechanics of random fields,  has to do with the statistics of local heights that parameterize a Gaussian random surface~\cite{Lebowitz_Saleur_1986,Bricmont_Lebowitz_Maes_1987,Rodriguez_Sznitman_2013,Cao_Santachiara_2021}. Here, the focus is on the part of the surface that lies at a height  of {\em at least}  $+\delta$ above the mean height of the surface as a whole.  In general, this part of the surface breaks up into multiple connected components. The geometry of these connected components depends strongly on the long-range height correlations characteristic of such Gaussian random surfaces. At large $\delta$, these components are of course all small in size since the corresponding fluctuations of height are rare. As $\delta$ is lowered, a giant component first makes its appearance at a sharply defined threshold that marks the location of an interesting percolation transition whose character depends on the correlations of the Gaussian height field~\cite{Lebowitz_Saleur_1986,Bricmont_Lebowitz_Maes_1987,Rodriguez_Sznitman_2013,Cao_Santachiara_2021}. 

Other examples of percolation, somewhat different in flavour, are associated with the nontrivial {\em dynamics} of evolving networks~\cite{Barabasi_Albert_1999,Barabasi_2000,Albert_Barabasi_2002}. 
For instance, in an Achlioptas process~\cite{Achlioptas_DSouza_Spencer_2009,daCosta_Dorogovtsev_Goltsev_Mendes_2010,
Grassberger_Christensen_Bizhani_Son_Paczuski_2011,Riordan_Warnke_2011,Riordan_Warnke_2011,
Riordan_Warnke_AnnApplProb2012,DSouza_Gardenes_Nagler_Arenas_2019}, one starts with an initial condition in which all the allowed links of a network are deleted. At each time step, a predetermined number of candidate links are randomly drawn from among the allowed links, and the vertices at the ends of these candidates examined. One of these candidate links is then chosen to be added, with this choice depending in a predetermined way on the sizes of the connected components to which each of these vertices belong. Such a {\em delayed choice protocol} introduces interesting correlations into the dynamics. As the system evolves, it hits a well-defined threshold at which a giant component first appears. For many choices of protocol in which the delayed choice serves to postpone the formation of a giant component, such a giant component appears very suddenly at the threshold. This somewhat unusual threshold behavior has therefore been dubbed ``explosive percolation'~\cite{Achlioptas_DSouza_Spencer_2009}, although, strictly speaking, the transition is continuous, albeit with unusual finite-size scaling~\cite{daCosta_Dorogovtsev_Goltsev_Mendes_2010,
Grassberger_Christensen_Bizhani_Son_Paczuski_2011,Riordan_Warnke_2011,Riordan_Warnke_2011,
Riordan_Warnke_AnnApplProb2012,DSouza_Gardenes_Nagler_Arenas_2019}.

Bootstrap percolation and its variants, motivated by questions as diverse as the physics of magnetic ordering and the dynamics of influence spreading in social networks, is yet another well-studied example with a similar dynamical flavour~\cite{Pollak_Reiss_1975,Chalupa_Leath_Reich_1979,Kogut_Leath_1981,
Adler_Aharony_1988,Adler_1991,Schonmann_1992,Sabhapandit_Dhar_Shukla_2002,Kempe_Kleinberg_Tardos_2003,
Baxter_Dorogovtsev_Goltsev_Mendes_2010}. This involves a different dynamical process in which one starts with an initial condition that corresponds to a randomly diluted lattice or network with some site-dilution probability $n_v$ for a site to be initially ``inactive''. Starting with this random initial condition, one sweeps through the lattice carrying out a culling procedure to delete or de-activate sites whose coordination number is smaller than a preset fixed threshold that defines the bootstrap dynamics. A complementary formulation is in terms of a growth process starting from an initial condition with dilution probability $1-n_v$ (of initially inactive sites), whereby one sweeps repeatedly through the lattice and activates previously inactive sites that have a certain minimum number of active neighbors. In either formulation, the end product of the culling or growth process is a final state that does not change further. The random geometry of this final state has an interesting percolation transition as a function of the dilution $n_v$ that defines the ensemble of initial conditions.  
\begin{figure}
	{ \includegraphics[width=\columnwidth]{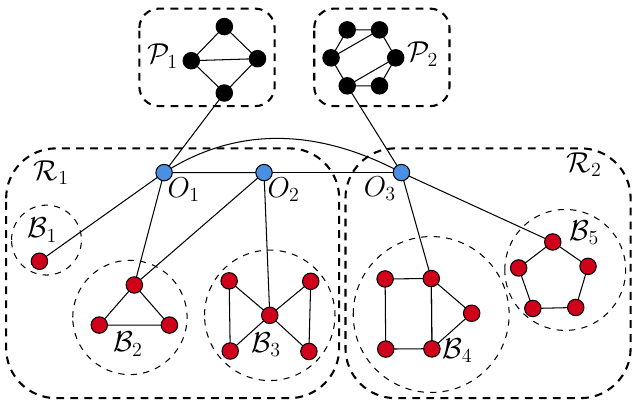}}
	\caption{A single site is the simplest ``blossom'' $\calB_1$. The next simplest blossom $\calB_2$ is of size three and can be constructed by attaching an ``ear'' of length three to a single site. $\calB_3$, $\calB_4$, and $\calB_5$ are blossoms of size five, respectively formed by attaching an ear of length three to a single site of $\calB_2$ or by attaching an ear of length three that starts and ends at two distinct sites of $\calB_2$, or by attaching an ear of length five to $\calB_1$. The Gallai-Edmonds theorem provides a well-defined and unique decomposition of the vertices of any graph into three groups, e-type vertices (colored red) that form blossoms and can host a monomer in some maximum matching, o-type vertices (colored blue) that are always perfectly matched in any maximum matching but have at least one e-type neighbor, and u-type vertices (colored black) that are matched to other u-type vertices in any maximum matching. The $\calP$-type regions defined in this work represent connected components of the subgraph defined by u-type vertices. The $\calR$-type regions defined in this work are the connected components of the subgraph comprising e-type and o-type vertices, in which all links between any two o-type vertices have been deleted. The motivation for and significance of this definition of $\calR$-type and $\calP$-type regions is discussed in Sec.~\ref{sec:GETheoryImplications}.	\label{fig:Blossoms}}
\end{figure}

Here, we identify and study some unusual percolation phenomena associated with the 
classic combinatorial problem of {\em maximum matchings}~\cite{Lovasz_Plummer_1986} of site-diluted lattices, or, equivalently, the classical statistical mechanics of {\em maximum-density dimer packings} on such lattices. Here, dimers are hard rods that live on  links of the lattice and touch the two sites at either end of the link. Two sites are said to be matched to each other if the intervening link hosts a dimer. The hard-core constraint on dimers requires that no site is touched simultaneously by more than one dimer; equivalently, a site can be matched with at most one of its neighbors. 

A maximum-density dimer packing is a dimer configuration with the maximum (subject to this hard-core constraint) possible number of dimers  on links of the lattice,  so that the largest possible number of sites are touched by dimers. The corresponding maximum matching therefore has the fewest number of unmatched sites. Each such unmatched site is said to host a monomer. If the maximum matchings have no monomers at all, they are said to be perfect matchings, and the corresponding dimer covers are said to be fully packed; these constitute the configuration space of fully-packed dimer models that have been extensively studied in statistical mechanics~\cite{Temperley_Fisher_1961,Fisher_1961,Kasteleyn_1961,Fisher_Stephenson_1963,Fisher_1966,
Youngblood_Axe_1980,Youngblood_Axe_McCoy_1980,Samuel_1980a,Samuel_1980b,Samuel_1980c,Henley_1997,
Fendley_Moessner_Sondhi_2002,Moessner_Sondhi_2003,Huse_Krauth_Moessner_Sondhi_2003,
Alet_Ikhlef_Jacobsen_Gregoire_2006,Papanikolaou_Luijten_Fradkin_2007,Desai_Pujari_Damle_2021,
Morita_Lee_Damle_Kawashima_2023} and probability theory~\cite{Kenyon_2000,Kenyon_2001,Kenyon_Okounkov_Sheffield_2006,Kenyon_2014,Nash_OConner_2017}, and are of interest in multiple contexts in condensed matter physics~\cite{Anderson_1973,Fazekas_Anderson_1974,Rokhsar_Kivelson_1988,
Fradkin_Huse_Moessner_Oganesyan_Sondhi_2004,Vishwanath_Balents_Senthil_2004,
Castelnovo_Chamon_Mudry_Pujol_2005,
Mambrini_2008,Henley_2010,Albuquerque_Alet_2010,Tang_Sandvik_Henley_2011,
Albuquerque_Alet_Moessner_2012,Damle_Dhar_Ramola_2012,Patil_Dasgupta_Damle_2014}. 
	\begin{figure}
		\begin{tabular}{cc}
			\includegraphics[width=0.45\columnwidth]{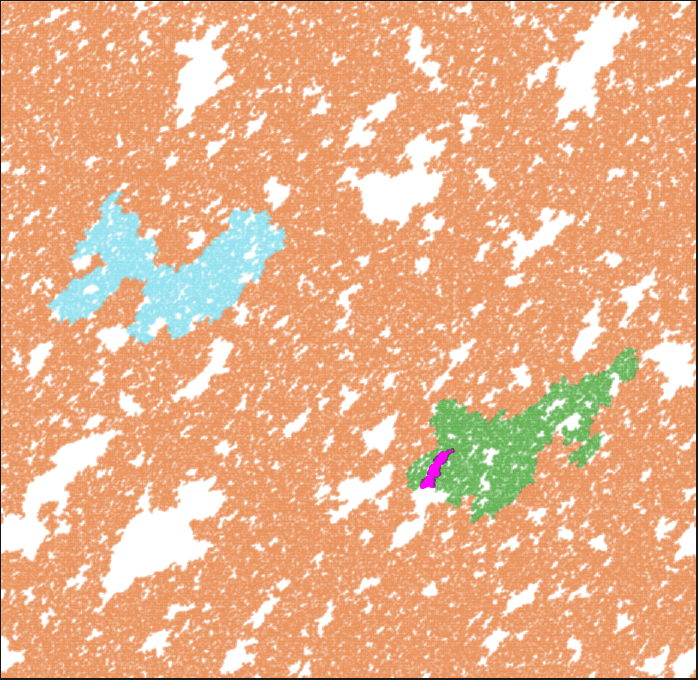} &
			\includegraphics[width=0.45\columnwidth]{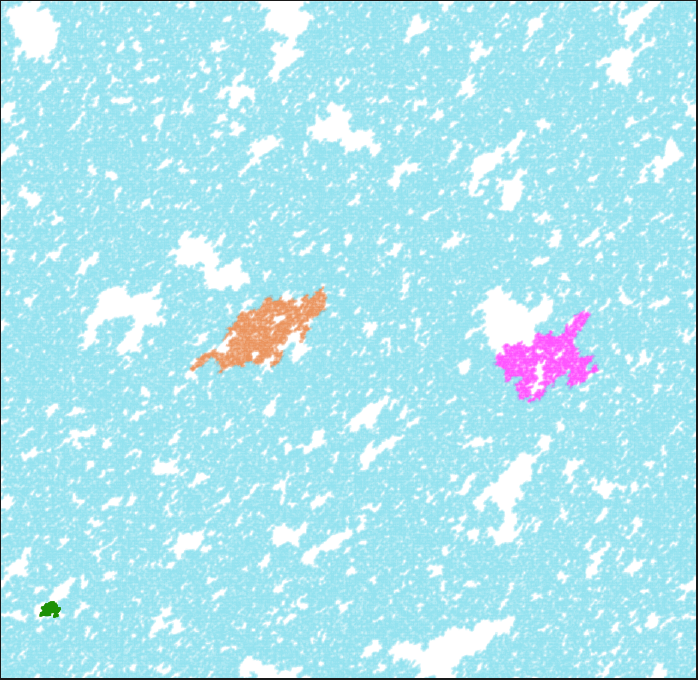}\\
		\end{tabular}
		\caption{The large-scale morphology of $\calR$-type and $\calP$-type regions of two identically-prepared disordered samples of a $L=5000$ diluted triangular lattice exhibits macroscopic differences. Both samples shown here were constructed using the same protocol for randomly deleting sites of the triangular lattice with probability $n_v = 0.48$. In each sample, the vertices belonging to the largest and second-largest $\calR$-type region are colored brown and pink respectively, those belonging to largest and second-largest $\calP$-type region are colored cyan and green respectively, and those not belonging to any of these four regions are left uncolored. The sample on the left has one very large $\calR$-type region that appears to almost span the whole sample. In comparison, the largest two $\calP$-type regions are both small. The second-largest $\calR$-type region is even smaller. The sample on the right has completely different morphology, with the largest $\calP$-type region almost spanning the whole lattice, the largest and second-largest $\calR$-type regions being rather small, and the second-largest $\calP$-type region being the smallest of the four regions displayed. See  Sec.~\ref{sec:GETheoryImplications} and Sec.~\ref{sec:ComputationalResults} for details.\label{fig:RtypePtypePictures2D}}		
	\end{figure}			

Generically, maximum matchings of disordered lattices tend to have a nonzero density of monomers (see, {\em e.g.}, Fig.~\ref{fig:w}) unless special features of the lattice~\cite{Faudree_1997} or the disorder ensemble guarantee the existence of perfect~\cite{Sumner_1974} or near-perfect matchings~\cite{Ansari_Damle_2024,Bhola_Damle_ClawFree2024}.
As we review in Sec.~\ref{sec:GETheoryImplications}, this density of monomers is predicted to be confined to certain sharply demarcated regions of the lattice~\cite{Lovasz_Plummer_1986,Gallai_1964,Edmonds_1965,Damle_2022}, which have been dubbed ${\mathcal R}$-type regions in previous work~\cite{Damle_2022}. The rest of the lattice can likewise be decomposed into connected ${\mathcal P}$-type regions whose sites are always perfectly matched (so that no monomers can live in these ${\mathcal P}$-type regions in any maximum matching)~\cite{Lovasz_Plummer_1986,Gallai_1964,Edmonds_1965,Damle_2022}. 
	
This decomposition of the lattice into ${\mathcal R}$-type and ${\mathcal P}$-type regions~\cite{Damle_2022} can be obtained using a computationally efficient method which relies on the structure theory of arbitrary graphs due to Gallai and Edmonds~\cite{Lovasz_Plummer_1986,Gallai_1964,Edmonds_1965}. Fig.~\ref{fig:Blossoms} shows a schematic representation of the predictions of this theory (which is reviewed in Sec.~\ref{sec:GETheoryImplications}). The percolation phenomena studied here have to do with the spatial extent of these ${\mathcal R}$-type and ${\mathcal P}$-type regions. These percolation phenomena occur relatively deep in the {\em geometrically percolated phase} of the diluted lattices we study, and as such, are not connected in any way with the well-studied geometric percolation transitions of these lattices.

Given that this setting is so far removed from any of the various percolation phenomena that have been previously of interest in the large literature on this subject, the reader may at this point legitimately wonder what led us in the first place to a study of the random geometry of these ${\mathcal R}$-type and ${\mathcal P}$-type regions? The answer is that the morphology of these regions is interesting and significant in multiple contexts. We digress here to elaborate on this before proceeding any further, since this digression helps provide the context for our study. 

 Part of our motivation had to do with the fact that the random geometry of these ${\mathcal R}$-type regions controls the spatial extent of topologically protected collective Majorana modes of Majorana networks,  whose vertices represent localized Majorana modes and edges correspond to bilinear mixing amplitudes between them~\cite{Affleck_Rahmani_Pikulin_2017,Li_Franz_2018}. Such network Hamiltonians are interesting because they provide a unified framework for describing the low energy physics of several interesting systems. Examples include the Kitaev model and its many generalizations~\cite{Kitaev_2006,Yao_Kivelson_2007,Yao_Lee_2011,Lai_Motrunich_2011,
 Chua_Yao_Fiete_2011,Fu_2019}, SU(2) symmetric Majorana spin liquids~\cite{Biswas_Fu_Laumann_Sachdev_2011,Sanyal_Damle_Chalker_Moessner_2021}, and superconductors with strong spin-orbit scattering~\cite{Senthil_Fisher_2000,Motrunich_Damle_Huse_2001}; in all these systems, the bulk low-energy excitations admit a description in terms of Majorana operators. Other examples include Majorana excitations associated with pinned vortices in the mixed state of topological superconductors~\cite{Read_Green_2000,Ivanov_2001,Laumann_Ludwig_Huse_Trebst_2012,Kraus_Stern_2011,Biswas_2013}, as well as proposed platforms for topological quantum computing using such Majorana modes associated with defects or edges~\cite{Kitaev_2001,Nayak_Simon_Stern_Freedman_2008,Sau_Tewari_Lutchyn_Stanescu_2010,
Potter_Lee_2010,Alicea_Oreg_Rafael_Fisher_2011,Stern_Linder_2013,McGinley_Knolle_Nunnenkamp_2017}. This is reviewed briefly in Sec.~\ref{sec:GETheoryImplications}, along with additional discussion in Sec.~\ref{sec:DiscussionandOutlook} of the relevance of our results in some of these contexts.

Another part of our motivation came from at-first-sight unrelated questions about the spatial form-factor of vacancy-induced emergent local moments~\cite{Ansari_Damle_2024} in short ranged resonating valence bond spin liquid states~\cite{Anderson_1973,Fazekas_Anderson_1974,Rokhsar_Kivelson_1988,
Moessner_Sondhi_RVB_2001,Moessner_Sondhi_Fradkin_2001,Albuquerque_Alet_2010,Tang_Sandvik_Henley_2011,
Damle_Dhar_Ramola_2012,Wildeboer_Seidel_2012,Patil_Dasgupta_Damle_2014,
Wildeboer_Seidel_Melko_2017} of frustrated antiferromagnets on non-bipartite lattices. It turns out that the existence of such vacancy-induced local moments is associated with the presence of monomers in maximum matchings of the diluted lattice~\cite{Ansari_Damle_2024}. Moreover, the spatial profile of these emergent local moments is controlled by the geometry of ${\mathcal R}$-type regions, and is expected to be crucial in determining the actual ground state of the system whenever such short range RVB states are rendered unstable by a vacancy-induced local moment instability~\cite{Ansari_Damle_2024} of this kind. Additionally, although the mechanism is quite different and involves the topologically protected collective Majorana modes alluded to earlier, the morphology of $\calR$-type regions is also expected to control the low temperature physics of vacancy induced local moments in Majorana spin liquids. In Sec.~\ref{sec:GETheoryImplications} and Sec.~\ref{sec:DiscussionandOutlook}, we provide a brief discussion of possible implications of our results in the context of such local moment instabilities.

Finally, there is also another more direct motivation for the present study. This has to do with previous work~\cite{Bhola_Biswas_Islam_Damle_2022} that identified interesting percolation phenomena exhibited by the random geometry of monomer-carrying $\calR$-type regions of site-diluted {\em bipartite} lattices. Such bipartite lattices admit a decomposition into two sublattices $A$ and $B$, with vertices belonging to the $A$ sublattice only having nearest neighbors that belong to the $B$ sublattice and vice versa. In this bipartite case, it is {\em local sublattice imbalance}, that is, an imbalance in the {\em local density} of $A$-sublattice sites relative to $B$-sublattice sites, that is responsible for maximum matchings having a nonzero density of monomers~\cite{Bhola_Biswas_Islam_Damle_2022}. As a result, ${\mathcal R}$-type regions are associated with regions of the lattice with such imbalance and come in two flavors, $\calR_A$-type regions with a local excess of $A$-sublattice sites, and $\calR_B$-type regions with a local excess of $B$-sublattice sites~\cite{Bhola_Biswas_Islam_Damle_2022}. Clearly, there is no analog of this in the non-bipartite case. Indeed, in the non-bipartite case, it turns out that monomer-carrying $\calR$-type regions are instead composed of motifs  built out of odd cycles in the nearest neighbor connectivity of the disordered lattice~\cite{Lovasz_Plummer_1986}; these are the ``blossoms'' described in Fig.~\ref{fig:Blossoms}. Given these key differences, it is very natural to ask just how they affect the random geometry of ${\calR}$-type and $\calP$-type regions in this general nonbipartite case. We discuss our results from this point of view in Sec.~\ref{sec:DiscussionandOutlook}.

In our computations, we study maximum matchings of the site-diluted triangular and Shastry-Sutherland lattices in two dimensions, as well as the site-diluted stacked-triangular and corner-sharing octahedral lattices in three dimensions. These disordered lattices provide us an appropriate setting for our study since they are {\em generic} in the sense alluded to earlier, {\em i.e.}, maximum matchings of these diluted lattices exhibit a nonzero density of monomers in the bulk for any nonzero dilution; this is confirmed by our results for the monomer density displayed in Fig.~\ref{fig:w}.  Our focus is the random geometry of ${\mathcal R}$-type and ${\mathcal P}$-type regions of these diluted lattices, and its evolution as a function of the density $n_v$ of vacancies. Our study uncovers rather unusual {\em Gallai-Edmonds percolation} phenomena deep in the low-$n_v$ geometrically percolated phase of the disordered lattice. Although the critical behavior in the vicinity of these Gallai-Edmonds percolation transitions falls  within the usual scaling framework for critical points~\cite{Stauffer_Aharony_1992,Cardy_1996}, the percolated phases themselves are very unusual in character.

Specifically, on the two-dimensional triangular and Shastry-Sutherland lattices, we find two phases separated by a critical point: a phase in which all $\calR$-type and $\calP$-type clusters are small, and a low-$n_v$ percolated phase. This low-$n_v$ phase is very unusual since it displays  a striking lack of self-averaging  in the thermodynamic limit: 
Each sample has a single percolating region which is of type 
$\calR$  with probability  $f_{\calR}$, and type $\calP$ with probability $1-f_{\calR}$, where $f_\calR \approx 0.50(2)$ (within our statistical errors) {\em independent of} $n_v$ away from transition region, {\em including  deep in the phase at very small $n_v$}. Fig.~\ref{fig:RtypePtypePictures2D}, which contrasts the morphology of two samples with identical dilution $n_v$ in this phase, provides a rather striking visual representation of this anomalous behavior. In the vicinity of the Gallai-Edmonds percolation threshold, we find universal scaling behavior that fits into the standard scaling framework, but is characterized by exponents that are appreciably different from the corresponding exponents of the usual geometric percolation transition~\cite{Stauffer_Aharony_1992,Nijs_1979,Nienhuis_Riedel_Schick_1980,Pearson_1980}
 in two dimensions.	
	
On the three-dimensional stacked triangular and corner-sharing octahedral lattices,
 apart from a phase with small ${\calR}$-type and ${\calP}$-type clusters, the thermodynamic limit is characterized by {\em four distinct} percolated phases separated by critical points at successively lower $n_v$.  In the first such phase, each sample has one percolating ${\mathcal R}$-type region and no such ${\mathcal P}$-type region. In the second, each sample simultaneously has one percolating region of each type (${\mathcal R}$ and ${\mathcal P}$).
 The third phase exhibits another interesting violation of self-averaging in the thermodynamics limit, different from the behavior in the percolated phase in two dimensions: $\calR$-type regions percolate with unit probability while $\calP$ type regions percolate with probability $1-f_{\calR} \approx 0.50(3)$ within our statistical errors throughout this (admittedly narrow) phase, except in the close vicinity of the phase transitions that define its extent. 
Finally, the lowest-$n_v$ phase is identical in character to the unusual percolated phase found in $d=2$, again with $f_\calR \approx 0.50(2)$ {\em independent of} $n_v$ within our statistical errors  (away from the transition region), including deep in the phase. As in the two-dimensional case described above, the various transitions between these phases exhibit universal scaling behavior that fits into the standard framework of critical scaling. Further, in contrast to the two-dimensional case, the exponents themselves are also consistent, within our statistical errors, with the corresponding exponents of the usual geometric percolation transition~\cite{Stauffer_Aharony_1992,Wang_Zhou_Zhang_Garoni_2013} in three dimensions.


Since geometric percolation phenomena 
of various kinds have been the focus of many decades of work in the physical sciences as well as in probability theory~\cite{Broadbent_Hammersley_1957,Stauffer_Aharony_1992,DuminilCopin_2019review,
Isichenko_1992,Sahimi_Hunt_2021,Sahimi_2023}, it is interesting to 
pause here and ask where and how the Gallai-Edmonds percolation phenomena identified here fit into this body of 
knowledge.
With this in mind, we devote the remainder of this introduction to a discussion that contrasts our results with the well-known results on geometric percolation and its classical variants. In addition, we contrast the phenomena identified here with the corresponding monomer percolation phenomena studied recently~\cite{Bhola_Biswas_Islam_Damle_2022} on bipartite lattices.
		
To this end, we first note that since the underlying quenched disorder is completely uncorrelated, it is tempting to view the rather unusual lack of self-averaging identified here from the vantage point of Bernoulli percolation ({\em i.e.}, geometric percolation with uncorrelated disorder).
When viewed in this way, our percolated phase in two dimensions {\em appears} to violate the Kolmogorov zero-one law that applies to Bernoulli percolation, whereby the probability of having an infinite cluster in the thermodynamic limit cannot take on any value other than zero or one at any dilution~\cite{DuminilCopin_2019review,Bollobas_Riordan_book}. In addition, the probability $P^{\calP}$ ($P^{\calR}$) for $\calP$-type ($\calR$-type) regions to percolate in $d=3$ exhibits non-monotonic behavior as a function of $n_v$, which is again something that is ruled out in Bernoulli percolation~\cite{DuminilCopin_2019review,Bollobas_Riordan_book}.

However, such an analogy misses an important feature of Gallai-Edmonds percolation: Although the disorder is indeed uncorrelated, the ${\mathcal R}$ or  ${\mathcal P}$ labels of the vertices are certainly not uncorrelated. In fact, since these labels are determined by the structure of maximum matchings of the disordered lattice~\cite{Lovasz_Plummer_1986,Damle_2022}, their correlations are related in a complicated way to the monomer and dimer correlation functions of maximum-density dimer packings on the disordered lattice. 
	
	Another instructive comparison is with Dulmage-Mendelsohn percolation~\cite{Bhola_Biswas_Islam_Damle_2022}, {\em i.e.} percolation phenomena of $\calR$-type regions of site-diluted {\em bipartite} lattices. As mentioned earlier, these are constructed~\cite{Bhola_Biswas_Islam_Damle_2022} using Dulmage and Mendelsohn's structure theory~\cite{Lovasz_Plummer_1986,Dulmage_Mendelsohn_1958,Pothen_Fan_1990,
Irving_Telikepalli_Mehlhorn_Michail_2006} and come in two flavors, $\calR_A$-type regions in which all the monomers live only on $A$-sublattice sites of the lattice, and $\calR_B$-type regions in which all the monomers live only on $B$-sublattice sites of the lattice. 	In two dimensions with site dilution, there seems to be no percolated phase. Instead, there is an incipient percolation transition of $\calR$-type regions as $n_v \to 0$, causing their size to grow as $n_v^{-\nu}$  in this limit; here $\nu \approx 5.1$ is the estimated universal correlation length exponent characterizing this incipient criticality~\cite{Bhola_Biswas_Islam_Damle_2022}. 

In contrast, on the bipartite three-dimensional cubic lattice, there are two percolated phases at successively lower values of $n_v$, with all $\calP$-type clusters being small in both phases~\cite{Bhola_Biswas_Islam_Damle_2022}. In the first of these phases, each sample has two infinite clusters in the thermodynamic limit, one $\calR_A$-type and the other $\calR_B$-type. In the second of these phases, at still lower dilutions, each sample has only one infinite $\calR$-type cluster in the thermodynamic limit, and this cluster spontaneously breaks the statistical bipartite symmetry of the random lattice by being either $\calR_A$-type or $\calR_B$-type with equal probability~\cite{Bhola_Biswas_Islam_Damle_2022}.  

Thus, this phase of three-dimensional bipartite lattices also exhibits something akin to a lack of self-averaging in the thermodynamic limit. 
But this lack of self-averaging in the bipartite case is of a more familiar kind that is associated with all instances of spontaneous symmetry breaking in the thermodynamic limit of many-body systems.  Indeed, one can plausibly make an analogy between this phase and the Fortuin-Kasteleyn cluster representation of the Z$_2$ symmetry-breaking low-temperature phase of the usual nearest neighbor ferromagnetic Ising model~\cite{Fortuin_Kasteleyn_1969,Fortuin_Kasteleyn_1972,Wu_1978,Swendsen_Wang_1987,Chayes_1997,
Bouabci_2000,Graham_Grimmett_2006,Grimmett_book2006}. In this representation of the low temperature spontaneously magnetized phase, there is exactly one infinite cluster, but the spin label of this infinite cluster can be either $+1$ or $-1$ with equal probability. Here, the only (slightly) subtle aspect of this analogy is that the symmetry being spontaneously broken in Dulmage-Mendelsohn percolation is a {\em statistical} bipartite symmetry of the ensemble of diluted lattices.

In sharp contrast to this bipartite case, the behavior identified here in our study of Gallai-Edmonds percolation is qualitatively different from the behavior in the cluster representation of the usual Ising model, since there is no (statistical) symmetry between ${\mathcal R}$-type and ${\mathcal P}$-type clusters at the {\em microscopic} level.  This makes it even more puzzling that $f_{\calR} \approx 0.50(2)$ {\em independent of $n_v$} within our statistical errors deep in the lowest-$n_v$ percolated phases in both two and three dimensions, since the simplest explanation for this value of $f_\calR$ would be some kind of emergent symmetry between $\calR$-type and $\calP$-type regions. 
Of course, one might also wonder if the analogy with Fortuin-Kasteleyn percolation could be stretched further to accommodate the phenomena seen here, perhaps by considering the cluster representation of the classical spin $S=1$ Ising model in which the Ising spins take on three values $\sigma  = \pm 1, 0$~\cite{Chayes_1997,
Bouabci_2000,Graham_Grimmett_2006,Grimmett_book2006}. However, even in the cluster representation of the classical spin $S=1$ Ising model, there is no phase in which the infinite cluster has a nonzero probability for being in the $\sigma=0$ state in addition to nonzero probabilities for being in the  $\sigma = \pm 1$ states.

The Achlioptas processes described earlier provide another interesting comparision. In some of these processes that correspond to delayed choice protocols of a very specific kind, the percolated phase does have anomalously large run-to-run fluctuations that survive the thermodynamic limit~\cite{Riordan_Warnke_2012}. Although these do result in macroscopically significant fluctuations in the {\em size} of the giant component, they do not change the {\em fact} of its existence. Clearly, this is not quite the same thing as one sample having a percolating $\calP$-type region and another nominally identical sample not having one.
From the perspective of percolation theory, these comparisons with previous results on percolation of $\calR$-type regions of site-diluted bipartite lattices, with the usual expectations for Bernoulli percolation and Fortuin-Kasteleyn percolation, with the unusually large run-to-run fluctuations in some Achlioptas processes thus underscore challenging questions about how these unusual behaviors fit into the existing and well-developed theoretical understanding of percolation.

Indeed, since two nominally identical  samples can be in two different thermodynamic phases in the low-dilution regimes identified here, this failure of self-averaging in the thermodynamic limit is also qualitatively different from the mere existence of broad distributions of various observables in other random systems such as random antiferromagnetic spin chains~\cite{Fisher_PRB1,Fisher_PRB2}. Although such broad distributions do lead to differences between typical and average behavior of some physical observables, there is no {\em dichotomy} of behavior in such random systems. In other random systems such as spin glass models with all-to-all couplings~\cite{SpinGlassReview}, there is a sense in which there are multiple macroscopically distinct free energy minima. However, even in this case, there is no sense in which two nominally identical samples belong to different thermodynamic phases unless each free energy minimum is viewed as a different phase of matter. Thus,  this aspect of the Gallai-Edmonds percolation phenomena identified here truly appears to be without a known parallel.

The fact that two nominally-identical samples prepared using the same protocol can be in two entirely different phases also suggests that the phase in which a sample finds itself depends extremely sensitively on the microscopic vacancy configuration in a sample. We have explored this aspect for one case, that of the diluted triangular lattice in the percolated phase of Gallai-Edmonds regions, by studying a simple and illustrative dynamics for the vacancies, and found that very small microscopic changes in the vacancy configuration cause deterministic yet unpredictable, that is, chaotic, switching of a sample between these two qualitatively different macroscopic behaviors. This aspect is discussed further in Sec.~\ref{subsec:Chaos}.

What about experimental implications? The most dramatic consequence has to do with the heat transported by Majorana excitations associated with cores of vortices in the disordered triangular vortex lattice state of a class of topological superconductors~\cite{Read_Green_2000,Ivanov_2001,Laumann_Ludwig_Huse_Trebst_2012,Kraus_Stern_2011,Biswas_2013}. In this context, the lack of thermodynamic self-averaging of the large-scale geometry (in the percolated phase on the diluted triangular lattice) is expected to lead to observable $O(1)$ sample-to-sample fluctuations of the heat conductance if the vortex lattice has a small density of vacancies (missing vortices) in addition to positional disorder.  

Another implication has to do with the fate of short-range resonating valence bond spin liquids~\cite{Anderson_1973,Fazekas_Anderson_1974,Rokhsar_Kivelson_1988,
Moessner_Sondhi_RVB_2001,Moessner_Sondhi_Fradkin_2001,Albuquerque_Alet_2010,Tang_Sandvik_Henley_2011,
Damle_Dhar_Ramola_2012,Wildeboer_Seidel_2012,Patil_Dasgupta_Damle_2014,
Wildeboer_Seidel_Melko_2017} in the presence of nonmagnetic impurities, which lead to a vacancy-induced local moment instability~\cite{Ansari_Damle_2024} of such states. Here, the unusual nature of the small-$n_v$ percolated phase is expected to reflect itself directly in anomalous (macroscopic) sample-to-sample fluctuations in the magnetic susceptibility at low temperatures, well-below the scale at which the vacancy-induced local moments are formed. Similarly, weak vacancy disorder is expected to lead to observable effects in $SU(2)$ symmetric Majorana spin liquids~\cite{Biswas_Fu_Laumann_Sachdev_2011,Sanyal_Damle_Chalker_Moessner_2021} as well as non-bipartite lattice analogues of Kitaev magnets~\cite{Kitaev_2006,Yao_Kivelson_2007,Yao_Lee_2011,Lai_Motrunich_2011,
 Chua_Yao_Fiete_2011,Fu_2019}.  Note that these effects all owe their existence to the unusual nature of the entire percolated phase, and are therefore unaffected by the difficulty involved in fine-tuning the vacancy density to access the percolation transition in any particular experimental realization. A more detailed discussion of these aspects of our results is in Sec.~\ref{sec:GETheoryImplications} and Sec.~\ref{sec:DiscussionandOutlook}.

The remainder of the article is devoted to a more detailed presentation of our results and is organized as follows:
In Sec.~\ref{sec:GETheoryImplications}, we provide a summary of the structure theory of  Gallai and Edmonds~\cite{Lovasz_Plummer_1986,Gallai_1964,Edmonds_1965} and review how it can be used to decompose the lattice into a complete set of non-overlapping ${\mathcal R}$-type and ${\mathcal P}$-type regions~\cite{Damle_2022}. We also discuss the implications of this construction for the statistical mechanics of maximum-density dimer packings on such diluted lattices, provide a summary of recent work that uses this theory to identify regions of the graph that host topologically protected collective Majorana modes of the corresponding Majorana network Hamiltonian, and review the recently established connection between monomers of maximum matchings and vacancy-induced emergent local moments in short-range resonating valence bond spin liquid phases of frustrated magnets~\cite{Ansari_Damle_2024}. Sec.~\ref{sec:ComputationalMethodsObservables} describes our computational methods and introduces the various observables we track in our study.
Sec.~\ref{sec:ComputationalResults} is devoted to the results of our detailed computational study of the random geometry of $\calR$-type and $\calP$-type regions in spatial dimensions $d=2$ and $d=3$. We conclude in Sec.~\ref{sec:DiscussionandOutlook}  with a discussion of the implications of our work, as well as interesting directions for future work.

	\section{Gallai-Edmonds theory and physical implications}
	\label{sec:GETheoryImplications}
	The Gallai-Edmonds theorem~\cite{Lovasz_Plummer_1986,Gallai_1964,Edmonds_1965} provides useful information on the structure of a disordered lattice. This information is obtained from {\em invariant} aspects of the structure of maximum matchings (maximum-density dimer packings) of the lattice, {\em i.e.}, properties of the {\em ensemble} of maximum matchings. For us, its chief utility lies in the fact that it can be used to formulate a prescription~\cite{Damle_2022} for decomposing the lattice into a complete and non-overlapping set of ``${\mathcal R}$-type'' regions guaranteed to host the monomers of any maximum matching,  and ``${\mathcal P}$-type'' regions whose sites are guaranteed to be perfectly matched by any maximum matching. 
	
	This decomposition into ${\mathcal R}$-type and ${\mathcal P}$-type regions was originally motivated~\cite{Damle_2022} by the fact that these ${\mathcal R}$-type regions can be shown to host topologically protected collective Majorana modes of the corresponding Majorana network Hamiltonian. Below we provide a quick review of this theoretical background. We also review recent work~\cite{Ansari_Damle_2024} that establishes a link between vacancy-induced local moment instabilities of resonating valence bond states on frustrated lattices such as the triangular lattice, and the existence of a nonzero density of monomers in maximum matchings of the corresponding diluted lattice. In addition, we review how the collective Majorana modes alluded to above lead to dominant contributions to the low-temperature magnetic response of Majorana spin liquids~\cite{Sanyal_Damle_Chalker_Moessner_2021}; thus, the morphology of $\calR$-type regions is also expected to control this low-temperature behavior. Finally, we argue that the random geometry of $\calR$ and $\calP$ type regions has interesting implications for the statistical mechanics of maximum-density dimer packings and the corresponding quantum dimer models on disordered lattices. 

\subsection{Gallai-Edmonds theory}
\label{subsec:GETheory}
The key idea is to start with any one maximum matching (maximum-density dimer packing) of the disordered lattice, obtained in practice using, say,  Edmonds' ``Blossoms'' algorithm~\cite{Edmonds_1965}, and split the sites of the lattice into three groups. To do this, one has to think in terms of the forest of {\em alternating paths}~\cite{Lovasz_Plummer_1986} that start from each unmatched site (monomer) in this particular maximum matching. Starting from a monomer, such an alternating path traverses links that alternate between  being unmatched (unoccupied by a dimer) and being matched (occupied by a dimer) in this maximum matching. Sites of the disordered lattice are now labeled unreachable (u-type), odd (o-type) or even (e-type) based on the following criterion: If a site cannot be reached by any such alternating path starting from one of the monomers, it is labeled u-type. If a site can be reached in an even number of steps along some alternating path starting from some monomer, it is labeled e-type. All other sites are labeled o-type---these o-type sites can thus be reached starting from some monomer along some alternating path, but all such routes to an o-type site necessarily involve an {\em odd} number of steps. 

The theorem of Gallai and Edmonds~\cite{Lovasz_Plummer_1986,Gallai_1964,Edmonds_1965} now guarantees that this labeling is unique, {\em i.e.}, it is independent of the maximum matching one uses to assign these labels. This is an easy consequence of the fact that one is using a maximum matching to classify the sites. In other words, the label of a site is a statement about the full ensemble of maximum matchings, and provides some structural information about the disordered lattice itself. For instance, it is quite straightforward to prove that an e-type site can never be adjacent to any u-type site. Other immediate consequences of these definitions include the fact that monomers of any maximum matching can only live on e-type sites, and that each u-type site is necessarily matched with one of its u-type neighbors in any maximum matching. In addition, it is easy to see that each o-type site is always matched with one of its e-type neighbors in any maximum matching. 
		
The real utility of this classification stems from somewhat deeper (and less obvious) consequences of this classification of sites, which we turn to next. These consequences have to do with the structure of the subgraph formed by e-type sites, as well as properties of the nearest-neighbor links between o-type and e-type vertices. To formulate and understand these consequences, it is useful to imagine deleting all edges (nearest-neighbor links) between e-type and o-type vertices. This would cause the e-type vertices to organize into distinct connected components.  Clearly, two different connected components obtained in this way have no links between them, so that each such connected component only has external links connecting it to some o-type sites.

Crucially, it can be shown that each such connected component contains an odd number of vertices. In the remainder of this paper we slur over the (unimportant for our purposes) minor distinction between the ``blossoms'' that appear in most descriptions of Edmonds' matching algorithm~\cite{Edmonds_1965,Lovasz_Plummer_1986,Moret_Shapiro_1991,Tarjan_1983,Kececioglu_1998} and these odd-cardinality components (which are, in a certain fairly precise sense, comprised of {\em nested} blossoms), and use ``odd-cardinality components'' and ``blossoms'' interchangeably to always refer to the connected components defined by the edge-deletion protocol given here. Thus, each blossom, as defined here, is one such connected subgraph made up of an odd number of  e-type sites and the nearest-neighbor links between them. 

The structure theory of Gallai and Edmonds~\cite{Gallai_1964,Edmonds_1965,Lovasz_Plummer_1986} now guarantees that these blossoms have a very specific structure: The simplest (trivial) blossom is a single site. All other blossoms are either an odd cycle, or built from an odd cycle by attaching odd-length ``ears''. The simplest nontrivial blossom is thus a single triangle, which is the smallest odd cycle; it can also be thought of as being built from a single site by adding an ``ear'' comprising an odd-length path that starts and ends at a single site. Larger blossoms have a similar structure, that is, they are built up by attaching more odd-length ears, including attaching such odd-length ears onto previously attached ears.  Some examples of blossoms are shown in Fig.~\ref{fig:Blossoms}, which also provides a schematic depiction of the structural information provided by the Gallai-Edmonds decomposition, as well as a depiction of our construction of $\calR$-type and $\calP$-type regions (that we review in Sec.~\ref{subsec:RtypePtypeConstruction}). 

Crucially, the Gallai-Edmonds theorem guarantees~\cite{Gallai_1964,Edmonds_1965,Lovasz_Plummer_1986} that each such blossom is, in itself, a {\em factor critical graph}, which means it has a fully-packed dimer cover (perfect matching) if we delete {\em any one site} from it. Another implication of this theory~\cite{Gallai_1964,Edmonds_1965,Lovasz_Plummer_1986} is the following crucial structural property of the disordered lattice as a whole: Any arbitrarily chosen subset of o-type sites is always ``outnumbered'' by the distinct blossoms whose e-type sites have nearest-neighbor links to  this subset of o-type sites. This is also illustrated by the schematic representation provided in Fig.~\ref{fig:Blossoms}. 

This structural information has interesting implications for the location of monomers of any maximum matching of the disordered lattice. For instance, the number of monomers in any given blossom can only be zero or one in any maximum matching. 
Indeed, a blossom hosts a single monomer in a maximum matching only if no o-type site is matched in that maximum matching to any site of the blossom. On the other hand, if a maximum matching places a dimer connecting an o-type site to one of the sites of a blossom, then that blossom is perfectly matched by that maximum matching. 

Another related fact is that two o-type sites can never be matched to e-type sites in the same blossom in any maximum matching. In other words, a given blossom either has no dimers connecting any of its sites to any o-type site, or it has exactly one dimer connecting one of its sites to one o-type site. Finally, one can always find a maximum matching that places a monomer on any chosen e-type site of any blossom.


\subsection{Construction of ${\mathcal R}$-type and $\calP$-type regions}
\label{subsec:RtypePtypeConstruction}
We now review the construction~\cite{Damle_2022} of $\calR$-type regions and their complements, the $\calP$-type regions, from the site labels (u-type, e-type or o-type) obtained from Gallai-Edmonds theory. Consider deleting all links between u-type sites and o-type sites to obtain one graph ${\mathcal U}$ comprised of only the u-type sites and the nearest-neighbor links between them, and another graph ${\mathcal E}{\mathcal O}$ that contains all the o-type and e-type sites and all the edges connecting any two such sites in the original lattice. 

The connected components of ${\mathcal U}$ give us the ${\mathcal P}$-type regions. 
To construct the ${\mathcal R}$-type regions, we start with the graph ${\mathcal E}{\mathcal O}$ and delete all the edges that connect any o-type site to another o-type site. In addition, we ``shrink the blossoms'' as follows: All the e-type vertices that belong to a given blossom are replaced by a single super-vertex that inherits all the links that originally connected these e-type sites to o-type sites in ${\mathcal E}{\mathcal O}$ (multiple links to the same o-type vertex are collapsed into a single link in this process). 

Once this is done, we are left with an auxiliary bipartite graph ${\mathcal A}$ which has two sublattices, one comprising all the blossom super-vertices, and the other comprising all the o-type vertices. Next, we decompose ${\mathcal A}$ into its connected components. Finally, to obtain the ${\mathcal R}$-type regions of the original lattice from these connected components, we expand the blossoms to recover their constituent $e$-type sites and reinstate all the original edges incident on these $e$-type vertices (see Fig.~\ref{fig:Blossoms}). Each ${\mathcal R}$-type region identified in this way is thus comprised of some number $N_{{\mathcal B}}({\mathcal R})$ of blossoms (odd-cardinality components)  containing only e-type sites and links between them, and some number $N_{\rm o}({\mathcal R})$ of o-type sites  to which these blossoms are connected (as noted earlier, the original disordered lattice is guaranteed not to have any direct links between two distinct blossoms). 

From the discussion in the previous section, it is clear that $\calR$-type regions defined in this way have the following crucial property: The Gallai-Edmonds theorem guarantees that the {\em imbalance}, defined as $\calI(\calR) \equiv N_{\mathcal B}({\mathcal R}) - N_{\rm o}({\mathcal R})$, is positive for any $\calR$-type region, and that the blossoms of each $\calR$-type region together host a total of exactly $\calI (\calR)$ monomers in any maximum matching of the graph.   Additionally, in any maximum matching, it is clear that each $\calP$-type region defined in this way always hosts a perfect matching, in which each u-type vertex inside a $\calP$-type region is matched to another vertex inside the same region. 

What, apart from these properties, motivates our choice of construction? There are essentially two independent reasons: On the one hand, for the linear algebra of collective Majorana modes of the corresponding disordered Majorana network, it turns out that these $\calR$-type regions host a complete set of irreducible linear subsystems that one needs to analyze in order to construct these collective modes~\cite{Damle_2022}; we review this in Sec.~\ref{subsec:CollectiveMajoranaModes}. On the other hand, the classical or quantum dynamics of maximum-density dimer packings on such disordered lattices is also completely determined by the physics within these $\calR$-type and $\calP$-type regions, in the strong sense that the partition function of these systems factorizes into a product of partition functions corresponding to each individual $\calR$-type and $\calP$-type region; this is reviewed in Sec.~\ref{subsec:MaximallyPackedDimerModels}. Additionally, for the physics of vacancy-induced local moment formation in short-range resonating valence bond (RVB) spin liquid states, the random geometry of this decomposition into $\calR$-type and $\calP$-type regions is expected to determine the spatial form factor of vacancy-induced local moments and the nature of the interaction between them; a quick summary of this is in Sec.~\ref{subsec:VacancyInducedMoments}.
 
\subsection{Collective Majorana modes}
\label{subsec:CollectiveMajoranaModes}
As noted earlier in Sec.~\ref{sec:IntroductionOverview}, the bulk low-energy excitations of many condensed matter systems are best described as a system of Majorana fermions. Examples include the Kitaev model and its many generalizations~\cite{Kitaev_2006,Yao_Kivelson_2007,Yao_Lee_2011,Lai_Motrunich_2011,
 Chua_Yao_Fiete_2011,Fu_2019}, SU(2) symmetric Majorana spin liquids~\cite{Biswas_Fu_Laumann_Sachdev_2011}, and superconductors with strong spin-orbit scattering~\cite{Senthil_Fisher_2000,Motrunich_Damle_Huse_2001}. Other systems exhibit zero energy Majorana modes associated with sufficiently isolated defects or boundaries; examples include pinned vortices in the mixed state of topological superconductors~\cite{Read_Green_2000,Ivanov_2001,Laumann_Ludwig_Huse_Trebst_2012,Kraus_Stern_2011,Biswas_2013} and defects in Kitaev-like magnets~\cite{Udagawa_2018,Sanyal_Damle_Chalker_Moessner_2021}. Such modes~\cite{Kitaev_2001} have been of interest in the context of proposed platforms for topological quantum computing~\cite{Nayak_Simon_Stern_Freedman_2008,Sau_Tewari_Lutchyn_Stanescu_2010,
Potter_Lee_2010,Alicea_Oreg_Rafael_Fisher_2011,Stern_Linder_2013,McGinley_Knolle_Nunnenkamp_2017}. Their presence is usually taken to be a signature of the non-trivial topology of the underlying many body ground state~\cite{Wen_book2004}. When there are many such Majorana modes associated with defects, or when the bulk excitations admit a description in terms of Majorana fermions, it is natural to model the resulting low energy physics in terms of a {\em Majorana network}~\cite{Laumann_Ludwig_Huse_Trebst_2012,Kraus_Stern_2011,Biswas_2013,Affleck_Rahmani_Pikulin_2017,Li_Franz_2018}, {\em i.e.}, a Hamiltonian of the following generic form: 
	\begin{eqnarray}
		H_{\rm Majorana} &=& i\sum_{\langle r r' \rangle} a_{r r'} \eta_{r} \eta_{r'}\; .
		\label{eq:HMajorana}
	\end{eqnarray}
	Here $H$ is defined on a graph whose vertices $r$ host  Majorana operators $\eta_r$ and edges $\langle r r' \rangle$  correspond to purely imaginary antisymmetric mixing amplitudes $ia_{rr'}$ between these modes.  Do such networks as a whole exhibit collective zero-energy Majorana excitations that are robust to small perturbations of the network? The answer to this can have important implications in multiple physical contexts.  

For instance,  such collective Majorana modes are associated with emergent local moments and low-temperature Curie tails in the susceptibility of SU(2) symmetric Majorana spin liquids~\cite{Sanyal_Damle_Chalker_Moessner_2021}. 
Another context in which this question is important is the physics of a class of two-dimensional topological superconductors belonging to the so-called symmetry class D. In the presence of disorder, the vortex excitations in these systems are expected to form a disordered (typically triangular) lattice, with each vortex core associated with a Majorana mode~\cite{Read_Green_2000,Ivanov_2001,Biswas_2013,Laumann_Ludwig_Huse_Trebst_2012,Kraus_Stern_2011}. The mixing of these Majorana modes is again described by the Majorana network Hamiltonian Eq.~\ref{eq:HMajorana}, with $a_{r r'}$ representing the mixing amplitude for modes on neighboring vortices. Since the ``chemical potential'' is pinned to zero in these systems, the contribution of these Majorana modes to thermal transport in such a system depends crucially on the presence of collective zero energy Majorana modes of this network Hamiltonian. At a somewhat more speculative level, in the context of various proposed platforms~\cite{Nayak_Simon_Stern_Freedman_2008,Sau_Tewari_Lutchyn_Stanescu_2010,
Potter_Lee_2010,Alicea_Oreg_Rafael_Fisher_2011,Stern_Linder_2013,McGinley_Knolle_Nunnenkamp_2017} for Majorana quantum computing, such collective Majorana modes that arise due to the mixing of localized Majorana modes could themselves potentially serve as a resource for quantum computing.

	It is perhaps fortuitous that there is a surprisingly precise answer to this question about collective zero energy Majorana modes:
A Majorana network has collective topologically protected zero-energy Majorana excitations whenever the maximum matchings of the corresponding graph have unmatched vertices~\cite{Tutte_1947,Lovasz_Plummer_1986,Damle_2022}, as is typically the case if the graph is disordered. Indeed, one can say much more: The number of linearly independent collective Majorana modes of this type is exactly equal to the number of monomers in any maximum-density dimer packing of the graph~\cite{Damle_2022,Lovasz_1979,Anderson_1975}. 

	Recent work \cite{Damle_2022} has also used the Gallai-Edmonds theory~\cite{Gallai_1964,Edmonds_1965,Lovasz_Plummer_1986} to characterise the spatial profile of these collective Majorana modes by demonstrating that they live in the monomer-carrying ${\mathcal R}$-type regions defined in Sec.~\ref{subsec:RtypePtypeConstruction}; indeed, this is one of the motivations for the definition of $\calR$-type regions given there. This may be viewed as a generalization to the non-bipartite case of the previously-established connection between topologically protected zero energy states in the spectrum of the tight-binding model on disordered bipartite lattices and such monomer-carrying ${\mathcal R}$-type regions of a bipartite lattice~\cite{Bhola_Biswas_Islam_Damle_2022}. The basic result of Ref.~\cite{Damle_2022} is that each such ${\mathcal R}$-type region hosts exactly ${\mathcal I}({\mathcal R}) = N_{{\mathcal B}}({\mathcal R}) - N_{\rm o}({\mathcal R})$  linearly independent collective topologically protected Majorana modes of the corresponding Majorana network, thereby providing a local version of the global result of Lovasz~\cite{Lovasz_1979} and Anderson~\cite{Anderson_1975} (as noted in Sec.~\ref{subsec:RtypePtypeConstruction}, this imbalance ${\mathcal I}$ is guaranteed to be greater than zero for any $\calR$-type region).

	These modes have wavefunctions that are supported on all the e-type sites of the ${\mathcal R}$-type region. One way to see why this is the case is as follows: If we restrict the network Hamiltonian to an individual blossom, the matrix of mixing amplitudes is a pure imaginary antisymmetric matrix of  odd dimension. Such a matrix generically must have one topologically protected zero eigenvector. Thus, each blossom, if disconnected from the rest of the graph, hosts one topologically protected collective Majorana mode. Many of these modes will mix with each other and be destroyed when the nonzero mixing amplitudes connecting the blossoms to the o-type vertices are reinstated. However, one can look for linear combinations that survive this mixing. To do this, one has to impose $N_{\rm o}({\mathcal R})$ constraints on $N_{{\mathcal B}}({\mathcal R})$ variables, which allows $\calI(\calR)$ combinations to survive. 
	
This construction may be viewed as a procedure for obtaining a {\em maximally-localized basis} for the zero modes of the Majorana network Hamiltonian in Eq.~\ref{eq:HMajorana}. Having such a basis at our disposal is of course useful, since one can obtain information about basis independent quantities by evaluating them in this basis. An important example of such a basis independent quantity is the on-shell zero-energy Green function $\Delta G(r, r')$, defined as
\begin{eqnarray}
\Delta G(r,r') &=& \sum_z (\psi_z^{*}(r) \psi_z(r') + h.c)
\label{eq:DeltaG}
\end{eqnarray}
where the sum over $z$ runs over all zero modes, $\psi_z$ are the corresponding wavefunctions in any orthonormal basis, and $h.c.$ refers to the hermitean conjugate of the term displayed explicitly. The utility of our basis is now clear in this example: if all $\calR$-type regions are finite and small, we can evaluate Eq.~\ref{eq:DeltaG} in our maximally localized basis and deduce that $\Delta G(r, r')$ must vanish for large enough separations $|r-r'|$ independent of the actual details of the zero mode basis wavefunctions within each $\calR$-type region. This suggests that our results on the random geometry of $\calR$-type and $\calP$-type regions of diluted lattices are expected to have direct implications for such systems whose low-energy physics is described by the Majorana network Hamiltonian Eq.~\ref{eq:HMajorana}. We discuss this in some more detail in Sec.~\ref{subsec:PhysicalImplications}.

We close this review with comments about two special cases, namely the Majorana representation of the quasiparticle Hamiltonian of disordered superconductors with strong spin-orbit scattering (see {\em e.g.} Ref.~\cite{Senthil_Fisher_2000} and Ref.~\cite{Motrunich_Damle_Huse_2001}), and the special case of a bipartite Majorana network Hamiltonian, {\em i.e.}, when the graph corresponding to $H_{\rm Majorana}$ of Eq.~\ref{eq:HMajorana} has a bipartition of vertices into two classes $A$ and $B$, such that vertices belonging to $A$ only have edges connecting them to vertices belonging to $B$ and vice-versa. In the former case, one is rewriting the usual lattice-level mean-field description of such quasiparticles in terms of Majorana fermions, and this places additional restrictions on the network: each physical site corresponds to a pair of vertices of the network, giving rise to a bilayer network with strong constraints on the inter-layer connectivity. This also means that the natural disorder ensemble has strong correlations, in sharp contrast to the uncorrelated vacancy disorder that we focus on here.

In the special case of a bipartite network, the wavefunctions of topologically protected collective Majorana modes of the network can be constructed from  the zero modes of an equivalent tight-binding Hamiltonian with purely real hopping amplitudes on the same graph~\cite{Bhola_Biswas_Islam_Damle_2022}.  Such topologically protected zero modes of bipartite hopping models also have an interesting description in graph theoretical terms, constructed using the Dulmage-Mendelsohn decomposition of bipartite graphs. This has been explored in some detail in previous work~\cite{Bhola_Biswas_Islam_Damle_2022}, and our results can be viewed as a natural generalization of these ideas to the case of general non-bipartite Majorana networks. From this point of view, our results in the triangular lattice case make for an interesting contrast with the previous results on the square lattice. For one may  view the triangular lattice as a square lattice in which one has introduced additional nearest neighbor links corresponding to one set of diagonals. This point of view leads to some interesting  questions that are discussed briefly in Sec.~\ref{subsec:GallaiEdmondsPercolation}. 

\subsection{Maximum-density dimer packings and corresponding quantum dimer models}
\label{subsec:MaximallyPackedDimerModels}
As a result of the various structural guarantees provided by Gallai-Edmonds theory, the geometry of these ${\mathcal R}$-type and $\calP$-type regions also has very direct implications for the structure of dimer and monomer correlations of maximum-density dimer packings and the corresponding quantum dimer models on disordered lattices. 
To see this, first consider the overlap of two different maximum-density dimer configurations. This overlap can be represented as a configuration of overlap loops and open strings. Here, an open string starts at a monomer of one configuration and ends at a monomer of the other configuration, with all intermediate vertices being touched by a dimer in both configurations. On the other hand, all vertices on an overlap loop are touched by a dimer in both configurations. A ``trivial'' overlap loop of length two corresponds to a link that is occupied by a dimer in both maximum-density dimer configurations.  The links encountered while traversing any nontrivial overlap loop or open string are alternately occupied by dimers belonging to the first and second maximum-density dimer configuration. 

From the structural information provided by the Gallai-Edmonds theorem (see Sec.~\ref{subsec:GETheory}), we see that all such overlap loops and open strings live {\em entirely within a single $\calR$-type region or a single $\calP$-type region}. An immediate consequence of this is that the partition function of any such maximally-packed classical dimer model factorizes into a product of smaller partition functions, each of which is associated with a single $\calR$-type or  $\calP$-type region.   Interestingly, this remains valid in the presence of commonly-studied interaction terms that assign a potential energy to each occurrence of ``flippable loops'' of a specified type ({\em e.g.} the interactions considered in Refs.~\cite{Alet_Ikhlef_Jacobsen_Gregoire_2006,Papanikolaou_Luijten_Fradkin_2007,
Damle_Dhar_Ramola_2012,Patil_Dasgupta_Damle_2014,Sreejith_Powell_2014,Desai_Pujari_Damle_2021}), since all such flippable loops lie entirely within a single $\calR$-type or $\calP$-type region. In addition, since two e-type sites cannot be neighbors or share a common neighbor unless they belong to the same $\calR$-type region, this factorization also remains valid in the presence of inter-monomer interaction terms that do not extend beyond next-nearest neighbor.  Finally, since dimer kinetic energy terms associated with ring-exchange processes as well as monomer hopping terms are also constrained to act within a single $\calR$-type or $\calP$-type region, this factorization also remains valid for such maximally-packed quantum dimer models.
[In this paragraph, and in other such contexts that follow, we occasionally use ``maximally-packed'' as a convenient alternative to ``maximum-density'' and no distinction between the two, along the lines of the subtle distinction between ``maximal matchings'' and ``maximum matchings'' in the graph theoretic literature,  is being implied by this usage in this article.]

Thus, we see that the factorization established earlier for  the bipartite case~\cite{Bhola_Biswas_Islam_Damle_2022} remains valid in the general non-bipartite case considered here. As has been emphasized earlier~\cite{Bhola_Biswas_Islam_Damle_2022} in the context of such models on bipartite lattices, this factorized structure of the density matrix also implies that the entanglement entropy of any eigenstate of such a quantum dimer model obeys an area law if all $\calR$ and $\calP$ regions remain finite in extent when the thermodynamic limit of the disordered lattice is taken.  
Specifically, if the $\calR$-type and $\calP$-type regions have a typical (linear) size $\xi$ and if the eigenstates of the individual subsystem Hamiltonians of each $\calR$-type and $\calP$-type region have volume law entanglement in the middle of their respective energy spectrum, one expects the entanglement entropy of an arbitrary state in the middle of the many-body spectrum to scale as $\xi^{d} \times (L^{d-1}/\xi^{d-1}) \sim \xi L^{d-1}$ in $d$ dimensions.

From the vantage point of studies of many-body localization~\cite{Abanin_2019MBLReview,Alet_2018MBLReview} in disordered two-dimensional quantum dimer models~\cite{Pietracaprina_Alet_2021,Theveniaut_Lan_Meyer_Alet_2020}, as well as recent studies of Hilbert space fragmentation phenomena caused by kinematic constraints on the quantum dynamics~\cite{Moudgalya_2022FragReview},  we note that our arguments demonstrate that the Hilbert space of such maximally-packed quantum dimer models is {\em not} fragmented (even in the ``weak'' sense), nor are such quantum dimer models expected to display any of the other signatures of many-body localization. Notwithstanding this, states in the middle of their many-body spectrum {\em do have area-law entanglement}, and this is clearly due to the {\em kinematic constraints} imposed by the maximum matching condition operating in conjunction with the effects of quenched disorder.

Finally, we note that these conclusions are also significant in the context of the disorder effects in short-range resonating valence bond (RVB) spin liquid states of frustrated antiferromagnets on non-bipartite lattices, since the low-energy description of such states is in terms of such a maximally-packed quantum dimer model~\cite{Ansari_Damle_2024}. We discuss implications of our results in this context in the next section and in Sec.~\ref{subsec:PhysicalImplications}.

\subsection{Vacancy-induced local moments}
\label{subsec:VacancyInducedMoments}
In most magnetic insulators, the constituent spins order at low temperature. However, it has long been recognized~\cite{Anderson_1973,Fazekas_Anderson_1974,Rokhsar_Kivelson_1988} that an alternative scenario, whereby the ground state lacks spin or singlet order of any kind, would constitute a very interesting many-body state~\cite{Balents_Nature_2010,Savary_Balents_Reports_on_Progress_in_Physics_2016,
Norman_RevModPhys_2016,Zhou_Kanoda_Ng_RevModPhys_2017,
Broholm_Cava_Kivelson_Nocera_Norman_Senthil_Science_2020,
Clark_Abdeldaim_Annual_Review_of_Material_Research_2021}.  Anderson's original proposal~\cite{Anderson_1973,Fazekas_Anderson_1974} for such a spin-liquid state of the $S=1/2$ triangular lattice antiferromagnet was written as a singlet wavefunction constructed as a superposition of all possible nearest-neighbor singlet pairings of $S=1/2$ spins. This is in a class of short-range resonating valence bond (RVB) wavefunctions expected to exhibit $Z_2$ topological order~\cite{Wen_RevModPhys_2017} on nonbipartite lattices such as the triangular lattice~\cite{Moessner_Sondhi_RVB_2001}.
		  	\begin{figure*}
		\begin{tabular}{cc}
			a)\includegraphics[width=\columnwidth]{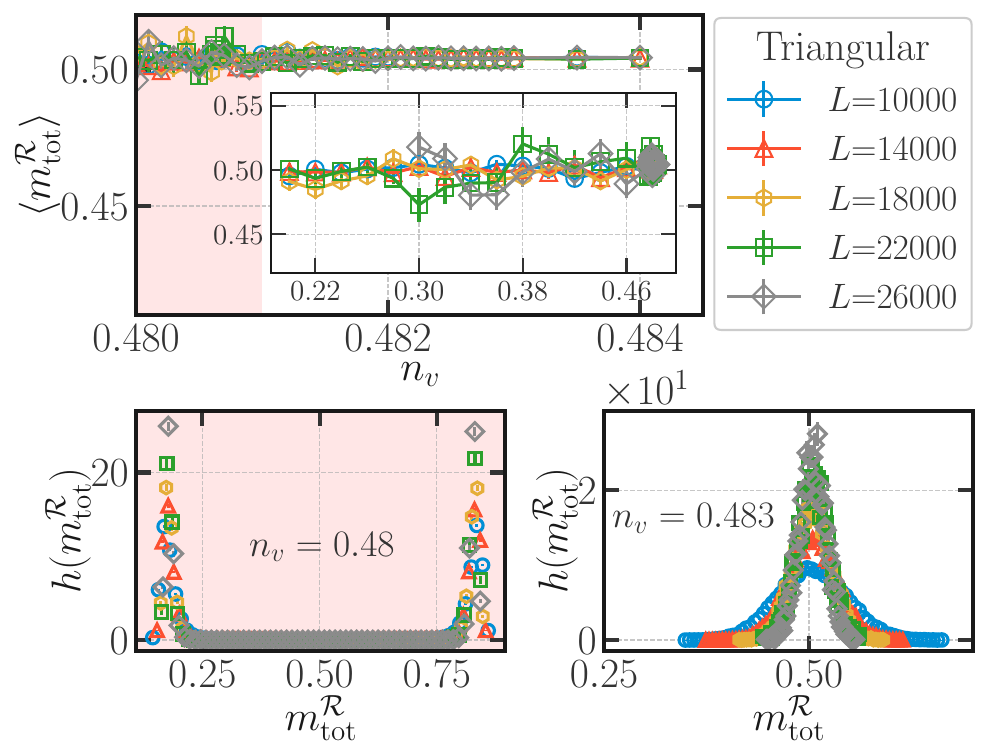}&	b)\includegraphics[width=\columnwidth]{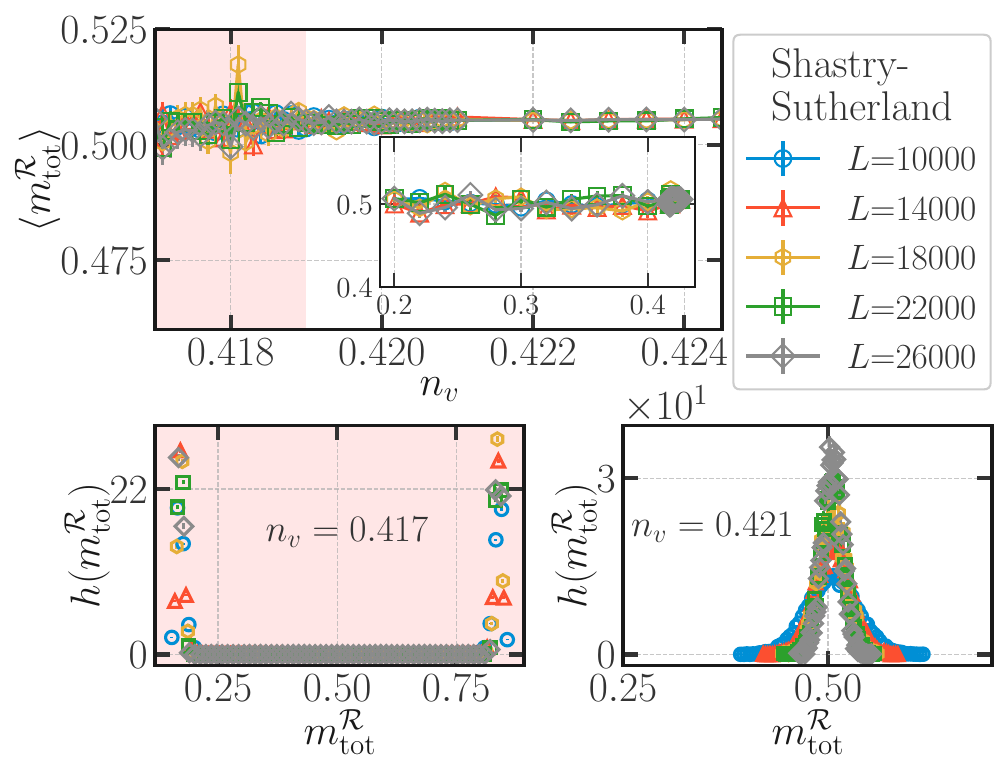}\\			
		\end{tabular}
		\caption{Top panels of (a) and (b): On both the triangular lattice and the Shastry-Sutherland lattice, the sample-averaged total mass density of ${\mathcal R}$-type regions, $\langle \mtotR \rangle$, is nonzero and size-independent in the large $L$ limit. However, it develops large sample-to-sample fluctuations below a threshold value of $n_v$. The bottom panels of (a) and (b) display the corresponding probability distribution $h(m^{\calR}_{\rm tot})$ at representative values of $n_v$ above and below this threshold. From these histograms, it is clear that the large sample to sample fluctuations seen below the threshold represent a bimodality in this probability distribution: $h(m^{\calR}_{\rm tot})$ has two well-defined peaks at macroscopically different values of $m_{\rm tot}^{\calR}$; these peaks become narrower with increasing size $L$ but do not shift in position. Below this threshold in $n_v$, we find that  $f_{\calR} \approx 0.50(2)$ independent of $n_v$, where $f_{\calR}$ is the weight in the right peak of this normalized distribution $h(m^{\calR}_{\rm tot})$. This bimodality should be contrasted with the single peak seen above the threshold, which narrows with increasing $L$ (as expected for a self-averaging thermodynamic density).  See Sec.~\ref{sec:ComputationalMethodsObservables} and Sec.~\ref{sec:ComputationalResults} for details.		\label{fig:mtotR2D}}
	\end{figure*}

In recent work~\cite{Ansari_Damle_2024} that used generalizations with $O(N)$ symmetry~\cite{Kaul_PRL_2015,Block_DEmidio_Kaul_PRB_2020} and a $1/N$ expansion framework~\cite{Read_Sachdev_1989} to analyze the effect of non-magnetic impurities, it was argued that a small density of non-magnetic impurities (which substitute for the magnetic ions and correspond to static vacancies in the spin model) is expected to give rise to a nonzero density of emergent local moments at intermediate energy scales in a short-range RVB state on the triangular lattice. 

Indeed, this is expected to be the case whenever the unperturbed ground state is of this short-range RVB type {\em and the maximum matchings of the diluted lattice have a nonzero density of monomers}~\cite{Ansari_Damle_2024}. 
When this is the case, the density of vacancy-induced local moments is set by this density of monomers, and the nature of the ground state in the presence of disorder is then controlled by the residual interactions between these local moments~\cite{Ansari_Damle_2024}. As a result, the original RVB liquid state is generically unstable to a small density of non-magnetic impurities. 

One way to think about this effect is as follows: In the pure case, the low-energy physics of a short-range RVB liquid state is expected to be accurately described by a quantum dimer model that acts within the subspace of nearest-neighbor singlet bonds and endows them with dynamics and interactions~\cite{Rokhsar_Kivelson_1988}. 
If the disordered lattice lacks perfect matchings, this description of the low-energy physics of short-range RVB liquids is modified in a fundamental way~\cite{Ansari_Damle_2024}: One now has a quantum monomer-dimer model that acts within the subspace of states represented by {\em maximum matchings of the disordered lattice}, with each dimer corresponding to a singlet bond as before, and each monomer of the maximum matching representing an emergent disorder-induced local moment. 

From this description, it becomes clear that the ground state of the disordered system is expected to change in a fundamental way whenever maximum matchings of the disordered lattice are characterized by a nonzero density of monomers.  When this is the case, it also stands to reason that the ultimate fate of the system at the lowest energies will depend sensitively on the spatial locations of these vacancy-induced local moments. Our study of the random geometry of $\calR$-type regions of site-diluted lattices is thus of direct relevance to the low-temperature physics of such frustrated magnets. In Sec.~\ref{subsec:PhysicalImplications}, we mention some interesting questions that arise in this connection.
 	\begin{figure*}
		\begin{tabular}{cc}
			a)\includegraphics[width=\columnwidth]{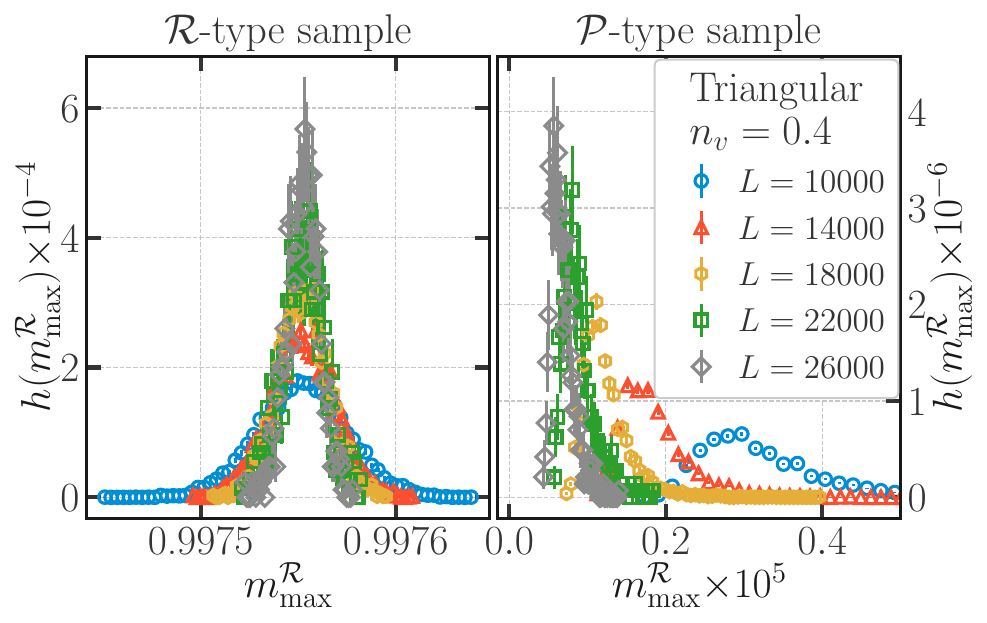} &	
			b)\includegraphics[width=\columnwidth]{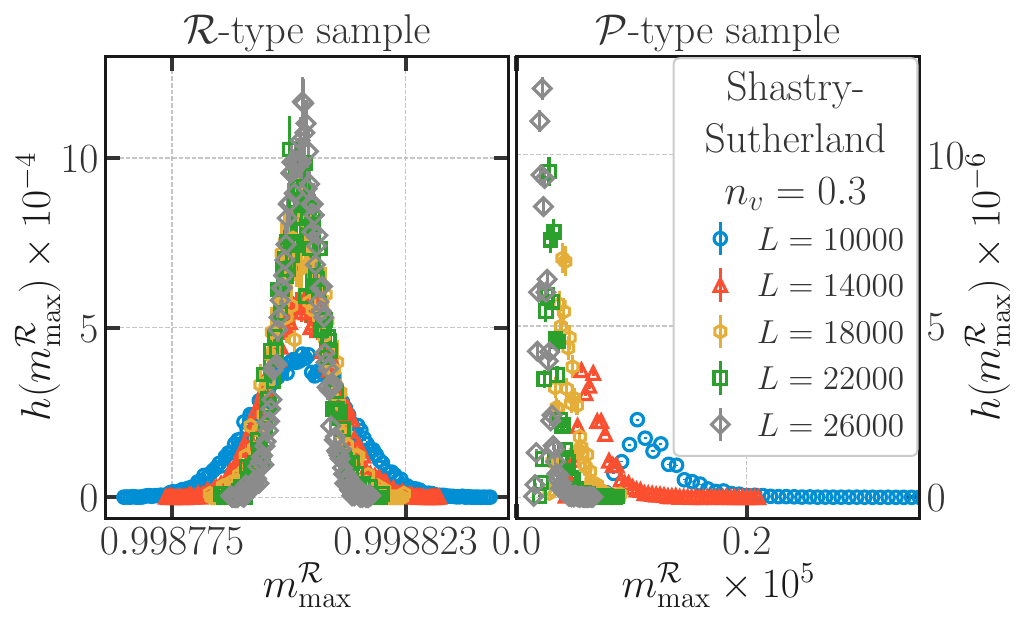}\\
			c)\includegraphics[width=\columnwidth]{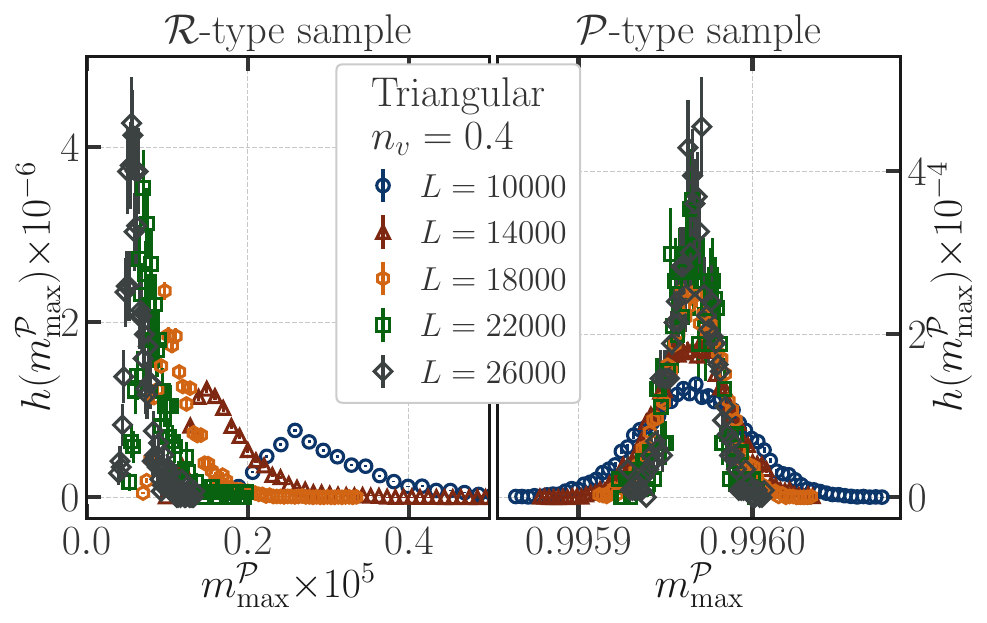} &	
			d)\includegraphics[width=\columnwidth]{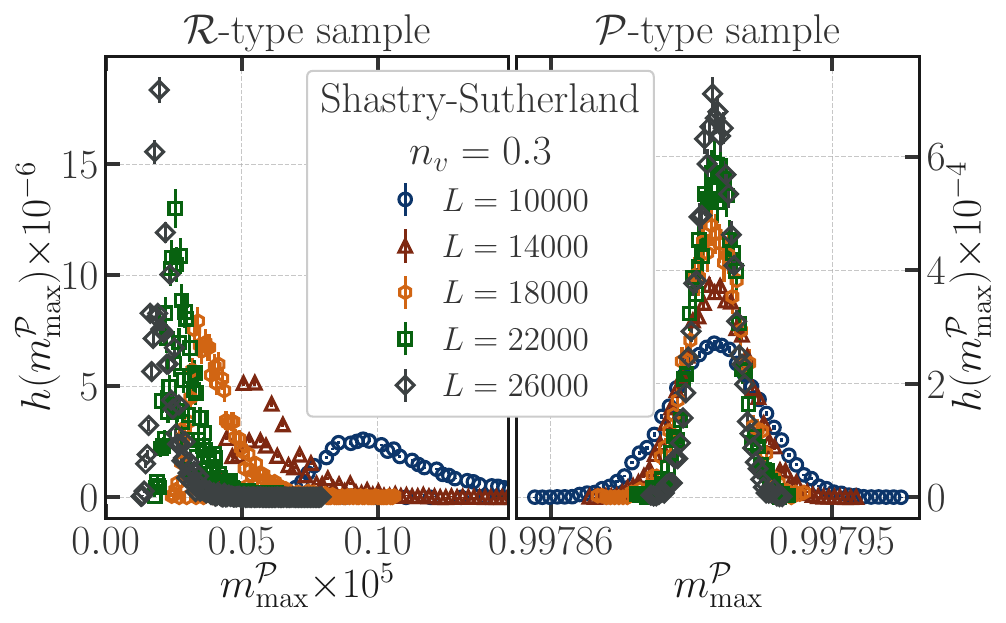}\\
		\end{tabular}
		\caption{
			Each sample of (a) the site-diluted triangular lattice with $n_v = 0.40$ (b) the Shastry-Sutherland lattice with $n_v = 0.30$ is labeled $\calR$-type or $\calP$-type based on whether it contributes to the right peak or left peak respectively in the histogram of $\mtotR$ in Fig.~\ref{fig:mtotR2D}. The histogram of $\mmaxR$ ($\mmaxP$) in the sub-ensemble of $\calR$-type ($\calP$-type) samples shows a single peak at a nonzero $L$-independent value of $\mmaxR$ ($\mmaxP$), with the width of this peak narrowing with increasing $L$. In sharp contrast to this behavior, the histogram of $\mmaxR$ ($\mmaxP$) in the sub-ensemble of $\calP$-type ($\calR$-type) samples has a single peak that shifts continuously with increasing $L$ to smaller and smaller values of $\mmaxR$ ($\mmaxP$), consistent with the fact that $\langle \mmaxR \rangle$ ($\langle \mmaxP \rangle$) tends to zero with increasing $L$ when the average taken over the sub-ensemble of $\calP$-type ($\calR$-type) samples. See Sec.~\ref{sec:ComputationalMethodsObservables} and Sec.~\ref{sec:ComputationalResults} for details.\label{fig:hist_mmaxpmax2D}}	
	\end{figure*}

	\section{Computational methods and observables}
	\label{sec:ComputationalMethodsObservables}
	Our focus here is the random geometry of Gallai-Edmonds clusters, i.e. $\calR$-type and $\calP$-type regions, of two-dimensional and three-dimensional site-diluted lattices. The dilution is completely uncorrelated, and parameterized by a dilution fraction $n_v$ that represents the density of static (quenched) vacancies. The range of sizes needed for a reliable extrapolation to the thermodynamic limit depends on the dilution $n_v$, with very small values of $n_v$ requiring the study of much larger lattices. This, in conjunction with limits set by available computational resources, sets the lower limit of $n_v$ up to which we can extend our results. In all cases, we therefore limit ourselves to not-too-small values of $n_v$ ($n_v \gtrsim 0.2$). Fortunately, this restriction does not affect our ability to delineate the complete structure of the phase diagram in both two and three dimensions.

		\subsection{Computational methods}
	\label{subsec:ComputationalMethods}	
 Our numerical results in two dimensions are for $L \times L$ triangular and Shastry-Sutherland lattices with periodic boundary conditions and $L$ ranging from $L=10000$ to $L=26000$ for both these lattices. In three dimensions, we study $L \times L \times L$ stacked triangular and corner-sharing octahedral lattices with periodic boundary conditions and $L$ ranging from $L=300$ to $L=900$ for the stacked triangular case and $L=200$ to $L=650$ in the corner-sharing octahedral case. Here, $L$ refers in each case to the number of unit cells in each periodic direction.
 Although our results are all within the geometrically percolated phase of the original lattice, the diluted lattice nevertheless has multiple distinct connected components in general. To further minimize finite-size effects, we focus attention on the largest connected component $\calG$ of each sample, and focus on the random geometry of the Gallai-Edmonds decomposition of this component. 
 		\begin{figure*}
		\begin{tabular}{cc}
			a)\includegraphics[width=\columnwidth]{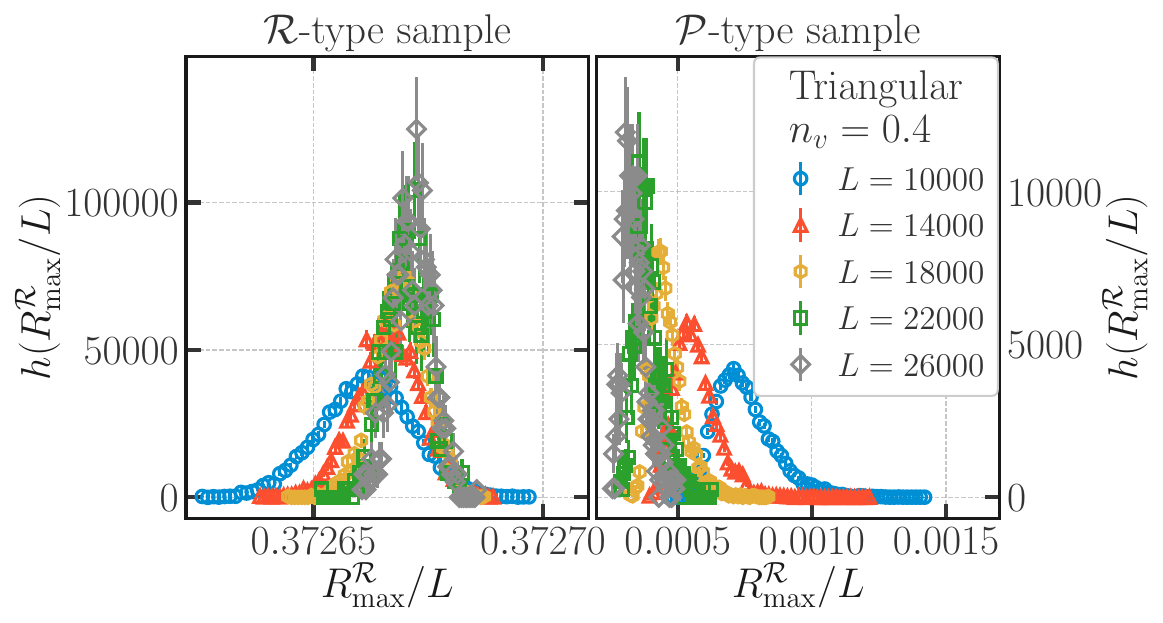} &	
			b)\includegraphics[width=\columnwidth]{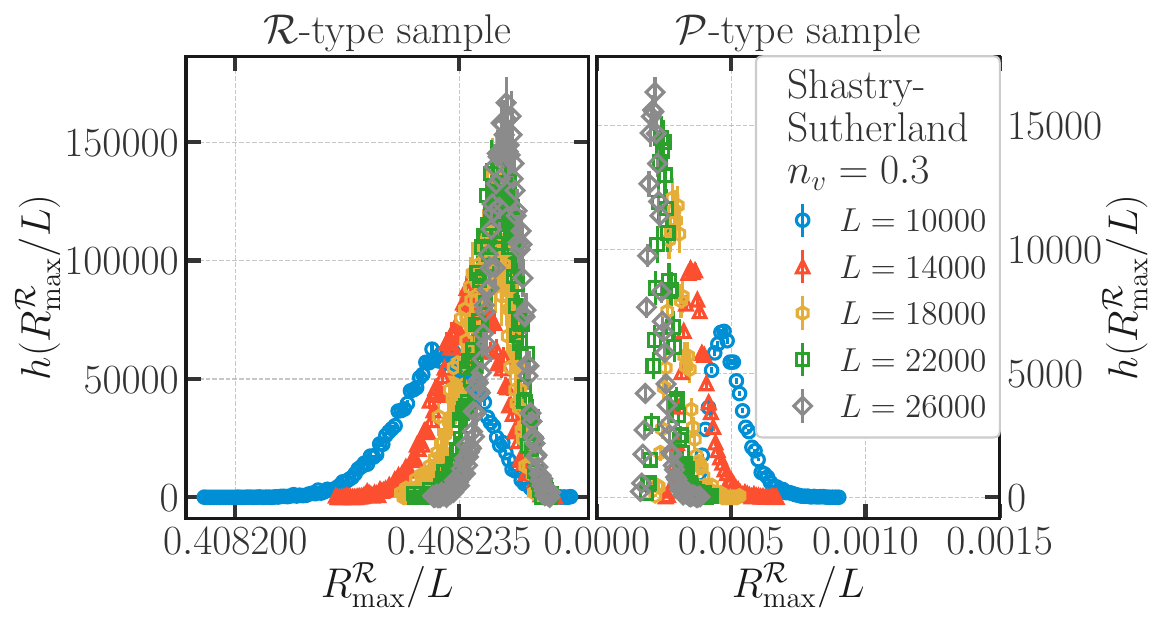}\\
			c)\includegraphics[width=\columnwidth]{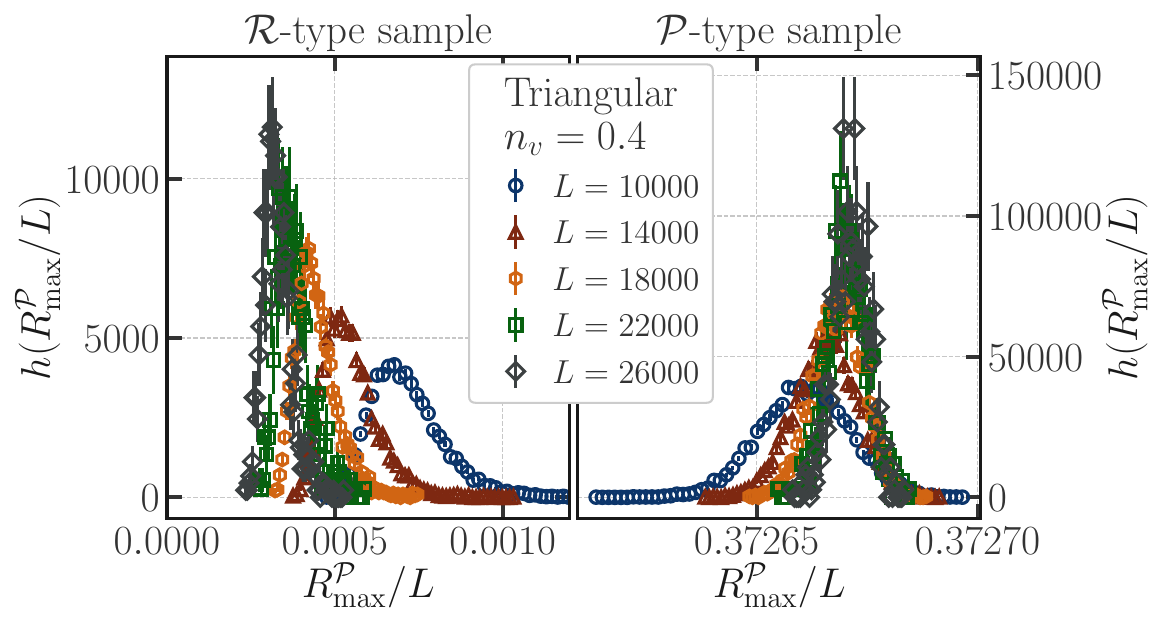} &	
			d)\includegraphics[width=\columnwidth]{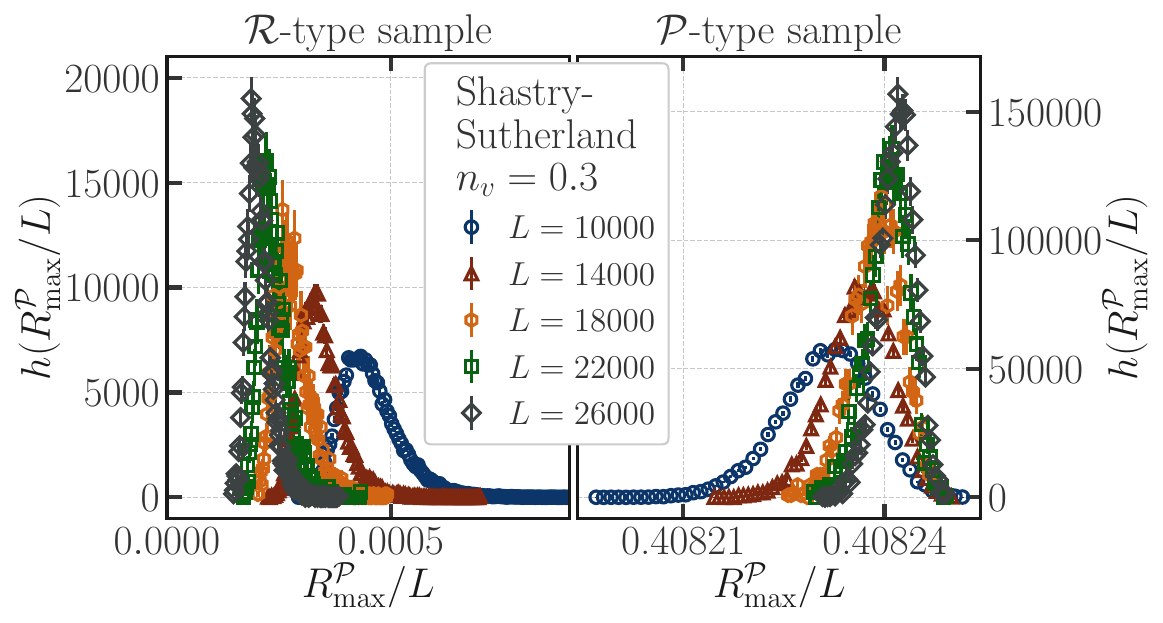}\\
		\end{tabular}
		\caption{
		Each sample of (a) the site-diluted triangular lattice with $n_v = 0.40$ (b) the Shastry-Sutherland lattice with $n_v = 0.30$ is labeled $\calR$-type or $\calP$-type based on whether it contributes to the right peak or left peak respectively in the histogram of $\mtotR$ in Fig.~\ref{fig:mtotR2D}. The histogram of $\RmaxR/L$ ($\RmaxP/L$) in the sub-ensemble of $\calR$-type ($\calP$-type) samples shows a single peak at a nonzero $L$-independent value of $\RmaxR/L$ ($\RmaxP/L$), with the width of this peak narrowing with increasing $L$. In sharp contrast to this behavior, the histogram of $\RmaxR/L$ ($\RmaxP/L$) in the sub-ensemble of $\calP$-type ($\calR$-type) samples has a single peak that shifts continuously with increasing $L$ to smaller and smaller values of $\RmaxR/L$ ($\RmaxP/L$), consistent with the fact that $\langle \RmaxR/L\rangle$ ($\langle \RmaxP/L \rangle$) tends to zero with increasing $L$ when the average taken over the sub-ensemble of $\calP$-type ($\calR$-type) samples. See Sec.~\ref{sec:ComputationalMethodsObservables} and Sec.~\ref{sec:ComputationalResults} for details.\label{fig:hist_RmaxRRmaxP2D}}	
	\end{figure*}

	We have typically averaged over $3000$ to $10000$ such samples at each value of $n_v$ (with a smaller number of samples at larger sizes) in $d=2$. In the three-dimensional case, we have averaged over $8000$ to $12000$ samples (with more extensive averaging at smaller sizes, and near the various critical points) in order to reliably resolve some of the more unusual aspects of the three-dimensional phase diagram. In addition to yielding accurate estimates of various self-averaging quantities, this also allows us to obtain reliable information about the {\em probability distributions} of quantities that have anomalously large sample-to-sample fluctuations and violate self-averaging expectations in the thermodynamic limit. 
		
The key subtlety in non-bipartite maximum matching (relative to its simpler bipartite cousin) is that sites belonging to the blossoms (alluded to earlier in Sec.~\ref{subsec:GETheory}) can be reached by both even and odd length alternating paths starting from a monomer of the matching. This makes it substantially more difficult to find {\em augmenting paths}, {\em i.e.}, alternating paths that start and end at two different monomers and allow one to grow the size of the matching by shifting the dimers along this path towards one end, and then adding a dimer at the other end to remove the two monomers at either end and grow the matching.	

	This is the difficulty that is overcome by Edmonds' matching algorithm~\cite{Edmonds_1965} for identifying blossoms and then constructing augmenting paths. For the diluted lattices considered here, and particularly in the range of $n_v$ that is our focus, we find that a breadth-first search (BFS) based implementation of Edmonds' matching algorithm performs better than a depth-first search (DFS) based implementation when we prune branches of the search tree to increase efficiency. Our BFS implementation relies on the version of Edmonds' matching algorithm given by Moret and Shapiro~\cite{Moret_Shapiro_1991} and uses the union-find data structure of Tarjan~\cite{Tarjan_1983}. 

	We have found that it is more efficient to organize the search for augmenting paths by starting with one monomer at a time rather than multiple monomers. In addition, we have also incorporated a few heuristics for speeding up the implementation; the underlying ideas are drawn from the implementation published by Kececioglu and Pecqueur~\cite{Kececioglu_1998}. For instance, to increase the efficiency, an array of unmatched vertices is maintained at all times to help begin the search for augmenting paths. In order to test our implementation, we  have compared the output of our implementation against the output from Kececioglu and Pecqueur's implementation~\cite{Kececioglu_1998}, and found consistent results for all test cases.

At intermediate steps in the algorithm, the union-find structure used in our implementation provides us a particularly convenient way of keeping track of the blossoms (here, and only in this paragraph, we use the term in its original algorithmic sense~\cite{Edmonds_1965}) identified during the search for augmenting paths. This is done via pointers that point to a single representative ``root'' vertex from all the vertices in such an ``algorithmic'' blossom. Of course, this information at intermediate steps depends on the order in which the search is conducted, and carries no invariant structural information about the underlying diluted lattice. 
	 		\begin{figure*}
		\begin{tabular}{cc}
			a)\includegraphics[width=\columnwidth]{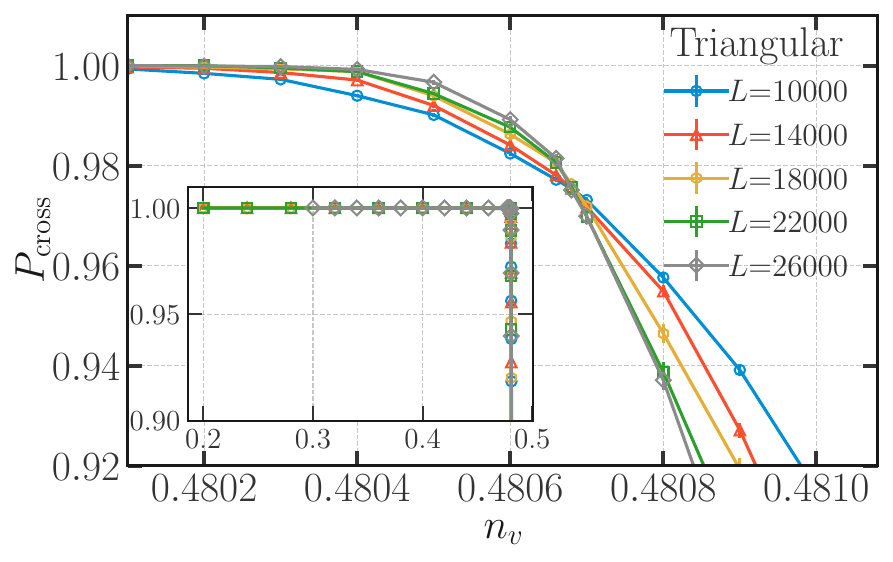}&
			b)\includegraphics[width=\columnwidth]{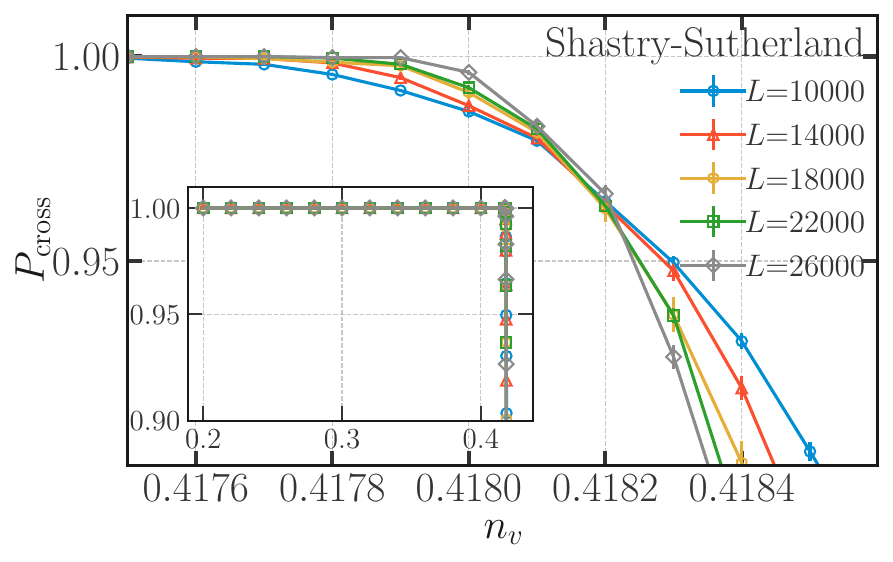}\\
		\end{tabular}
		\caption{ Inset: $P_{\rm cross}$, the probability that there exists a Gallai-Edmonds cluster (either a $\calR$-type region or a $\calP$-type region) in $\calG$ (the largest geometrically connected component  of the diluted lattice) which wraps around the torus simultaneously in two independent directions, shows threshold behavior on (a) the triangular lattice and (b) the Shastry-Sutherland lattice. Main figure: This threshold becomes sharper with increasing system size on both lattices. On the low-dilution side of this threshold,  $P_{\rm cross}$ saturates to a size-independent value of $P_{\rm cross} \approx 1$. On the high-dilution side of this threshold, $P_{\rm cross}$ decreases monotonically with increasing size $L$. In the vicinity of the threshold, curves corresponding to various sizes $L$  all cross at a critical dilution
$n_v^{\rm crit} \approx 0.48068(4)$ [$n_v^{\rm crit} \approx 0.41820(5)$] on the triangular [Shastry-Sutherland] lattice. See Sec.~\ref{sec:ComputationalMethodsObservables} and Sec.~\ref{sec:ComputationalResults} for details.
		\label{fig:Pcross2D}
}
	\end{figure*}

However, and this is key, when we {\em rerun} our implementation of Edmonds' matching algorithm one additional time {\em after finding a maximum matching}, the information recorded in the union-find data structure after this final pass of the algorithm correctly provides information about the odd-sized connected components of the graph that we have referred to as blossoms in Fig.~\ref{fig:Blossoms} (and everywhere else in this article except in the preceding paragraph). More explicitly, we obtain for each e-type site a pointer that points to a ``root'' site that serves as a unique representative of the blossom to which that e-type site belongs, with the additional property that a root site points to itself in this data structure.

Thus, after identifying a maximum matching of the lattice at a given vacancy concentration, we simply rerun Edmonds' matching algorithm to obtain the labels (even, odd or unreachable) of all the sites as well as membership information for each blossom of the diluted lattice.
	 Using these labels, we identify all the $\calR$-type and $\calP$-type regions using a standard burning algorithm.	Note that this allows us to keep track of  the number of blossoms by counting the number of e-type sites that point to themselves. In addition, by counting the number of e-type sites that point to a given root site, we are also able to keep track of the size of the corresponding blossom, {\em i.e.}, the number of e-type sites that belong to it.

	Using this implementation of Edmonds' matching algorithm, we proceed to obtain data for a grid of vacancy concentrations $n_v$. At each concentration, we use the maximum matching found earlier at an adjacent value of $n_v$ to obtain the initial matching configuration from which the algorithm starts its search for augmenting paths. For the initial value of $n_v$ in our grid, a random matching was taken as an initial matching configuration. The two-dimensional data discussed in this paper was obtained by moving up in a grid of $n_v$ values. However, in three dimensions, we found it advantageous in terms of efficiency to move down a grid of $n_v$ values.  

		\subsection{Useful observables}
	\label{subsec:UsefulObservables}	
 As we have already reviewed, the physics of topologically protected collective Majorana modes of Majorana networks is controlled by the size and morphology of $\calR$-type regions of the diluted lattice, and the structure of $\calP$-type regions has no direct role in this physics. 
 
 However, the nature of vacancy-induced local moment instabilities in short-range resonating valence bond spin liquids is expected to depend sensitively on the structure of $\calP$-type regions in addition to $\calR$-type regions. Further, as will be clear from the subsequent discussion, it is in fact impossible to fully understand the interesting dilution dependence of the random geometry of $\calR$-type regions without at the same time understanding properties of $\calP$-type regions as well. 
 
 With this in mind, we study the $\calR$-type and $\calP$-type regions of each sample on an equal footing, measuring the disorder average of a number of observables that probe the geometry of these regions, as well as the probability distributions of some of these observables. We turn now to the definition of these quantities. 
 \begin{figure*}
	\begin{tabular}{cc}
		\includegraphics[width=\columnwidth]{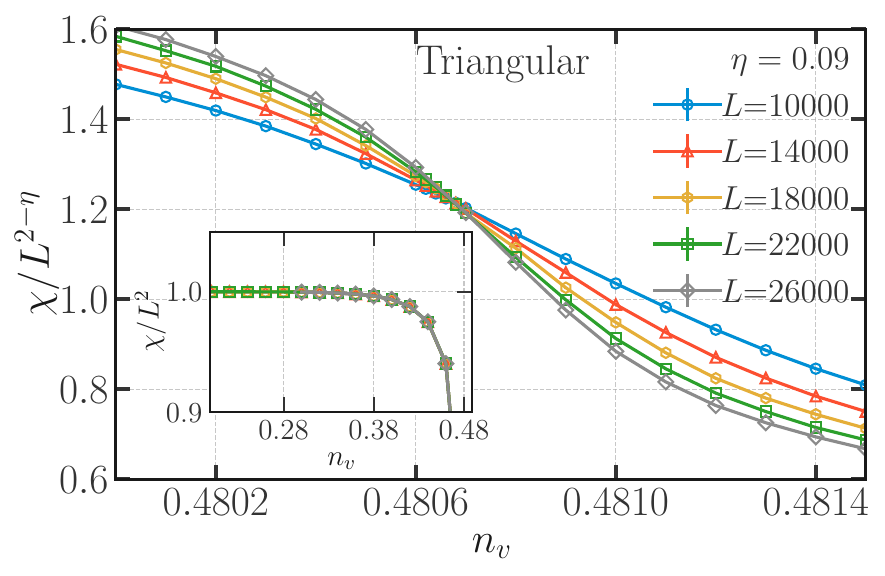} &
		\includegraphics[width=\columnwidth]{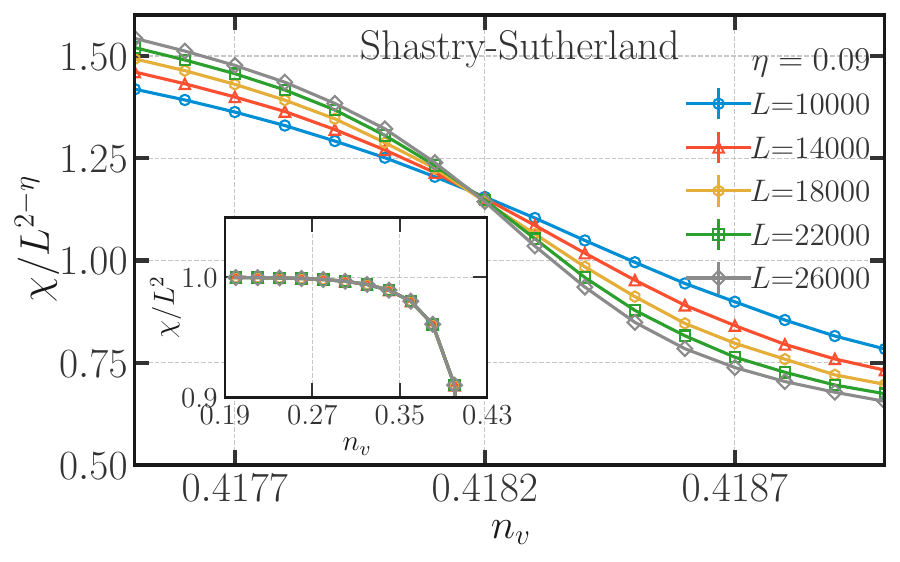}\\
		\end{tabular}
		\caption{Inset: $\chi/L^2$, where $\chi$ is the  susceptibility associated with sample-averaged geometric correlation function $C(r-r^\prime)$ in two dimensional $L \times L$ samples, shows threshold behavior on (a) the triangular lattice and (b) the Shastry-Sutherland lattice. Inset and main figure: On the low-dilution side of this threshold,  $\chi/L^2$ saturates to a size-independent value. On the high-dilution side of this threshold, $\chi/L^2$ decreases monotonically with increasing size $L$. Main figure: The threshold gets sharper with increasing system size on both lattices. In the vicinity of the threshold, curves of $\chi/L^{2-\eta}$ (with $\eta$ chosen to be $\eta = 0.09$) corresponding to various sizes $L$  all cross at a critical dilution
$n_v^{\rm crit} \approx 0.48068(4)$ [$n_v^{\rm crit} \approx 0.41820(5)$] on the triangular [Shastry-Sutherland] lattice. The choice for $\eta$ shown in the figure lies in the middle of a narrow band of values, all of which yield a crossing point that is comparably sharp and lies within our estimated range of $n_v^{\rm crit}$; this uncertainty feeds into the error bar on our estimated value of $\eta \approx 0.09(2)$. See Sec.~\ref{sec:ComputationalMethodsObservables} and Sec.~\ref{sec:ComputationalResults} for details. \label{fig:chi2D}}		
	\end{figure*}

\subsubsection{Thermodynamic densities and distributions}
\label{subsubsec:ThermodynamicDensitiesDistributions}
The first observable we keep track of is of course the monomer number
\begin{eqnarray}
W &=& \sum_{\calR_i \in \calG} w(\calR_i)
\end{eqnarray}
that characterizes the ensemble of maximum matchings of the largest connected component $\calG$ of the diluted lattice. In the above, the sum is over $\calR$-type regions $\calR_i$ belonging to $\calG$ and $w(\calR_i)$ is the number of monomers hosted by each region. Since $W$ is generically an extensive quantity (as already discussed in Sec.~\ref{sec:IntroductionOverview}), we also define the corresponding density $w$:
\begin{eqnarray}
		w &=& W/m(\calG) \; ,
\end{eqnarray}
where $m(\calG)$ is the mass (number of lattice sites) of $\calG$.
	Another simple overall characterization is in terms of the total number $N_{\calR}$ ($N_{\calP}$) of $\calR$-type ($\calP$-type) regions belonging to $\calG$. We define the corresponding number densities as
	\begin{eqnarray}
		n_{\calR} & = &  N_{\calR}/m(\calG) \; , \nonumber \\
		n_{\calP} & = &  N_{\calP}/m(\calG) \; .
		\label{eq:nR_nP}
	\end{eqnarray}

For the simplest gross characterization of the structure of Gallai-Edmonds clusters, it is useful to study the $n_v$ dependence of the fraction of vertices that carry $\calR$-type and $\calP$-type labels. To this end, we define.
	\begin{eqnarray}
		\mtotR &=& \frac{1}{m(\calG)}\sum_{\calR_i \in \calG} m({\calR_i})\; , \nonumber \\
		\mtotP &=& \frac{1}{m(\calG)}\sum_{\calP_i \in \calG} m({\calP_i}) \; ,
		\label{eq:mtotR_mtotP}
	\end{eqnarray}
	where  the sum in the definition of $\mtotR$ ($\mtotP$) is over $\calR$-type ($\calP$-type) regions $\calR_i$ ($\calP_i$) contained in $\calG$, $m({\calR_i})$ ($m({\calP_i})$) are their respective masses (total number of vertices contained in the cluster), and $\calG$ and $m(\calG)$ have already been defined above.

		
	Additional information about $\calR$-type regions is also provided by a separate measurement of the density $m^{\rm o}_{\rm tot}$ of o-type sites:
	\begin{eqnarray}
		\mtoto &=& \frac{1}{m(\calG)}\sum_{\calR_i \in \calG} m_{\rm o}({\calR_i})\; .
		\label{eq:mtoto}
	\end{eqnarray}
The corresponding density $m^{\rm e}_{\rm tot}$ of e-type sites can of course be deduced by using this in conjunction with the measured value of $\mtotR$ since 
	\begin{eqnarray}
		\mtote &=& \mtotR-\mtoto \; .
		\label{eq:mtote}
	\end{eqnarray}
Likewise, from knowledge of $m_{\rm o}(\calR_i)$ and $m(\calR_i)$, we also have access to the corresponding quantity $m_{\rm e}(\calR_i)$ at the level of each $\calR$-type region $\calR_i$. 
			\begin{figure*}
		\begin{tabular}{cc}
			a)\includegraphics[width=\columnwidth]{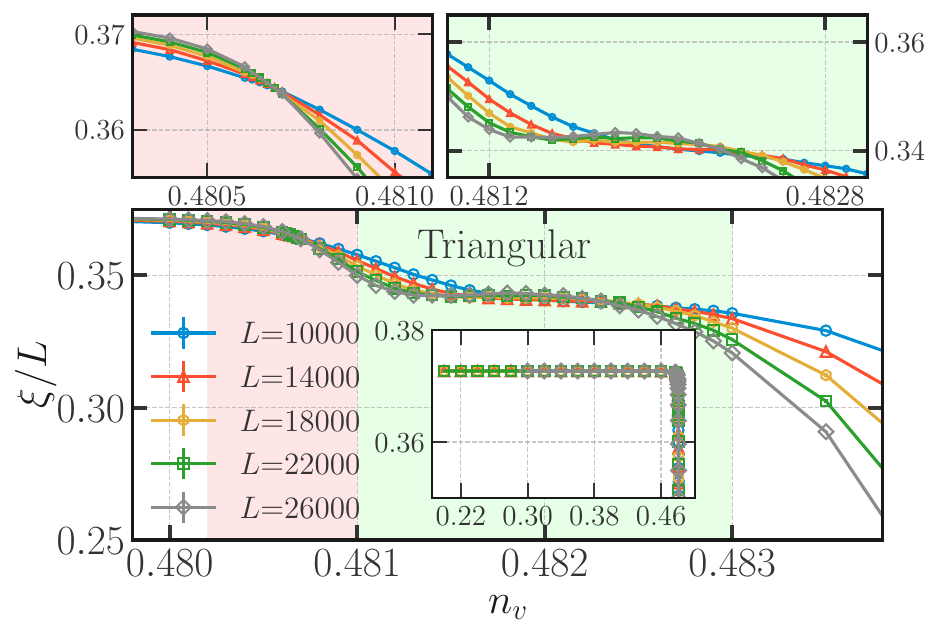}&	b)\includegraphics[width=\columnwidth]{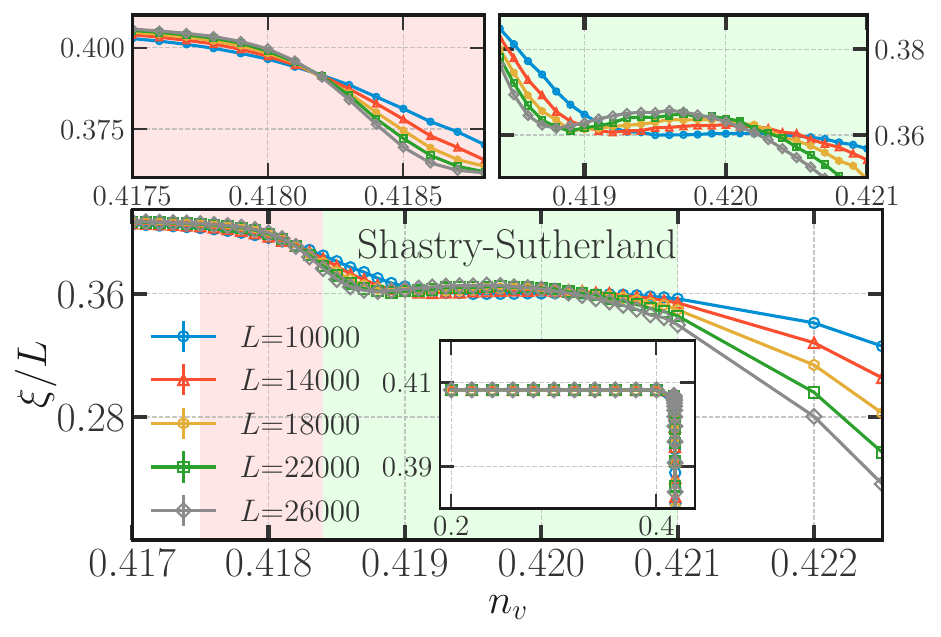}\\			
		\end{tabular}
		\caption{Inset to lower panels: $\xi/L$, where $\xi$ is the correlation length associated with sample-averaged geometric correlation function $C(r-r^\prime)$ in two dimensional  $L \times L$ samples, shows threshold behavior on (a) the triangular lattice and (b) the Shastry-Sutherland lattice. Lower panel main figures and upper left panels: This threshold gets sharper with increasing system size on both lattices and curves of $\xi/L$ corresponding to various sizes $L$  all cross at a critical dilution
$n_v^{\rm crit} \approx 0.48068(4)$ [$n_v^{\rm crit} \approx 0.41820(5)$] on the triangular [Shastry-Sutherland] lattice. On the low-dilution side of this threshold,  $\xi/L$ saturates to a size-independent value. Far away from this threshold, on the high-dilution side, $\xi/L$ decreases monotonically with increasing size $L$.  Lower panel main figures and upper right insets: In the vicinity of the threshold, slightly towards the high-dilution side, $\xi/L$ has non-monotonic behavior (as a function of $n_v$). This regime is characterized by a shallow maximum in $\xi/L$, with the position of the maximum shifting with increasing $L$. See Fig.~\ref{fig:Psingle2D} for an interpretation of this behavior, and Sec.~\ref{sec:ComputationalMethodsObservables} and Sec.~\ref{sec:ComputationalResults} for details. 		\label{fig:xi2D}}
\end{figure*}

	In addition to keeping track of $m_{\rm o}(\calR_i)$ and $m(\calR_i)$ for each $\calR$-type region and the corresponding densities $\mtoto$ and $\mtotR$ in $\calG$ as a whole, we also keep independent track of the number $n_{\calB}(\calR_i)$ of blossoms in each $\calR$-type region as well as the corresponding number density of blossoms in $\calG$ as a whole:
	\begin{eqnarray}
		\ntotB &=& \frac{1}{m(\calG)}\sum_{\calR_i \in \calG} n_{{\mathcal B}}({\calR_i})\; .
		\label{eq:ntotB}
	\end{eqnarray}	
This number $n_{\calB}(\calR_i)$, in conjunction with our knowledge of $m_{\rm e}(\calR_i)$, allows us to obtain the average mass (number of e-type sites) of a blossom either by the average of the ratio	
	\begin{eqnarray}
&&\frac{1}{N_{\calR}} \sum_{\calR_i \in \calG} \frac{m_{\rm e}(\calR_i)}{n_{\calB}(\calR_i)}
\end{eqnarray}
  or by the ratio
  \begin{eqnarray}
	&&\mtote/n^{\calB}_{\rm tot}  
  \end{eqnarray}
of the corresponding global densities. 

Since we separately measure both $m_{\rm o}(\calR_i)$ and $n_{\calB}(\calR_i)$ in addition to the monomer number $w(\calR_i)$,  we may define the imbalance
\begin{eqnarray}
\calI(\calR_i) &\equiv& n_{\calB}(\calR_i) - m_{\rm o}(\calR_i) \; 
\end{eqnarray}
of each ${\calR}$-type region $\calR_i$, and test for consistency by demanding that our data satisfy the following constraint:
\begin{eqnarray}
\calI(\calR_i) &=& w(\calR_i) \;  \forall i \;.
\end{eqnarray}
We have checked that this constraint is always satisfied at the level of each $\calR$-type region by our data. We will therefore slur over the distinction between these two quantities in our subsequent discussion.
	
		\subsubsection{Extremal properties}
\label{subsubsec:ExtremalProperties}	
	We also find it instructive to study the structure of $\calR_{\rm max}$ and $\calP_{\rm max}$,  the largest (in $\calG$, the largest connected component of the lattice) $\calR$-type and $\calP$-type regions respectively, by recording some of their properties separately. To this end, we record $R^{\calR}_{\rm max}$ and $R^{\calP}_{\rm max}$, their radii of gyration. Since we work with periodic boundary conditions in both directions, radii of gyration are tricky to measure efficiently unless defined in a suitable way. This issue has been discussed in the earlier literature in diverse contexts, and we follow the definition and procedure suggested in Ref.~\cite{Bai_Breen_2008}; this is also consistent with the approach used in an earlier study of the bipartite case, {\em i.e.}, Dulmage-Mendelsohn percolation~\cite{Bhola_Biswas_Islam_Damle_2022}.
	
	We also record $\mathcal{I}_{\rm max}^{\calR}$, the number of monomers that live in $\calR_{\rm max}$. In addition, we keep track of $m(\calR_{\rm max})$ and $m(\calP_{\rm max})$, the masses (total number of sites) of $\calR_{\rm max}$ and $\calP_{\rm max}$ respectively. In some regimes, we find it more convenient to focus on $m({\rm GE}_{\rm max})$, the mass of the largest Gallai-Edmonds cluster ${\rm GE}_{\rm max}$ in $\calG$, {\em i.e.}, the mass of the larger of these two regions.

	If these $\calR$-type or $\calP$-type regions percolate, we expect their mass to scale with the mass $m(\calG)$ of $\calG$. With this in mind, we define the corresponding densities
	\begin{eqnarray}
	\mmaxR &=& \frac{m(\calR_{\rm max})}{m(\calG)}\; ,\nonumber \\
		\mmaxP &=& \frac{m(\calP_{\rm max})}{m(\calG)}\; , \nonumber \\
		\mmax &=& \frac{m({\rm GE}_{\rm max})}{m(\calG)} \; .
	\end{eqnarray}	
	In addition, for a finer characterization of the structure of $\calR$-type regions, in particular, the manner in which e-type sites are organized into the blossoms, we also keep track of the ratio of $m({\mathcal B}_{\rm max})$, the mass of the largest blossom in the largest $\calR$-type region in $\calG$, to the mass $m({\rm GE}_{\rm max})$ of the largest Gallai-Edmonds cluster in $\calG$:
\begin{eqnarray}
\mmaxB &=& \frac{m({\mathcal B}_{\rm max})}{m({\rm GE}_{\rm max})} \; .
\label{eq:MaxBlossomDensity}
\end{eqnarray}

	\subsubsection{Correlation function and length scales}
\label{subsubsec:CorrelationFunctionLengthScales}
 For a somewhat more detailed characterization of the structure of $\calR$ and $\calP$ type regions, we also keep track of their individual radii of gyration $R_i$. These radii of gyration $R_i$ are again defined and measured as in Ref.~\cite{Bai_Breen_2008}. 
 					 		\begin{figure*}
		\begin{tabular}{cc}
			a)\includegraphics[width=\columnwidth]{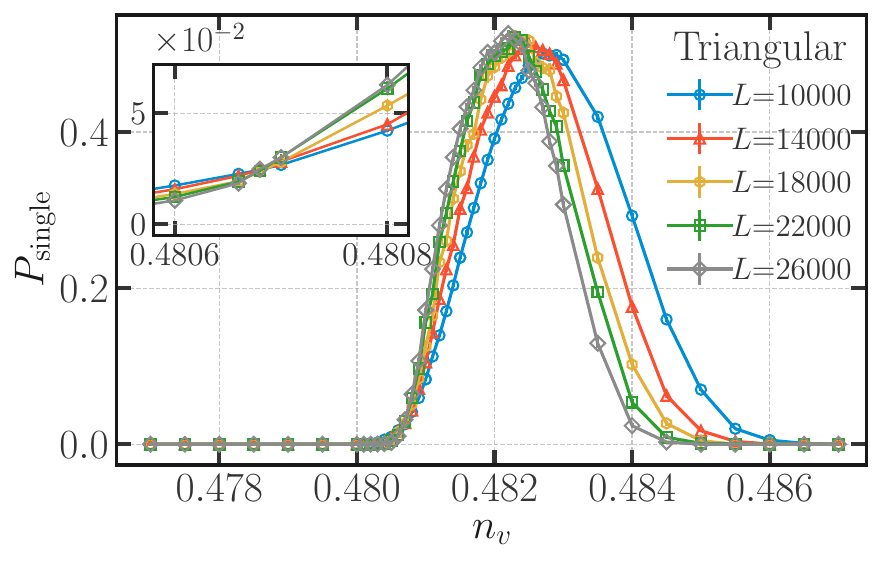}&
			b)\includegraphics[width=\columnwidth]{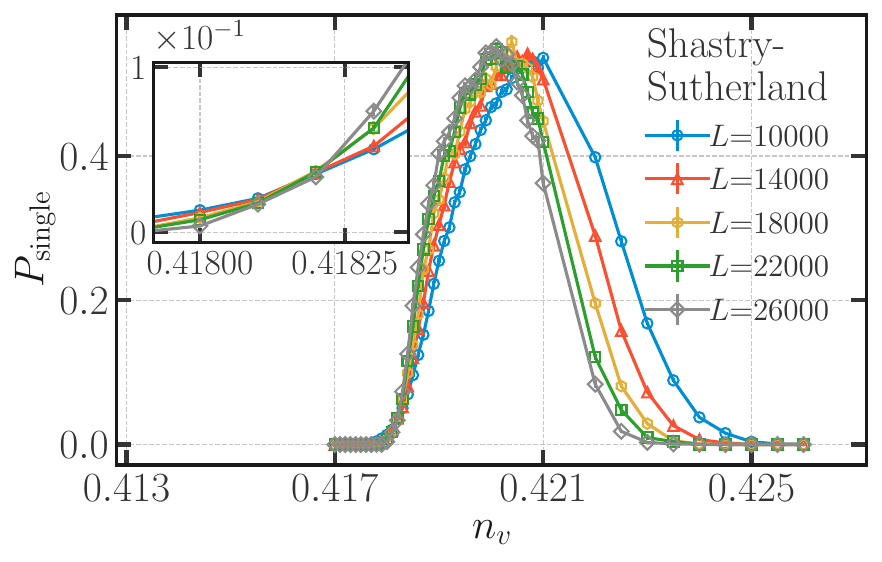}\\	
		\end{tabular}
		\caption{Main figures: On (a) the triangular lattice and (b) the Shastry-Sutherland lattice, $P_{\rm single}$, the probability that there exists a Gallai-Edmonds cluster (either an $\calR$-type region or a $\calP$-type region) which wraps around the torus in exactly one direction, displays complex non-monotonic behavior over a range of $n_v$ in which $\xi/L$ also shows similar behavior (Fig.~\ref{fig:xi2D}). On both the low-dilution and the high-dilution  side of this interesting regime, $P_{\rm single}$ vanishes with increasing size $L$. As the system traverses this regime from high to low dilution, $P_{\rm single}$ first develops a peak at an approximately $L$-independent value less than $1$. The position of this peak is $L$-dependent, and shifts to lower values of dilution with increasing size. By comparing this behavior to the behavior seen earlier in the probability $P_{\rm cross}$ for a cluster to wrap simultaneously in two independent directions (see Fig.~\ref{fig:Pcross2D}), we conclude that this crossover behavior is associated with the wrapping characteristics of large clusters when they first form: Below this crossover, large clusters are more likely to wrap simultaneously around two independent directions of the torus, while above this crossover, clusters are too small to wrap in even one direction. The actual critical point identified from the behavior of $P_{\rm cross}$ and $\chi$ (see Fig.~\ref{fig:Pcross2D} and Fig.~\ref{fig:chi2D}) is at lower values of $n_v$, and occurs in a regime in which $P_{\rm single}$ decreases with $n_v$.  The inset zooms in on the vicinity of this critical point, and shows that curves corresponding to various sizes $L$  all cross at a critical dilution $n_v^{\rm crit} \approx 0.48068(4)$ [$n_v^{\rm crit} \approx 0.41820(5)$] on the triangular [Shastry-Sutherland] lattice. See Sec.~\ref{sec:ComputationalMethodsObservables} and Sec.~\ref{sec:ComputationalResults} for details.		\label{fig:Psingle2D}}
	\end{figure*}

	In studies of geometric percolation, it is conventional to define a correlation function $C^{\rm geom}(r-r') \equiv \langle C^{\rm geom}(r, r') \rangle$, where the angular brackets denote an ensemble average, and $C^{\rm geom}(r, r') =1$ if $r$ and $r'$ belong to the same cluster, and $C^{\rm geom}(r, r')=0$ otherwise~\cite{Stauffer_Aharony_1992}. The corresponding correlation length $\xi$ is then related to the root mean square radius of gyration of the clusters, with each cluster of mass $m$ weighted by $m^2$ when calculating the mean~\cite{Stauffer_Aharony_1992}.  We use entirely analogous definitions for $C(r-r')$, $C^{\calR}(r-r')$ and $C^{\calP}(r-r')$, where $C(r-r')$ is defined with reference to {\em all} Gallai-Edmonds regions (of types $\calR$ {\em and} $\calP$), while $C^{\calR}$ ($C^{\calP}$) is defined using only $\calR$-type ($\calP$-type) regions of a sample. 
	
	The corresponding correlation lengths are also analogously defined:
	\begin{eqnarray}
		\xi^2 &=& \left\langle \frac{\sum_{i \in \calG}m_i^2R_i^2}{\sum_{i\in \calG}m_i^2} \right\rangle \; , \nonumber \\	
		\left(\xi^{\calR}\right)^2 &=&  \left\langle \frac{\sum_{\calR_i \in \calG}m_{\calR_i}^2R_i^2}{\sum_{\calR_i\in \calG}m_{\calR_i}^2} \right\rangle \; , \nonumber \\
		\left(\xi^{\calP}\right)^2 &=& \left\langle \frac{\sum_{\calP_i \in \calG}m_{\calP_i}^2R_i^2}{\sum_{\calP_i\in \calG}m_{\calP_i}^2} \right\rangle \; .
		\label{eq:xi}
	\end{eqnarray}
	Here, the angular bracket denotes an ensemble average, the sum over $i \in \calG$ in the definition of $\xi$ is over all Gallai-Edmonds clusters (both $\calP$-type and $\calR$-type) belonging to the largest geometric cluster $\calG$, while the corresponding sum in the definition of $\xi_\calR$ ($\xi_\calP$) is restricted to all $\calR$-type (all $\calP$-type) Gallai-Edmonds regions belonging to $\calG$.

	Since the radii of gyration $R_i$ that appear in the above are defined and measured as in Ref.~\cite{Bai_Breen_2008}, our definitions of the correlation lengths that characterize a sample do not correspond exactly to the correlation length one would have obtained from the corresponding correlation functions. However, when clusters are large, they behave in the same way, and the definition suggested in Ref.~\cite{Bai_Breen_2008} is computationally much more efficient~\cite{Bhola_Biswas_Islam_Damle_2022}. It is also useful to note that these definitions are in terms of an average of a ratio; this ratio defines the square of a correlation length that characterizes each sample, which is then averaged over the disorder ensemble. When there are large sample-to-sample fluctuations of the morphology of Gallai-Edmonds clusters (as will turn out to be the case in the low-dilution regimes we study), this can be different from other definitions that use a ratio of averages. We have tried both approaches, and found the quantities defined above to be more sensitive probes.

		Finally, we also keep track of the susceptibilities $\chi$, $\chi^{\calR}$ and $\chi^{\calP}$ corresponding to these correlation functions:
	\begin{eqnarray}
		\chi &=&\frac{1}{L^d}\sum_{r,r'}\langle C(r,r') \rangle \nonumber \\
		& =& \frac{1}{L^d}\left\langle \sum_{i \in \calG}m_i^2 \;\right\rangle \;, \nonumber \\
		\chi^{\calR} &=&\frac{1}{L^d}\sum_{r,r'}\langle C^{\calR}(r,r') \rangle \nonumber \\
		& =& \frac{1}{L^d}\left\langle \sum_{\calR_i \in \calG}m(\calR_i)^2 \;\right\rangle \;, \nonumber \\
		\chi^{\calP} &=&\frac{1}{L^d}\sum_{r,r'}\langle C^{\calP}(r,r') \rangle \nonumber \\
		& =& \frac{1}{L^d}\left\langle \sum_{\calP_i \in \calG}m(\calP_i)^2 \;\right\rangle \;, 			
		\label{eq:chi}
	\end{eqnarray}
	where the angular bracket again denotes an ensemble average and $d$ is the spatial dimension.
			\begin{figure*}
 \begin{tabular}{cc}
	(a)	\includegraphics[width=\columnwidth]{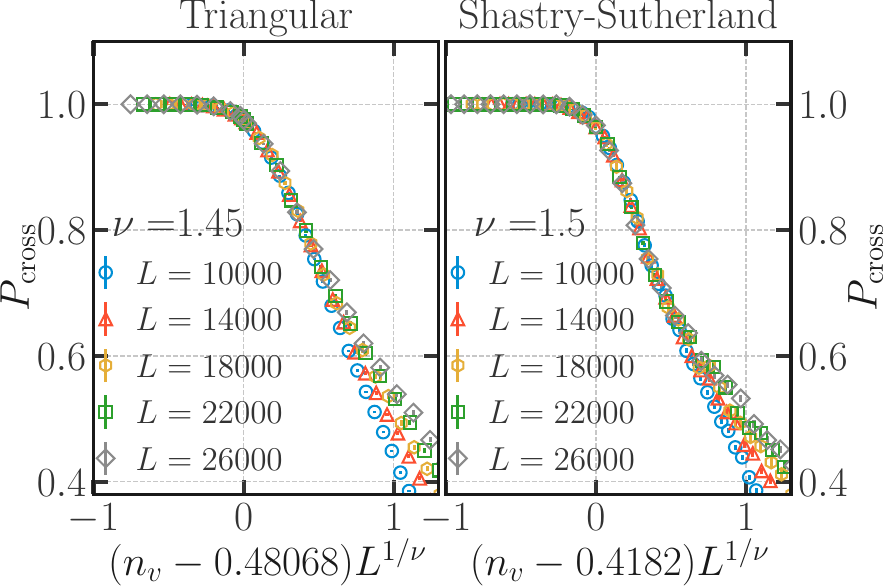} &
	(b)	\includegraphics[width=\columnwidth]{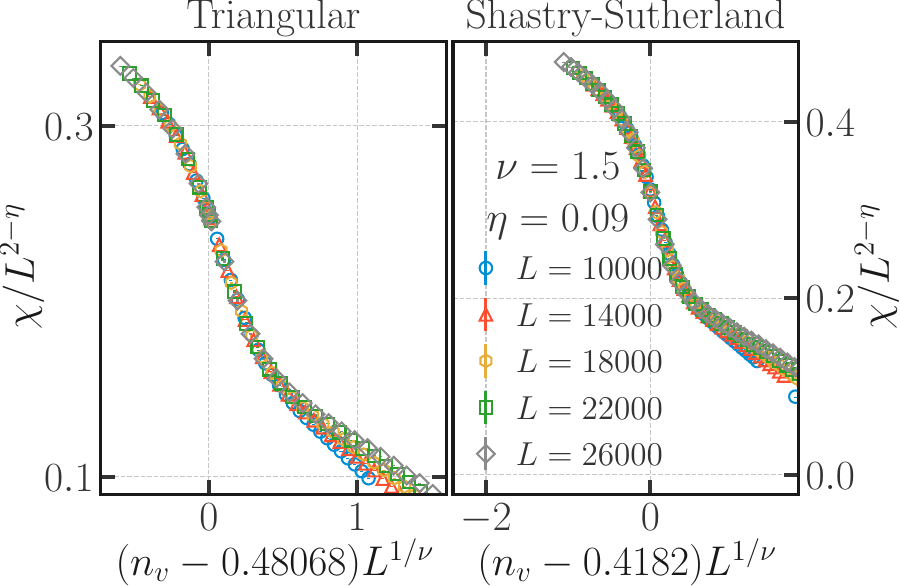}\\
 \end{tabular}
		\caption{(a) Data for the dilution dependence of $P_{\rm cross}(n_v, L)$, the probability that there exists a Gallai-Edmonds cluster that wraps around the torus in two independent directions for two dimensional $L\times L$ samples, collapses onto a single scaling curve in the vicinity of the critical point when plotted against the scaling variable $(n_v-n_v^{\rm crit})L^{1/\nu}$ for an appropriate choice of the exponent $\nu$. (b) Similarly, the susceptibility $\chi(L,n_v)$, associated with the sample-averaged correlation function $C(r-r')$, when scaled by $L^{2-\eta}$ and plotted against the scaling variable $(n_v - n_v^{\rm crit})L^{1/\nu}$, collapses onto a single curve in the vicinity of the critical point for appropriate choice of $\nu$ and $\eta$. The choices for $\nu$ and $\eta$ shown in the figure lie in a band of values, all of which yield scaling collapse of similar quality; this uncertainty feeds into the error on our estimate for these exponents: $\nu \approx 1.45(10)$, $\eta \approx 0.09(2)$.  See Sec.~\ref{sec:ComputationalMethodsObservables} and Sec.~\ref{sec:ComputationalResults} for details. \label{fig:scaling_Pcrosschi2D}}		
	\end{figure*}
	
	\subsubsection{Topology of clusters}
	\label{subsubsec:ClusterTopology}
In addition, we find it useful to keep track of the topology of each $\calR$-type and $\calP$-type region, that is, check whether it wraps round the torus in some way, and if yes, record the details.  In two dimensions, there are three possibilities for this topology: a given region does not wrap around the torus, or wraps around the torus in exactly one direction, or wraps around the torus simultaneously in two independent directions. In a similar way, there are four possibilities in three dimensions, since a region either does not wrap around the torus at all, or wraps around the torus in exactly one, two, or three different directions.  Using the efficient techniques developed in earlier studies of geometric percolation~\cite{Newman_Ziff_2001,Pruessner_Moloney_2003,Pruessner_Moloney_2004}, we gather this information on the fly, {\em i.e.}, during the construction of each Gallai-Edmonds cluster, with minimal overhead.

For each sample, this gives us the information needed to compute the number $n_{\rm wrap}$ of ``wrapping'' Gallai-Edmonds clusters, the number $n_{\rm cross}$ of ``crossing'' Gallai-Edmonds clusters, and the number $n_{\rm single}$ of ``single-wrapping'' Gallai-Edmonds clusters. Here, wrapping Gallai-Edmonds clusters are $\calR$-type or $\calP$-type regions that wrap around the torus in at least one direction, single-wrapping Gallai-Edmonds clusters are those that wrap around the torus in exactly one direction, and crossing Gallai-Edmonds clusters are those that wrap around the torus in all (two in $d=2$ and three in $d=3$) directions simultaneously. This information also allows us to measure the probability $P_{\rm single}$ that a sample has $n_{\rm single} > 0$ (which implies $n_{\rm cross} = 0$ and $n_{\rm wrap} = n_{\rm single}$ for planar lattices, but not in the three-dimensional case), the probability $P_{\rm cross}$ that a sample has $n_{\rm cross} > 0$, and the probability $P$ that $n_{\rm wrap} >0$. Naturally, we are also in a position to separately keep track of these wrapping numbers and probabilities for $\calR$-type and $\calP$-type regions, and we use this information to also measure $n_{\rm wrap}^{\calR / \calP}$, $n_{\rm single}^{\calR / \calP}$, $n_{\rm cross}^{\calR / \calP}$, as well as the corresponding probabilities $P^{\calR / \calP}$, $P_{\rm single}^{\calR / \calP}$, and $P_{\rm cross}^{\calR / \calP}$.

At this point, the reader may consider the rather large number of different observables described in the foregoing to be something of an overkill, with too much redundancy built into the array of quantities we measure. However, some of the phenomena we uncover are strikingly unusual and do not fit easily into any well-understood theoretical framework. We have therefore found a simultaneous analysis of this large set of quantities to be invaluable for obtaining a reliable overall picture. For instance, as will be clear from our discussion of the two-dimensional case, the actual percolation transition in two dimensions is preceded by a crossover visible in the $n_v$ dependence of the length scales $\xi$ and $R_{\rm max}$. The nature of this crossover only becomes clear when one studies the behavior of $P_{\rm single}$ in the same range of $n_v$. While we do not show the results for all these quantities in all regimes, the conclusions we present have been cross-checked by this kind of more detailed study in all regimes.
	\begin{figure*}
	\begin{tabular}{cc}
	(a) \includegraphics[width=\columnwidth]{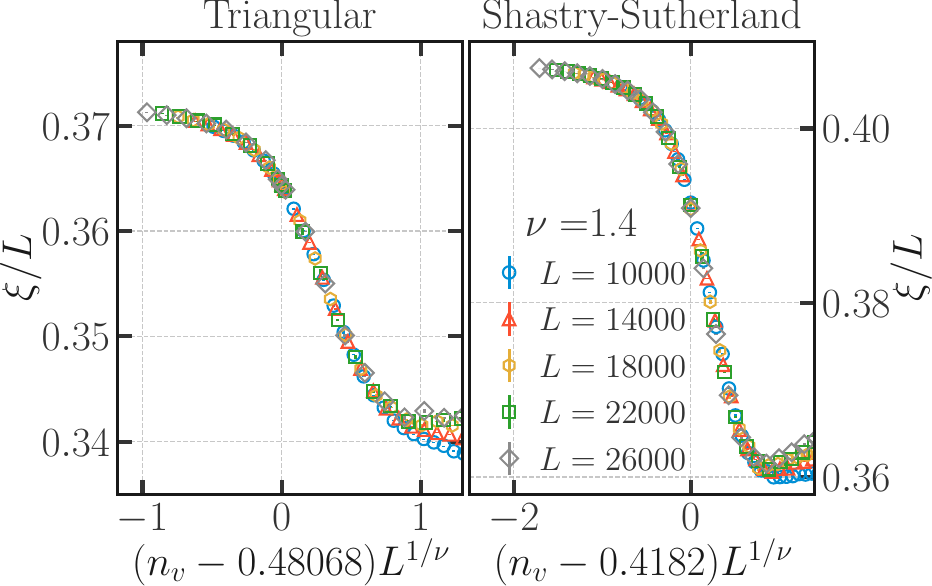} &
	(b) \includegraphics[width=\columnwidth]{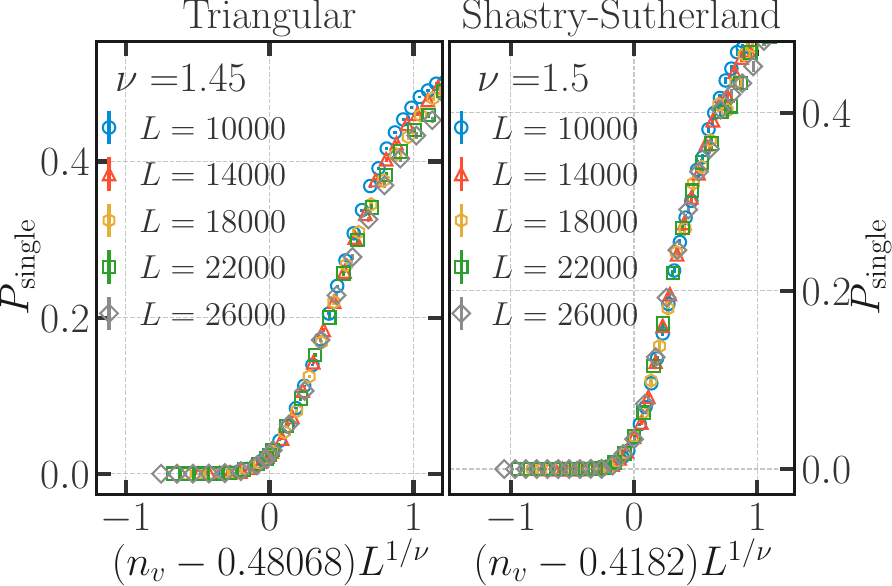}
	\end{tabular}
		\caption{ (a) Data for the dilution dependence of the dimensionless ratio $\xi(n_v,L)/L$, where $\xi(n_v,L)$ is the correlation length associated with the correlation function $C(r-r')$ in two dimensional $L\times L$ samples (for all $L$ studied here), collapses onto a single scaling curve when plotted against the scaling variable $(n_v-n_v^{\rm crit})L^{1/\nu}$. (b) The corresponding data for $P_{\rm single}(n_v, L)$, the probability that there exists a Gallai-Edmonds cluster that wraps around the torus in exactly one independent direction, also collapses onto a single scaling curve when plotted against the scaling variable $(n_v-n_v^{\rm crit})L^{1/\nu}$. The choices for $\nu$ shown in the figure lie in a band of values, all of which yield scaling collapse of similar quality; this uncertainty feeds into the error on our estimate for this exponent: $\nu \approx 1.45(10)$.  See Sec.~\ref{sec:ComputationalMethodsObservables} and Sec.~\ref{sec:ComputationalResults} for details.	\label{fig:scaling_xiPsingle2D}}
\end{figure*}

	\section{Computational results}
	\label{sec:ComputationalResults}
As noted earlier in Sec.~\ref{sec:IntroductionOverview}, the triangular and Shastry-Sutherland lattices that we study in two dimensions, as well as the stacked triangular and corner-sharing octahedral lattices that are our focus in the three-dimensional case, all have the property that site dilution with a nonzero vacancy density $n_v$ leads to a nonzero density $w = W/m(\calG) $ of monomers in maximum matchings of these lattices. 

The fact that $w$ is nonzero in the thermodynamic limit in {\em every sample} actually follows from simple constructions of monomer-carrying regions of the type discussed in Ref.~\cite{Damle_2022}. Such monomer-carrying regions are formed when multiple vacancies are located near each other at specific relative positions. As this construction is {\em local}, there is a nonzero probability for such a region to occur in any small patch of the sample, independent of the morphology of the rest of the sample. Therefore, the existence of any such construction implies that $w$ is nonzero in every sample, and provides a strictly positive lower bound on the ensemble-averaged density $\langle w \rangle$ in the thermodynamic limit.

This is expected to be the generic behavior, and in this sense, the examples we focus on here are generic. However, it is perhaps useful to note at this point that there do exist examples of well-known lattice geometries, for example, the two-dimensional kagome lattice and the three-dimensional pyrochlore lattice, in which a nonzero density $n_v$ of vacancies does {\em not} lead to a nonzero monomer density $w$~\cite{Faudree_1997,Sumner_1974,Ansari_Damle_2024,Bhola_Damle_ClawFree2024}.

Returning to the generic case studied here, such explicit constructions are of course too crude to obtain a quantitatively accurate estimate of $ w$, nor do they provide any information about the morphology of {\em typical} monomer-carrying regions at a given dilution $n_v$; rather they are useful only insofar as they suggest that $\langle w \rangle $ decreases monotonically with $n_v$. Our computational results demonstrate that $\langle w \rangle $ is indeed nonzero in the thermodynamic limit of large $L$, and that it does indeed decrease with $n_v$. This is displayed in Fig.~\ref{fig:w} for all the lattice geometries studied here.

With these basic facts established, we now turn to a study of the structure of $\calR$-type and $\calP$-type regions of these site-diluted lattices.
The phenomena we identify and study have some surprising and rather unusual features, and it is therefore important to ensure that we do not lose sight of the wood for the trees in our presentation of the facts. Therefore, we  focus in the main text on a minimal set of observables that lead us to unambiguous conclusions about the large-scale geometry, relegating many other pieces of supporting evidence to an Appendix that can be consulted for further details. 
		 		 \begin{figure*}[ht]
	\begin{tabular}{cc}
		\includegraphics[width=\columnwidth]{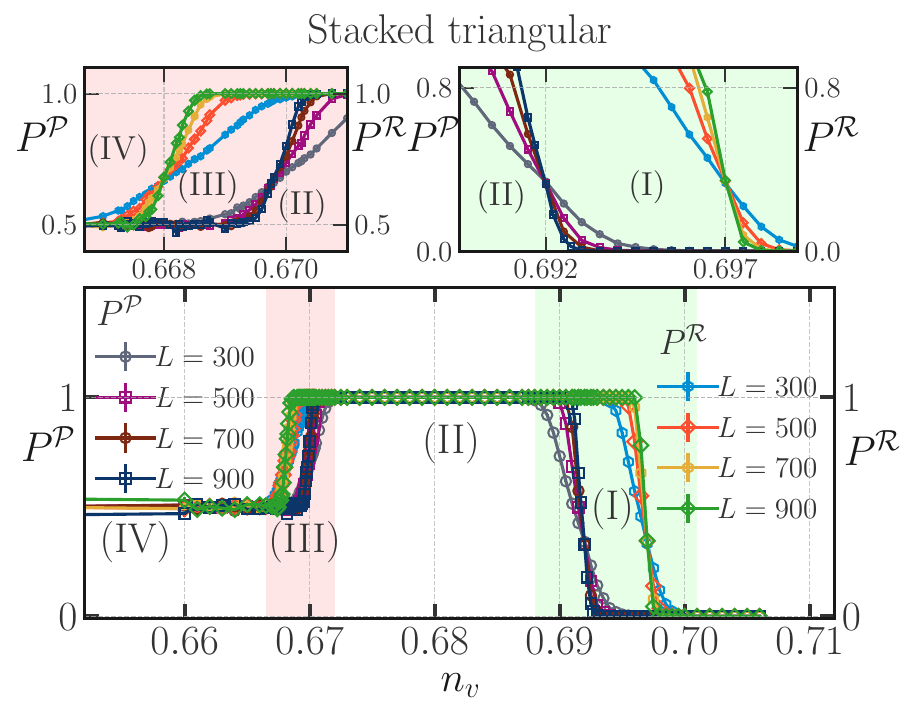}		
	&	\includegraphics[width=\columnwidth]{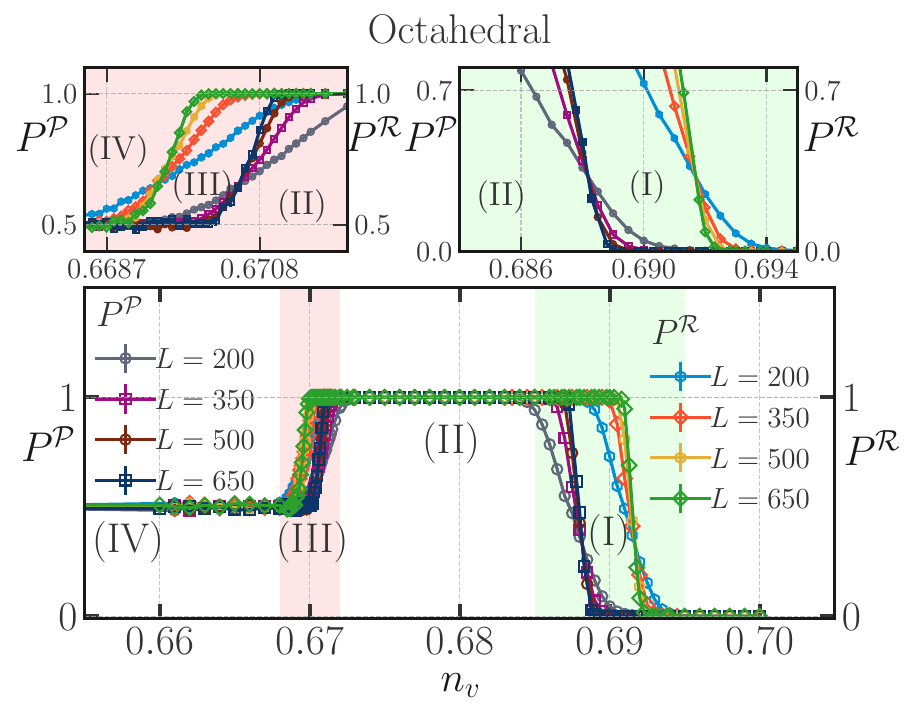}\\	
	\end{tabular}
	\caption{$P^{\calR}(n_v)$, the probability that a sample has an $\calR$-type region that wraps around the three-dimensional torus in some way, and $P^{\calP}(n_v)$, the corresponding probability that a sample has a $\calP$-type region that wraps in some way, both show clear signatures of two distinct transitions each as $n_v$ is lowered. In other words, we have a total of four distinct transitions: As $n_v$ is lowered, data for $P^{\calR}(n_v)$ corresponding to different sizes $L$ in the stacked triangular [octahedral] lattice case first cross at $n_v^{c_1}= 0.6970(5)$ [$n_v^{c_1}=0.6917(5)$], below which $P^{\calR}(n_v) =  1.0$ independent of $L$. This behavior of $P^{\calR}(n_v)$ persists up to a second crossing point at $n_v^{c_4}=0.6680(2)$ [$n_v^{c_4}=0.6694(4)$], below which $P^{\calR}(n_v)\approx 0.50(2)$. $P^{\calP}(n_v)$ shows similar behavior, but with different locations for the two crossing points:  On the stacked triangular [octahedral] lattice, the crossing point below which $P^{\calP}(n_v)$  saturates to $P^{\calP}(n_v) = 1.0$   is at  $n_v^{c_2}=0.6920(3)$ [$n_v^{c_2}=0.6882(3)$], while the second crossing point, at which curves for $P^{\calP}(n_v)$ drop down to a value of $P^{\calP}(n_v)\approx 0.50(3)$, is at $n_v^{c_3}= 0.66975(10)$ [$n_v^{c_3}=0.6705(4)$]. From the zoomed in insets, we see that the crossing points are all sharply defined; indeed, the error bars quoted here for their locations also fold in the slightly different crossing points displayed by other observables (analyzed separately). The data shown here thus indicates the presence of one unpercolated phase for $n_v > n_v^{c_1}$ and four distinct percolating phases separated from each other by transitions at $n_v^{c2}$, $n_v^{c_3}$, and $n_v^{c_4}$. See Sec.~\ref{sec:ComputationalMethodsObservables} and Sec.~\ref{sec:ComputationalResults} for details. \label{fig:POverview3D}
}		
\end{figure*}

	\subsection{Results in $d=2$}
	\label{subsec:Results2D}
 We structure our narrative to first understand these unusual behaviors in the simpler two-dimensional setting in this subsection, and then build on this in our discussion of the more intricate phase diagram in three dimensions in the next subsection.

	\subsubsection{Thermodynamic densities in $d=2$}
	\label{subsubsec:ThermodynamicDensities2D}
The random geometry of $\calR$-type and $\calP$-type regions is determined by the manner in which sites carrying e-type, u-type and o-type labels (obtained from  Edmonds' matching algorithm) are organized in a sample. As a preliminary step, we first study the density of each of these types of sites. We begin by examining $m_{\rm tot}^\calR$, the fraction of sites in $\calR$-type regions. From the upper panels of Fig.~\ref{fig:mtotR2D}, we see that $\langle m_{\rm tot}^\calR \rangle$ is nonzero in the thermodynamic limit. Below a threshold value of $n_v$, the data displayed in these upper panels of Fig.~\ref{fig:mtotR2D} are seen to be rather noisy. This noisy behavior is seen to extend down to the lowest $n_v$ accessible to our computations. In this rather wide range of $n_v$, we also find that $\langle m_{\rm tot}^\calR \rangle $ has essentially no $n_v$ dependence. This is shown in the insets of these upper panels of Fig.~\ref{fig:mtotR2D}. Thus, the most striking feature of our data for $\langle \mtotR \rangle$ is not its $n_v$ dependence. Rather, it is this noisy behavior at low $n_v$.

	We emphasize that data at all $n_v$ displayed in Fig.~\ref{fig:mtotR2D} were obtained by averaging over the same number of samples at each $n_v$ (with a larger number of samples at smaller sizes and vice versa). So the reason for the noisy character of the data at lower values of $n_v$ is not poorer sampling of the disorder ensemble. Rather, it is a dramatic change in the {\em distribution} of $m_{\rm tot}^\calR$. This is evident from the lower panels of Fig.~\ref{fig:mtotR2D}. Below a threshold value of $n_v$, $m_{\rm tot}^\calR$ has a bimodal distribution, with two well-separated peaks, each of which narrows and becomes more sharply-defined with increasing $L$. This separation between the peaks indicates a {\em macroscopic} difference in the morphology of the corresponding samples, {\em i.e.}, a difference that remains significant in the thermodynamic limit (since $m_{\rm tot}^\calR$, as defined by us, is an intensive fraction or density).

	To understand this better, we now separate samples by asking which of the two peaks each sample belongs to. This is unambiguous because the peaks are well-separated even at the smallest size we study.	
	 We label the samples belonging to the peak with larger 
	$m_{\rm tot}^\calR$ as $\calR$-type samples, and those in the peak with smaller $m_{\rm tot}^\calR$ are labeled $\calP$-type.  This terminology is well-motivated, since samples with a high value of $m_{\rm tot}^\calR$ always have a low value of $m_{\rm tot}^\calP$ and vice-versa; this follows from the fact that $m^{\calR}_{\rm tot} + m^{\calP}_{\rm tot} = 1$. We define $f_{\calR}$ to be the fraction of $\calR$-type samples (defined as described above) in the ensemble, with $1-f_{\calR}$ being the corresponding fraction of $\calP$-type samples. From Fig.~\ref{fig:mtotR2D}, we see that the weight of the two peaks appears nearly equal; indeed we find that $f_{\calR} \approx 0.50(2)$ independent of $n_v$ in this low-dilution regime. A more detailed discussion of this interesting aspect is postponed to Sec.~\ref{subsec:DueDiligence}.
							 \begin{figure*}[ht]
	\begin{tabular}{cc}
	    \includegraphics[width=\columnwidth]{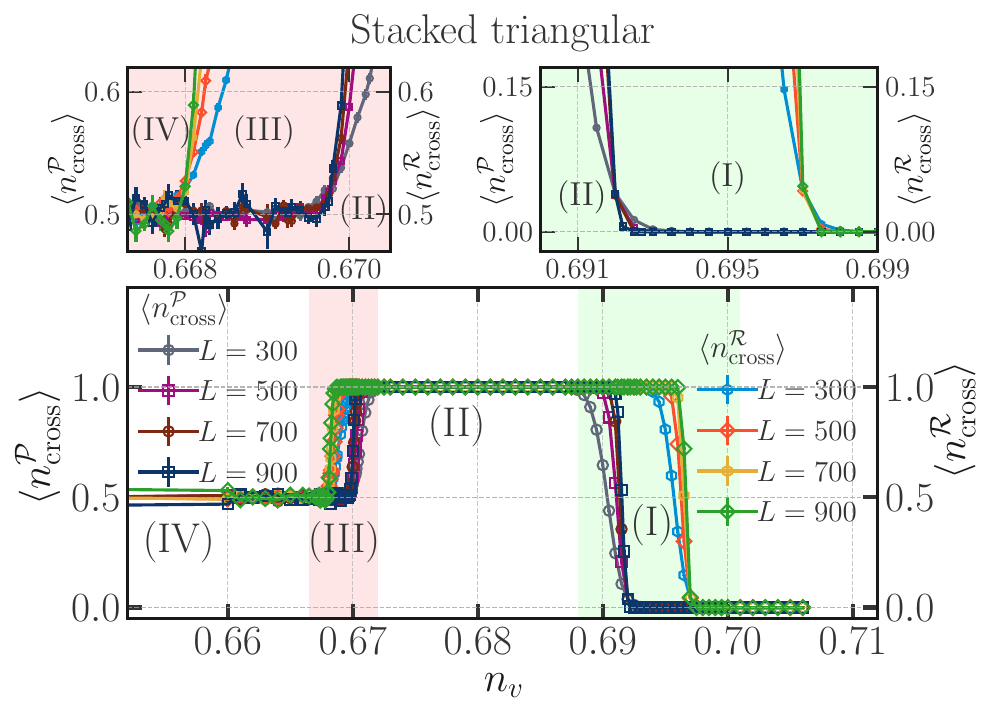}
	    		
	&	\includegraphics[width=\columnwidth]{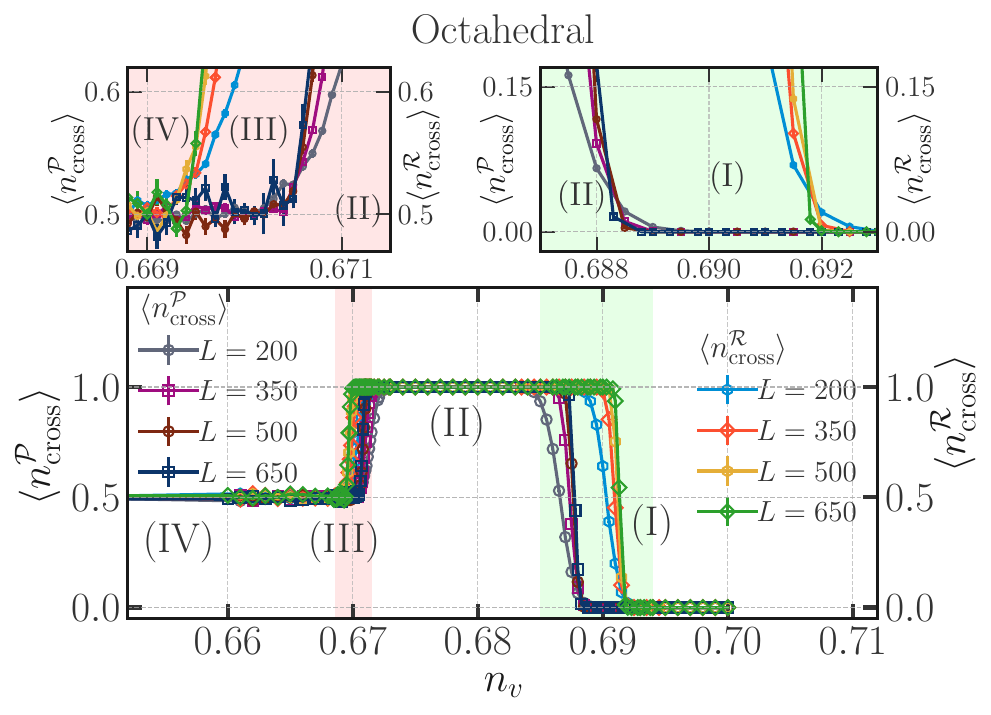}\\
	\end{tabular}
	\caption{The data shown here for $n_{\rm cross}^{\calR}$ and $n_{\rm cross}^{\calP}$, respectively the number of $\calR$-type and $\calP$-type regions in a sample that wrap around the three-dimensional torus in three independent directions, should be viewed in conjunction with Fig.~\ref{fig:POverview3D} since it provides additional corroboration regarding the nature of the four distinct percolated phases identified in Fig.~\ref{fig:POverview3D}. For $n_v \in (n_v^{c_2}, n_v^{c1})$ (except in the immediate vicinity of the transition points), $\langle n_{\rm cross}^{\calR}\rangle = 1$ even at the largest size studied, while $\langle n_{\rm cross}^{\calP} \rangle \to 0$ with increasing $L$; in the thermodynamic limit, there is thus exactly one $\calR$-type Gallai-Edmonds cluster that wraps around the torus, and it does so in all three directions simultaneously, while there is no $\calP$-type region that wraps around the torus in any way. For $n_v \in (n_v^{c_3}, n_v^{c2})$ (except in the immediate vicinity of the transition points), there is exactly one $\calR$-type and one $\calP$-type region that wraps around the torus even at the largest sizes, and both these regions do so in all three directions simultaneously, {\em i.e.}, $\langle n_{\rm cross}^{\calR}\rangle =\langle n_{\rm cross}^{\calP}\rangle = 1$ in the thermodynamic limit. For $n_v \in (n_v^{c_4}, n_v^{c3})$ (except in the immediate vicinity of the crossing points), $\langle n_{\rm cross}^{\calR}\rangle = 1$ while $\langle n_{\rm cross}^{\calP}\rangle \approx 0.50(3)$ in the thermodynamic limit; in this regime, we have checked that $\calR$-type samples have $n_{\rm cross}^{\calP}=0$, while $\calP$-type samples have  $n_{\rm cross}^{\calP}=1$. And finally, for $n_v < n_v^{c_4}$, $\langle n_{\rm cross}^{\calR}\rangle \approx \langle n_{\rm cross}^{\calP}\rangle \approx 0.50(2)$ in the thermodynamic limit; in this regime, we have checked that $n_{\rm cross}^{\calR}=1$ ($n_{\rm cross}^{\calP}=1$) in $\calR$-type ($\calP$-type) samples, while $n_{\rm cross}^{\calR}=0$ ($n_{\rm cross}^{\calP}=0$) in $\calP$-type ($\calR$-type) samples.  See Sec.~\ref{sec:ComputationalMethodsObservables} and Sec.~\ref{sec:ComputationalResults} for details. \label{fig:ncrossOverview3D}}
\end{figure*}

	When analyzed in this manner, the data paints a very clear picture which we now describe: Consider first the histograms of $m_{\rm max}^{\calR}$ and $m_{\rm max}^{\calP}$, the fraction of sites of $\calG$ that belong to the largest $\calR$-type and largest $\calP$-type region respectively. From Fig.~\ref{fig:hist_mmaxpmax2D} which displays these histograms at a representative value of $n_v$ in this bimodal regime, it is clear that these histograms show perfectly correlated bimodal behavior: In $\calR$-type samples, the histogram of $m_{\rm max}^{\calR}$ has a single peak at a fairly large $L$-independent value, with the peak becoming sharper with increasing $L$. The same is true of $m_{\rm max}^{\calP}$ in $\calP$-type samples.
However, and this is absolutely key, the histogram of $m_{\rm max}^{\calR}$ in $\calP$-type samples displays a peak at a very low value, which continuously shifts to even lower values with increasing $L$. The behavior of $m_{\rm max}^{\calP}$ in $\calR$-type samples is analogous.

In addition, we find that $n_{\calR}$ and $n_{\calP}$, respectively number densities of $\calR$-type and $\calP$-type regions in $\calG$, also have perfectly correlated bimodal distributions: $\calR$-type samples contribute exclusively to the left peak (at smaller values of $n_{\calR}$) of the distribution of $n_{\calR}$, while $\calP$-type samples contribute exclusively to the right peak (at larger values of $n_{\calR}$) of this distribution. The distribution of $n_{\calP}$ displays entirely analogous behavior, except for a reversal of the role of $\calR$-type and $\calP$-type samples. This is clear for instance from the data analyzed in Fig.~\ref{fig:App:hist_nR_nP_2D} in the Appendix. 

Thus, the picture that emerges is of fewer but much larger $\calR$-type regions in $\calR$-type samples compared to $\calP$-type samples (with the largest $\calR$-type region being macroscopic in size in $\calR$-type samples but not $\calP$-type samples), and entirely analogous behavior of $\calP$-type regions, except for an interchange of the roles of $\calR$-type and $\calP$-type samples.
 							 \begin{figure*}[t]
	\begin{tabular}{cc}
	    \includegraphics[width=\columnwidth]{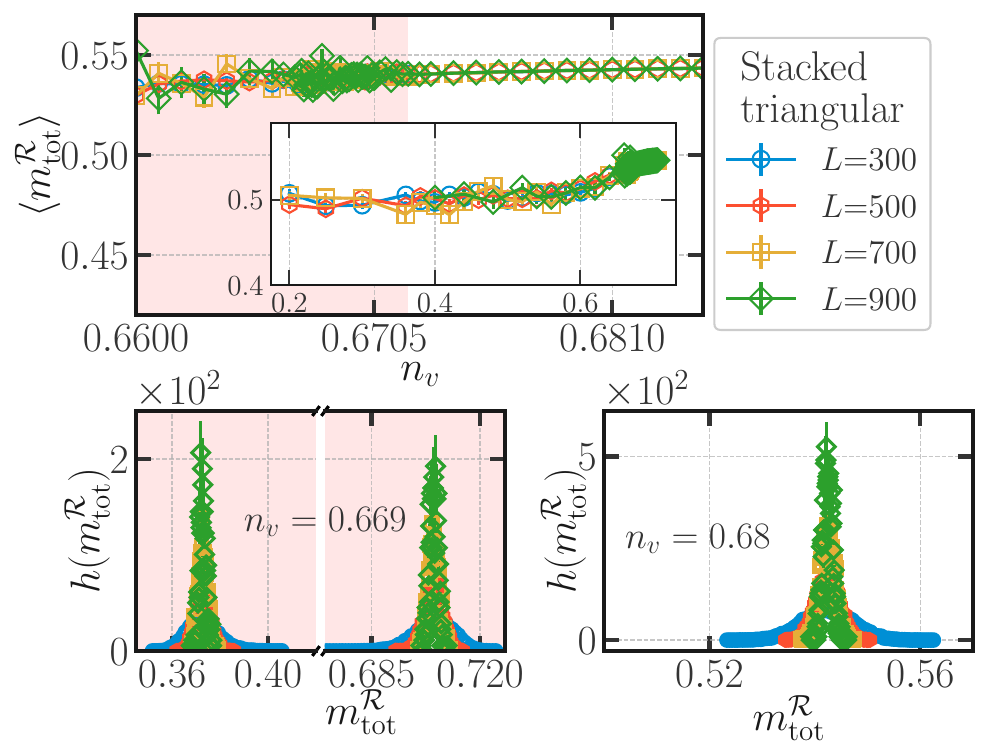}
	    		
	&	\includegraphics[width=\columnwidth]{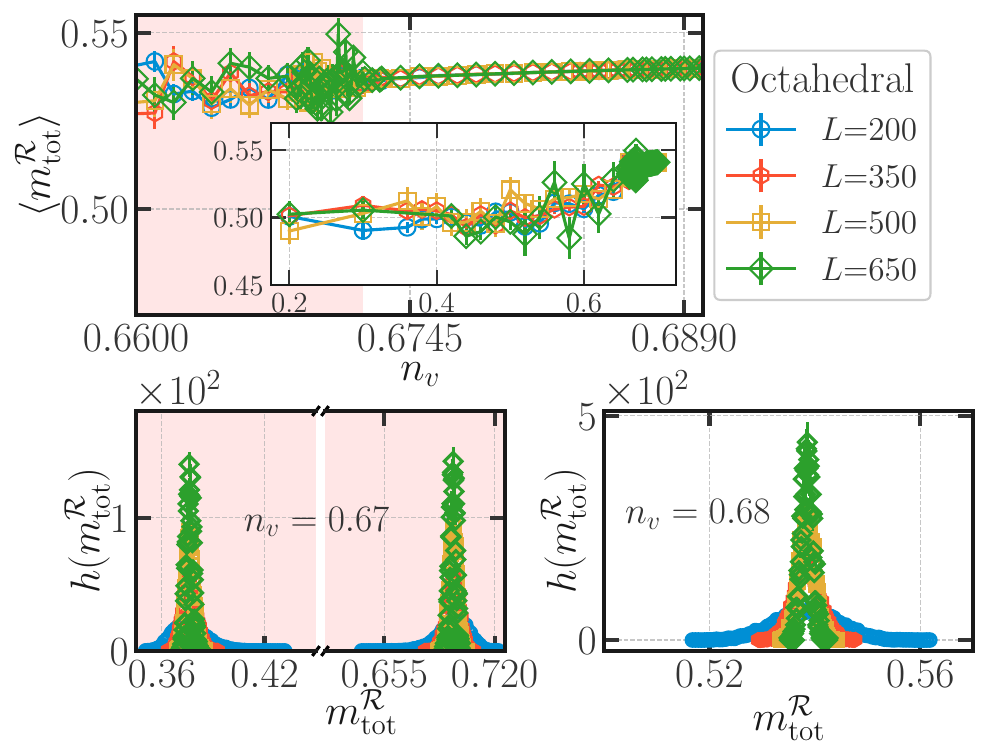}\\
	\end{tabular}
	\caption{In the regime $n_v \in (n_v^{c_4}, n_v^{c3})$ (identified in Fig.~\ref{fig:POverview3D} and Fig.~\ref{fig:ncrossOverview3D}), the large-scale geometry of Gallai-Edmonds regions on both the stacked triangular lattice and the Shastry-Sutherland lattice violates thermodynamic self-averaging. This is clear from the histogram of $\mtotR$ at a representative point in this regime, which shows two well-separated peaks which get more and more sharply defined with increasing $L$ without any shift in their position. Throughout this phase, we find that the weight in the right peak of this histogram is $f_{\calR} \approx 0.50(3)$ at large $L$. See Sec.~\ref{sec:ComputationalMethodsObservables} and Sec.~\ref{sec:ComputationalResults} for details. 
		\label{fig:hist_mtot3D}}
\end{figure*}		

Next we consider the histograms of $\RmaxR/L$ and $\RmaxP/L$,  the corresponding radii of gyration scaled by the system size $L$, in both groups of samples. From Fig.~\ref{fig:hist_RmaxRRmaxP2D}, which displays these histograms at a representative value of $n_v$ in this bimodal regime, we again see that these histograms also have perfectly correlated bimodal behavior: In $\calR$-type samples, the histogram of $\RmaxR/L$ has a single peak at a sizeable $L$-independent value, with the peak becoming sharper with increasing $L$.  However, in $\calP$-type samples, the histogram of $\RmaxR/L$ displays a single peak at a very low value which continuously shifts to even lower values with increasing $L$. The histograms of $\RmaxP$ show analogous behavior, with the role of $\calR$-type and $\calP$-type samples being reversed.

We find that the histograms of $\chi^{\calR}$ and $\chi^{\calP}$ show behavior similar to that of $\mmaxR$ and $\mmaxP$. Likewise, the histograms of $\xi^{\calR}$ and $\xi^{\calP}$ show behavior entirely analogous to that of $\RmaxR$ and $\RmaxP$. However, and this could perhaps have been anticipated from the lack of any noisy behavior in our data for the monomer density $\langle w \rangle$ at small $n_v$ (Fig.~\ref{fig:w}),  the histogram of the monomer density $w$ {\em does not show} any such distinction between $\calR$-type and $\calP$-type samples in the percolated phase. Thus, the monomer density $w$ is self-averaging despite the fact that large-scale morphology of monomer-carrying $\calR$-type regions is {\em not} self-averaging in the thermodynamic limit. We find that this is related to the fact that the local density of monomers inside individual $\calR$-type regions is higher in smaller $\calR$-type regions.

From this data analysis, we thus have a clear picture of a rather remarkable phenomenon: below a certain threshold value of the dilution $n_v$, the morphology of $\calR$-type and $\calP$-type regions shows a complete lack of self-averaging even in the thermodynamic limit. 
 However, the overall density $w$ of monomers does not show this anomalous behavior. The threshold at which this bimodality sets in appears to reflect some kind of unusual percolation transition, since $m_{\rm max}^{\calR}$ ($m_{\rm max}^{\calP}$) appears to saturate to a nonzero value in the thermodynamic limit of $\calR$-type ($\calP$-type) samples, as do $\RmaxR/L$ ($\RmaxP/L$). The detailed analysis in the next section confirms this expectation, and pinpoints the location of this percolation transition.

	\subsubsection{Gallai-Edmonds percolation in $d=2$}
	\label{subsubsec:PercolationTransition2D}	
	To confirm the existence of this putative percolation transition and pin down its precise location, we find it useful to study the probability $P_{\rm cross}$, the probability that there is at least one Gallai-Edmonds cluster (either an $\calR$-type region or a $\calP$-type region) that wraps simultaneously in two independent directions around the torus. We find, as shown in Fig.~\ref{fig:Pcross2D}, that there is a sharply-defined threshold value $n_v^{\rm crit}$ of the dilution at which the data curves for $P_{\rm cross}$ (as a function of $n_v$) corresponding to different sizes all cross each other. This is the expected behavior of a dimensionless quantity like $P_{\rm cross}$ at a critical point. Therefore, we identify this crossing point with the location of this {\em Gallai-Edmonds percolation} transition. We estimate $n_v^{\rm crit} = 0.48068(4)$ [$n_v^{\rm crit} = 0.41820(5)$] on the triangular [Shastry-Sutherland] lattice based on the crossing displayed in Fig.~\ref{fig:Pcross2D} and similar data for other quantities that we discuss below. 	
						\begin{figure*}
		\begin{tabular}{cc}
			a)\includegraphics[width=\columnwidth]{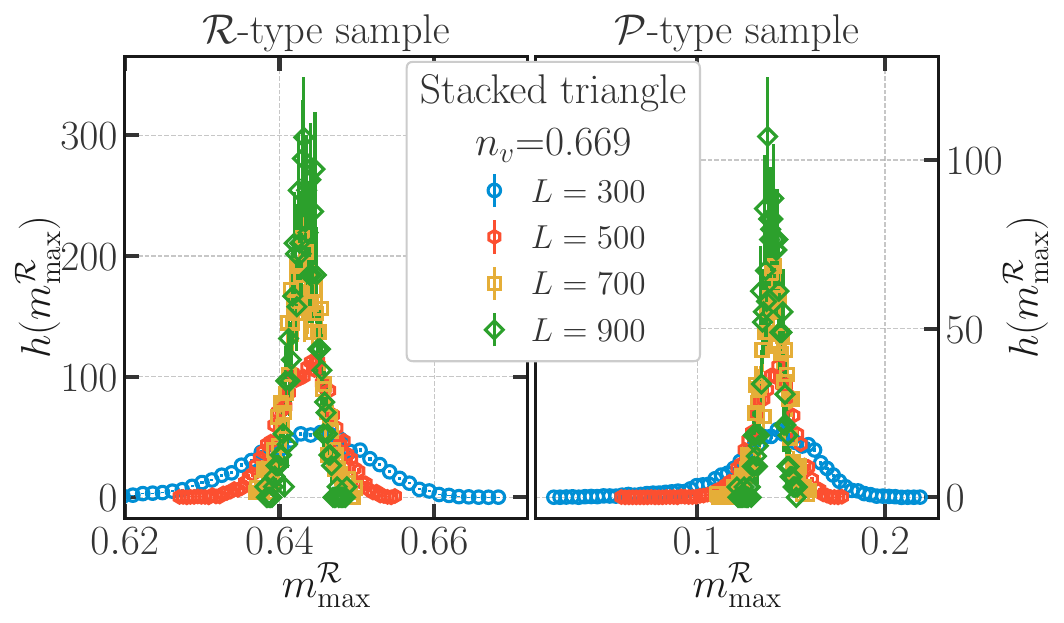} &	
			b)\includegraphics[width=\columnwidth]{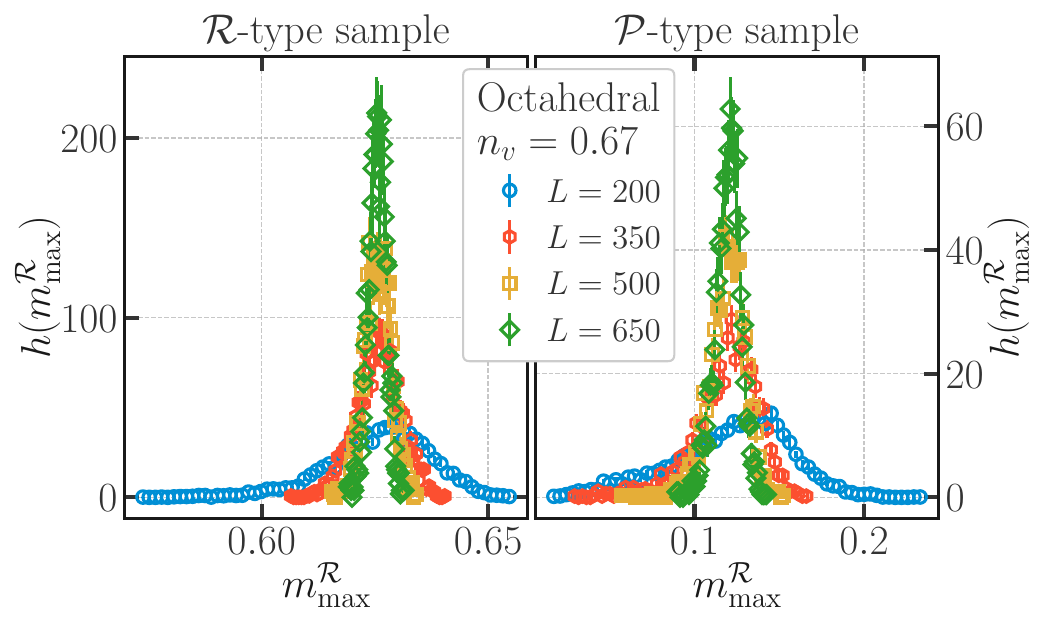}\\
			c)\includegraphics[width=\columnwidth]{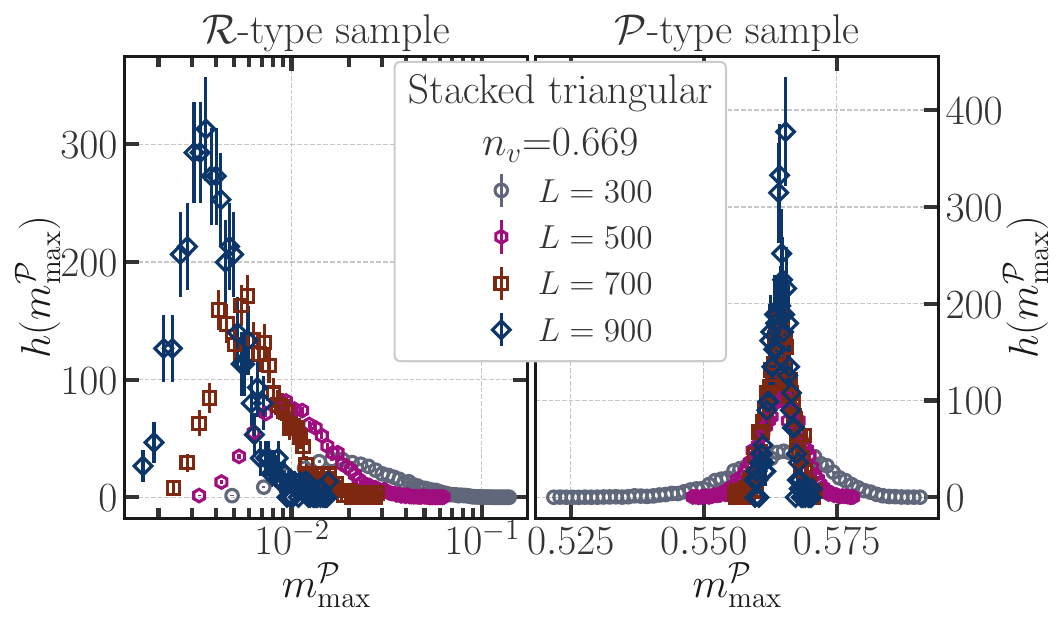} &	
			d)\includegraphics[width=\columnwidth]{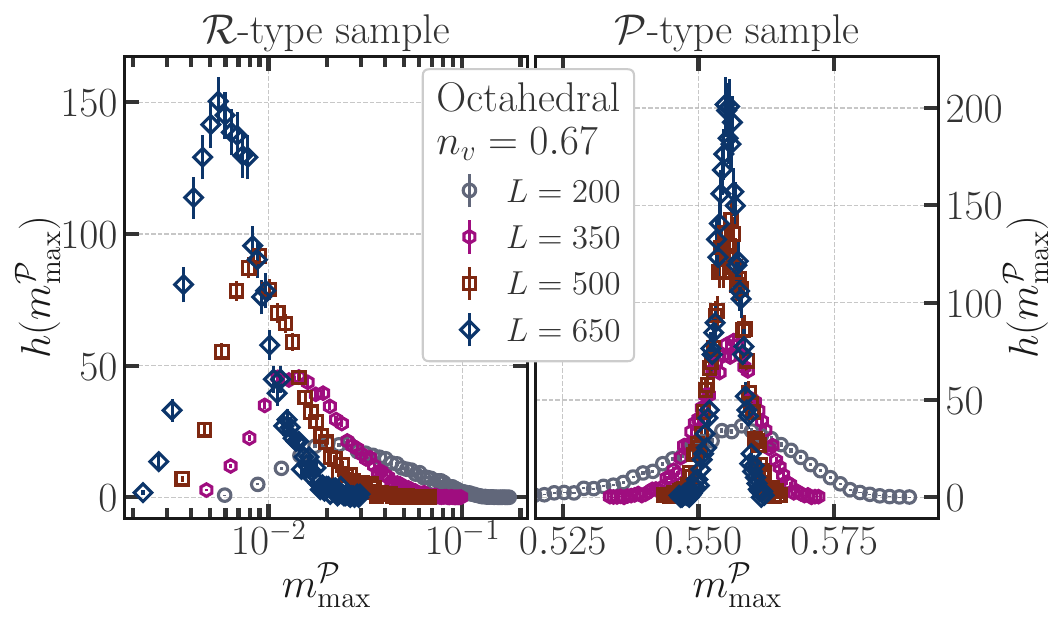}\\
		\end{tabular}
		\caption{
			At a representative value of $n_v$ that places the system in phase III, each sample is labeled $\calR$-type or $\calP$-type based on whether it contributes to the right peak or left peak respectively in the histogram of $\mtotR$ in Fig.~\ref{fig:hist_mtot3D}; this labeling is unambiguous since these two peaks in Fig.~\ref{fig:hist_mtot3D} are well-separated and sharply defined. (a) and (b) display histograms of $\mmaxR$, the mass of the largest $\calR$-type region, separately for each type of sample. For each type of sample, this histogram has a single peak at a nonzero $L$-independent value of $\mmaxR$, which becomes more and more sharply defined with increasing $L$. Thus, although the peak for $\calP$-type samples corresponds to a significantly lower value of $\mmaxR$ compared to the corresponding peak for $\calR$-type samples, $\calP$-type samples also have a nonzero $\mmaxR$ in the thermodynamic limit. (c) and (d) display histograms of $\mmaxP$, the mass of the largest $\calP$-type region, separately for each type of sample. In sharp contrast to the histogram of $\mmaxR$ in $\calP$-type samples, we see that the histogram of $\mmaxP$ in $\calR$-type samples has a single peak that moves to smaller and smaller values of $\mmaxP$ as $L$ is increased indicating that $\mmaxP$ tends to zero at large $L$ in $\calR$-type samples. In $\calP$-type samples on the other hand, $\mmaxP$ has a single peak at a nonzero value of $\mmaxP$, which gets more and more sharply defined with increasing $L$.  See Sec.~\ref{sec:ComputationalMethodsObservables} and Sec.~\ref{sec:ComputationalResults} for details. \label{fig:hist_mmaxpmax3D}}	
	\end{figure*}

	By way of additional confirmation, we also study the susceptibility $\chi$ associated with the correlation function $C(r-r')$ defined in Eq.~\ref{eq:chi}, and check if it shows signatures of a sharply-defined percolation transition at $n_v^{\rm crit}$. To test for this, we ask if $\chi/L^2$ tends to a nonzero $L$-independent value in the large $L$ limit for $n_v$ well below $n_v^{\rm crit}$, while tending to zero in this limit for $n_v$ well above this critical dilution. We find that this is indeed the case, as is clear from the data presented in the insets of Fig.~\ref{fig:chi2D}. Next, we estimate $\eta$, the anomalous exponent associated with the power-law behavior of $C(r-r')$ at criticality, by identifying the value of $\eta$ for which curves for $\chi/L^{2-\eta}$ as a function of $n_v$ (for different values of $L$) cross at a threshold value of dilution that is within errors the same as the previous estimate of $n_v^{\rm crit}$. We find that this is the case if $\eta$ lies in a relatively narrow band of values around $\eta = 0.09$, yielding the estimate $\eta \approx 0.09(2)$ for both the triangular and Shastry-Sutherland lattices. This is shown in Fig.~\ref{fig:chi2D}.
	
For a direct probe of the growing length scale associated with this percolation transition, we now study the $n_v$ dependence of $\xi $, the correlation length associated with the correlation function $C(r-r')$ defined in Eq.~\ref{eq:xi}. We find that data curves for $\xi /L$ (corresponding to different sizes $L$) display a sharp crossing at a value of $n_v$ consistent within our errors with the critical point $n_v^{\rm crit}$ determined from the crossing of $P_{\rm cross}$. For $n_v > n_v^{\rm crit}$ but close to the critical point, we also find that $ \xi $ has interesting non-monotonic $n_v$ dependence on both the triangular lattice and the Shastry-Sutherland lattice. In this range of $n_v$, $\xi/L$ for each $L$ has a local maximum, with the value of $n_v$ corresponding to this local maximum shifting to lower dilution with increasing size. This is evident from the data displayed in Fig.~\ref{fig:xi2D}.

	Since $\chi/L^2$ decreases with increasing $L$ in this range of $n_v$ in which $\xi/L$ displays this peak structure, it is clear that this regime does not exhibit percolation in the conventional sense ({\em i.e.}, does not have a giant Gallai-Edmonds cluster in the thermodynamic limit).  Nevertheless, the fact that the peak height at the local maximum of $\xi/L $ in Fig.~\ref{fig:xi2D} appears to be close to saturating (as a function of size $L$) at the largest value of $L$ accessed in our study suggests that this behavior is associated with some other large-scale property of the Gallai-Edmonds clusters in this regime.
	
To explore this further, we examine the behavior of $P_{\rm single}$, the probability that the sample has no Gallai-Edmonds cluster that wraps around the torus in two independent directions, but has a Gallai-Edmonds cluster that wraps around the torus in exactly one direction. We find that $P_{\rm single}$ also displays a well-defined maximum in the range of $n_v$ that corresponds to the non-monotonic behavior and local maximum of $\xi/L$. On both the triangular lattice and the Shastry-Sutherland lattice, this maximum in $P_{\rm single}$ also shifts to lower and lower values of $n_v$ as $L$ is increased. The corresponding data is displayed in Fig.~\ref{fig:Psingle2D}.

	From this behavior of $\xi/L$ and $P_{\rm single}$, we see that this maximum in both dimensionless quantities corresponds to a crossover and not a sharply defined transition. Since $\chi/L^2$ decreases with increasing $L$ in this range of $n_v$, we conclude that this crossover regime is characterized by clusters that are large in terms of their linear dimension (as characterized by the radius of gyration), which are able to wrap around the torus in one direction, but do not have a mass that scales with $L^2$. As we will see in Sec.~\ref{subsubsec:FinteSizeScaling2D}, it is this unusual non-monotonic dependence that appears to limit the range in $n_v$ for which finite-size scaling forms reliably describe the scaling collapse of the data in the critical regime for $n_v$ slightly greater than $n_v^{\rm crit}$. 
		 			\begin{figure*}
		\begin{tabular}{cc}
			a)\includegraphics[width=\columnwidth]{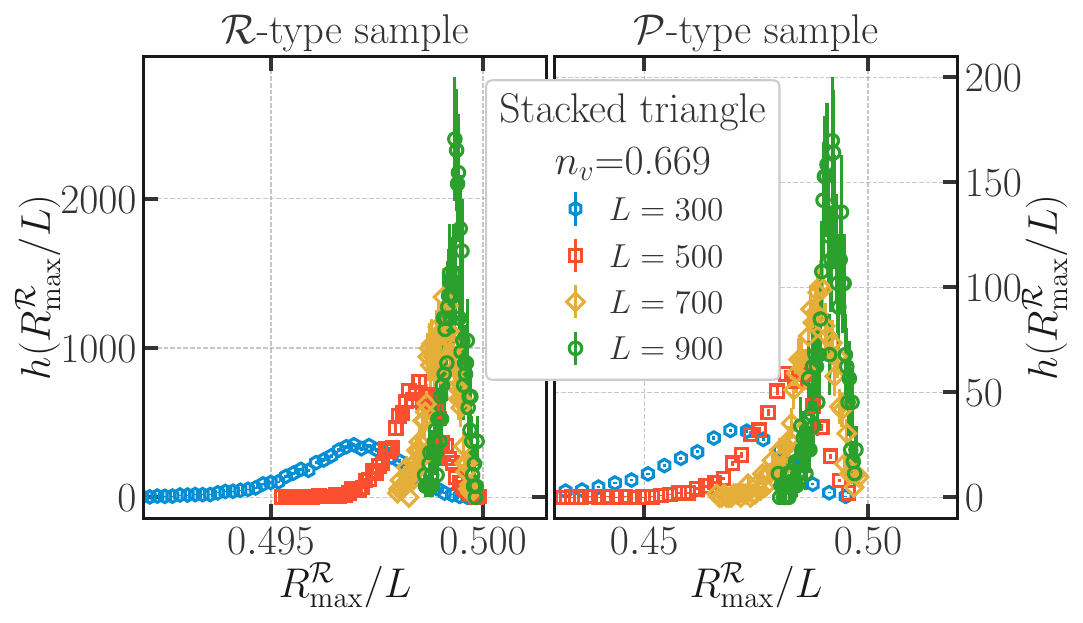} &	
			b)\includegraphics[width=\columnwidth]{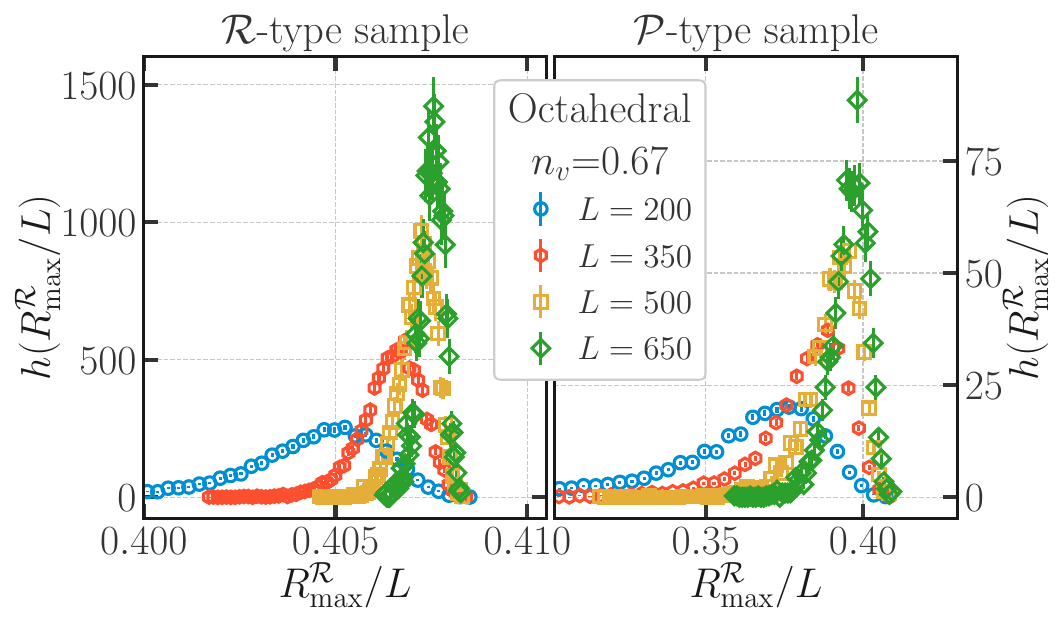}\\
			c)\includegraphics[width=\columnwidth]{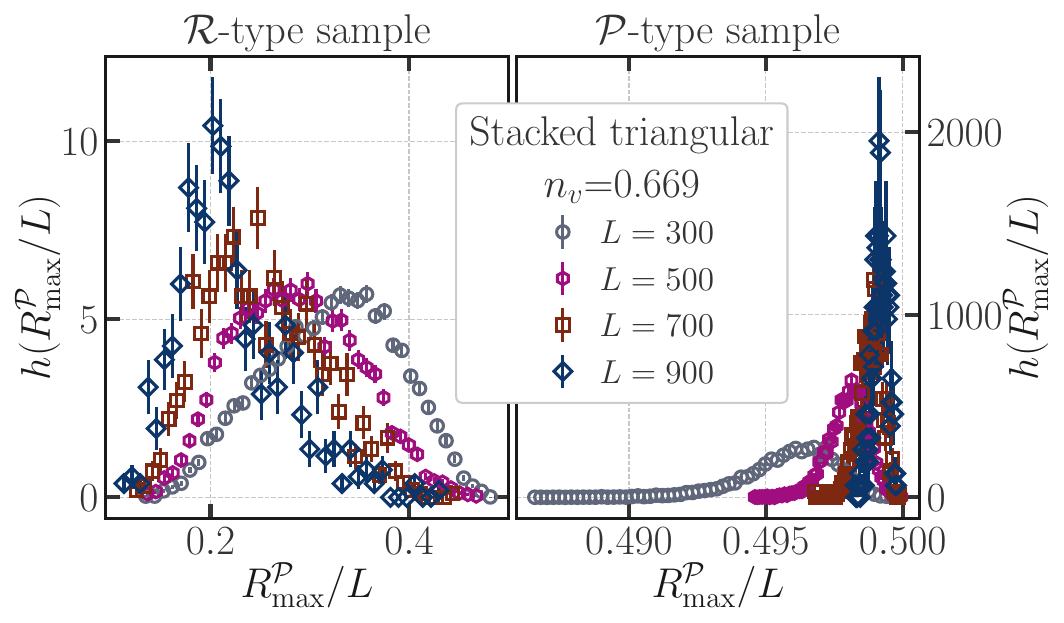} &	
			d)\includegraphics[width=\columnwidth]{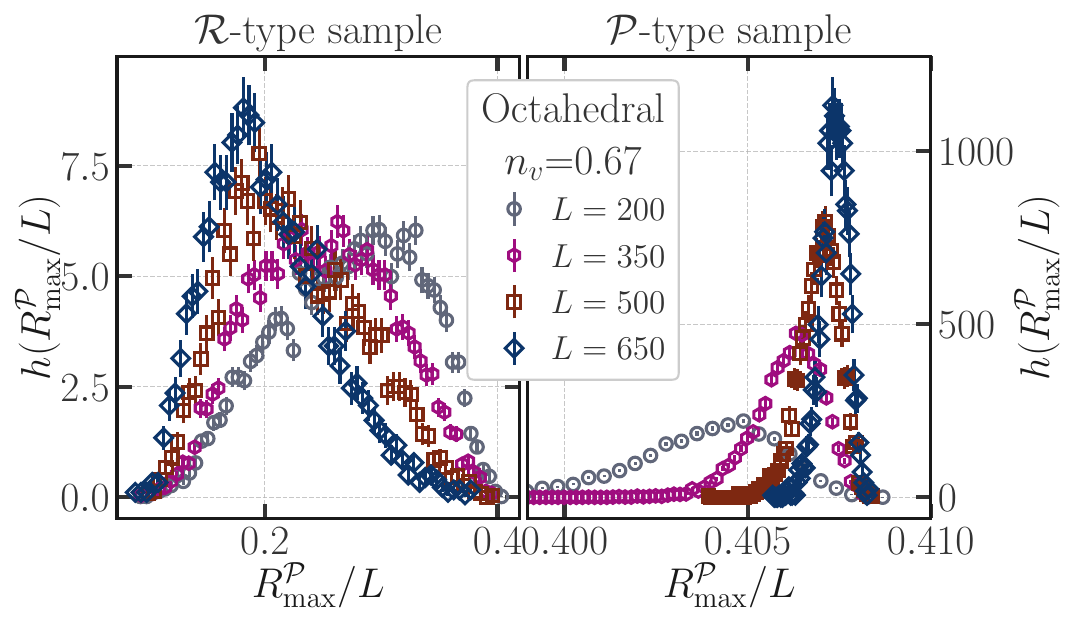}\\
		\end{tabular}
		\caption{
At a representative value of $n_v$ that places the system in phase III, each sample is labeled $\calR$-type or $\calP$-type based on whether it contributes to the right peak or left peak respectively in the histogram of $\mtotR$ in Fig.~\ref{fig:hist_mtot3D}; this labeling is unambiguous since these two peaks in Fig.~\ref{fig:hist_mtot3D} are well-separated and sharply defined. (a) and (b) display histograms of $\RmaxR/L$, the radius of gyration of the largest $\calR$-type region rescaled by $L$, separately for each type of sample. For each type of sample, this histogram has a single peak at a nonzero $L$-independent value of $\RmaxR/L$, which becomes more and more sharply defined with increasing $L$. Thus, although the peak for $\calP$-type samples corresponds to a significantly lower value of $\RmaxR/L$ compared to the corresponding peak for $\calR$-type samples, $\calP$-type samples also have a nonzero $\RmaxR/L$ in the thermodynamic limit. (c) and (d) display histograms of $\RmaxP/L$, the radius of gyration of the largest $\calP$-type region rescaled by $L$, separately for each type of sample. In sharp contrast to the histogram of $\RmaxR/L$ in $\calP$-type samples, we see that the histogram of $\RmaxP/L$ in $\calR$-type samples has a single peak that moves to smaller and smaller values of $\RmaxP/L$ as $L$ is increased, indicating that $\RmaxP/L$ tends to zero at large $L$ in $\calR$-type samples. In $\calP$-type samples on the other hand, $\RmaxP/L$ has a single peak at a nonzero value of $\RmaxP/L$, which gets more and more sharply defined with increasing $L$.  See Sec.~\ref{sec:ComputationalMethodsObservables} and Sec.~\ref{sec:ComputationalResults} for details. 				
\label{fig:hist_RmaxRRmaxP3D}}	
	\end{figure*}
	
Returning to our data for $P_{\rm single}$, we note that curves for different $L$ finally display a sharp crossing as a function of $n_v$ only at a {\em lower}  value of $n_v$ that is consistent, within error bars, with the previously determined location of the percolation threshold $n^{\rm crit}_v$ on both the triangular lattice and the Shastry-Sutherland lattice. This is clear from the inset of Fig.~\ref{fig:Psingle2D}.  	

	The picture is thus in terms of ribbon-like clusters that develop  as a precursor to the actual percolation transition, which occurs at a lower dilution $n_v^{\rm crit}$. Although all our results thus point to this non-monotonic regime being a crossover phenomenon that precedes (on the high-dilution side) the actual percolation transition of Gallai-Edmonds clusters, we caution that the range of sizes available to us does not make it possible to completely and unequivocally rule out the possibility that this is a distinct thermodynamic phase that flanks the low-diluted percolated phase in a very narrow range of $n_v$.

We have also examined the behavior of $\langle m_{\rm max} \rangle$, the ensemble average of the mass of the largest Gallai-Edmonds cluster (either $\calR$-type or $\calP$-type region) in a sample, as well as the behavior of $\langle \Rmax \rangle $, the ensemble average of the corresponding radius of gyration. We find that both these quantities also show a clear signature of the percolation transition at $n_v^{\rm crit} \approx 0.48068(4)$ [$n_v^{\rm crit} \approx 0.41820(5)$] on triangular [Shastry-Sutherland] lattice, with $\langle \Rmax \rangle $ additionally displaying the same crossover behavior as $\xi$ for $n_v > n_v^{\rm crit}$. For $\langle m_{\rm max} \rangle$, the value of $\eta$ obtained from this analysis is also consistent with the earlier estimate of $\eta=0.09(2)$ from the analysis of the susceptibility $\chi$. This is shown in Figs.~\ref{fig:App:mmax_2D} and \ref{fig:App:Rmax_2D} in the Appendix. 

Although the signatures of the onset of the low-dilution phase are thus quite conventional and entirely in line with expectations based on an analogy with well-studied geometric percolation phenomena, we emphasize that the observables we use to pinpoint the transition are in fact quite unconventional. These unconventional choices reflect the fact that the percolated phase itself is truly unusual: the lack of self-averaging that characterises this low-dilution regime is qualitatively different from the mere existence of broad distributions (that lead to different average and typical values of observables) or a free-energy landscape characterized by multiple valleys in other random systems. In our case, nominally identical samples can lie in two distinct thermodynamic phases, necessiating the use of rather unconventional measures to pinpoint the transition and study its scaling properties.
			 					\begin{figure*}
	\begin{tabular}{cc}
		a)\includegraphics[width=\columnwidth]{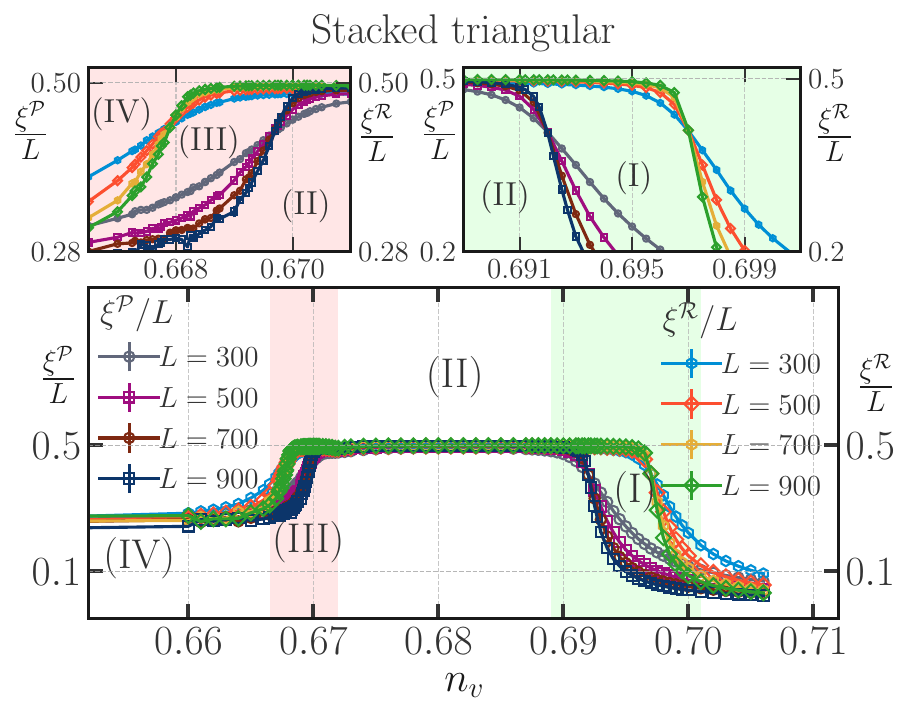}&
	   b)\includegraphics[width=\columnwidth]{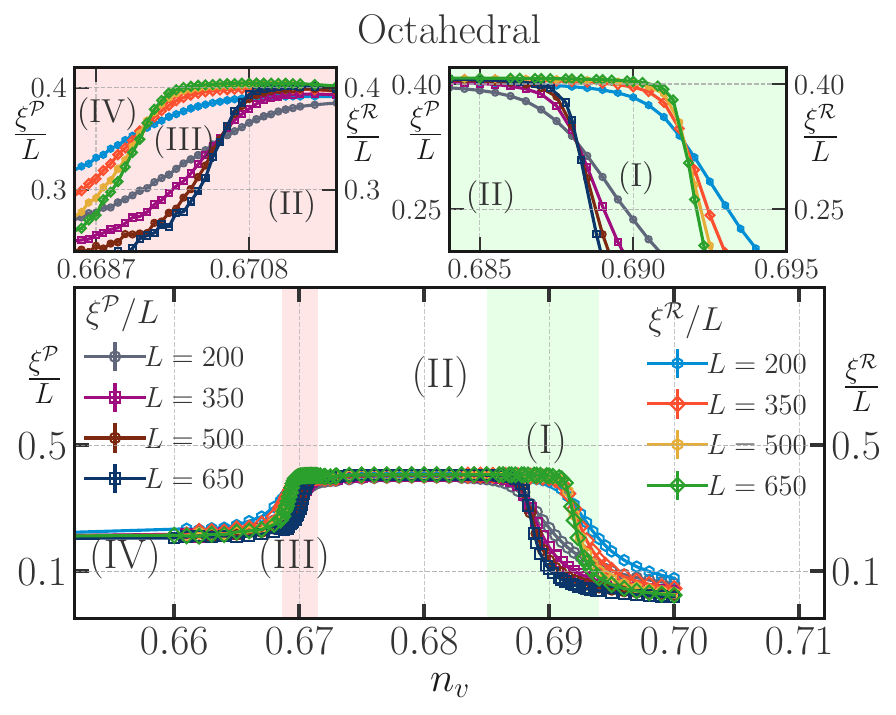}\\
	\end{tabular}
	\caption{$\xi^{\calR}$ and $\xi^{\calP}$, the correlation lengths corresponding to the correlation functions $C^{\calR}(r,r')$ and $C^{\calP}(r,r')$ respectively (see Eq.~\ref{eq:xi} and the discussion immediately preceding it) are finite in the thermodynamic limit for $n_v > n_v^{c_1}$. As the dilution $n_v$ is reduced, $\xi^{\calR}/L$ increases sharply in the vicinity of $n_v^{c_1}$ and curves corresponding to different $L$ display a sharp crossing at $n_v^{c_1}$, below which $\xi^{\calR}/L$ saturates to a nonzero $L$-independent value in the thermodynamic limit; thus, $\xi^{\calR}$ grows linearly with $L$ in phase I. In contrast, $\xi^{\calP}$ does not change character as $n_v$ is reduced below $n_v^{c_1}$, and remains finite in phase I. As we further lower $n_v$, $\xi^{\calP}/L$ increases sharply in the vicinity of $n_v^{c_2}$ and curves corresponding to different $L$ display a sharp crossing  at $n_v^{c_2}$, below which $\xi^{\calP}/L$ also saturates to a nonzero $L$-independent value in the thermodynamic limit. Thus, in phase II, both $\xi^{\calR}$ and $\xi^{\calP}$ grow linearly with $L$. Upon reducing $n_v$ further, $\xi^{\calP}/L$ decreases sharply in the vicinity of $n_v^{c_3}$ to a lower but still nonzero value that characterizes phase III, with curves corresponding to different lattice sizes displaying a sharp crossing at $n_v^{c_3}$. However, $\xi^{\calR}/L$ remains featureless in the vicinity of $n_v^{c_3}$. This is consistent with the overall picture of phase III that emerges from the data analyzed in Fig.~\ref{fig:hist_mmaxpmax3D} and Fig.~\ref{fig:hist_RmaxRRmaxP3D}. Upon lowering $n_v$ further, $\xi^{\calR}/L$ decreases sharply in the vicinity of $n_v^{c_4}$ and curves corresponding to different lattice sizes cross at $n_v^{c_4}$. This marks the begining of phase IV in which $\xi^{\calR}/L$ and $\xi^{\calP}/L$ both saturate to smaller nonzero values in the thermodynamic limit. The estimates for the critical values $n_v^{c_1}$, $n_v^{c_2}$, $n_v^{c_3}$, and $n_v^{c_2}$ obtained from the data analyzed here are consistent with those obtained from the analysis of wrapping probabilities shown in Fig.~\ref{fig:POverview3D}. See  Sec.~\ref{sec:ComputationalMethodsObservables} and Sec.~\ref{sec:ComputationalResults} for details. 		
	\label{fig:xi3D}}
\end{figure*}

	\subsubsection{Critical finite-size scaling in $d=2$}
	\label{subsubsec:FinteSizeScaling2D}
	We now ask if data for $P_{\rm cross}$ in the vicinity of the percolation threshold displays finite-size scaling behavior. To this end, we construct the scaling variable $(n_v - n_v^{\rm crit})L^{1/\nu}$ where $\nu$ is our estimate for the correlation length exponent, and ask if all our data for the $n_v$ dependence of $P_{\rm cross}$ at various sizes $L$ in the vicinity of this threshold collapses onto a single scaling curve for some choice of $\nu$.  We find that this is indeed the case for $\nu$ that lies within a reasonably small band of values around $\nu = 1.47$ on both the triangular and the Shastry-Sutherland lattice. Further confirmation of the scaling behavior is also provided by the data analysis of the susceptibility $\chi$. We find that the appropriately rescaled susceptibility $\chi/L^{2-\eta}$ (with $\eta \approx 0.09(2)$ on both the triangular lattice and the Shastry-Sutherland lattice) also collapses onto a single scaling curve in the vicinity of $n_v^{\rm crit}$ when plotted against the scaling variable $(n_v - n_v^{\rm crit})L^{1/\nu}$, with a choice of $\nu $ that lies within a reasonably small band of values around $\nu =1.5$ for both lattices. The corresponding scaling collapses for choices of $\nu$ within these bands are shown in Fig.~\ref{fig:scaling_Pcrosschi2D}. From this data, we obtain a consolidated estimate of $\nu \approx 1.45(10)$ for both the triangular lattice and the Shastry-Sutherland lattice, where the error bar now reflects the slightly different ranges of $\nu$ that yield scaling collapse of comparable quality for these two quantities on the two lattices studied here. 
								 \begin{figure*}[ht]
	\begin{tabular}{cc}
		\includegraphics[width=\columnwidth]{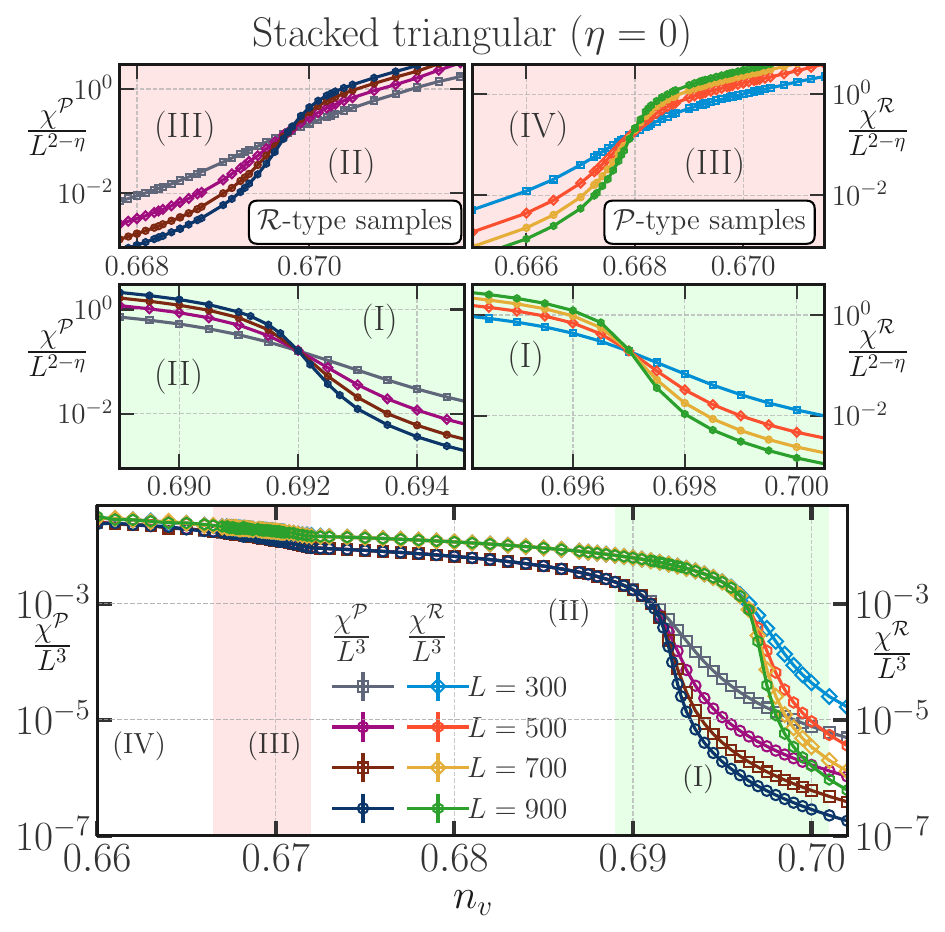}		
	&	\includegraphics[width=\columnwidth]{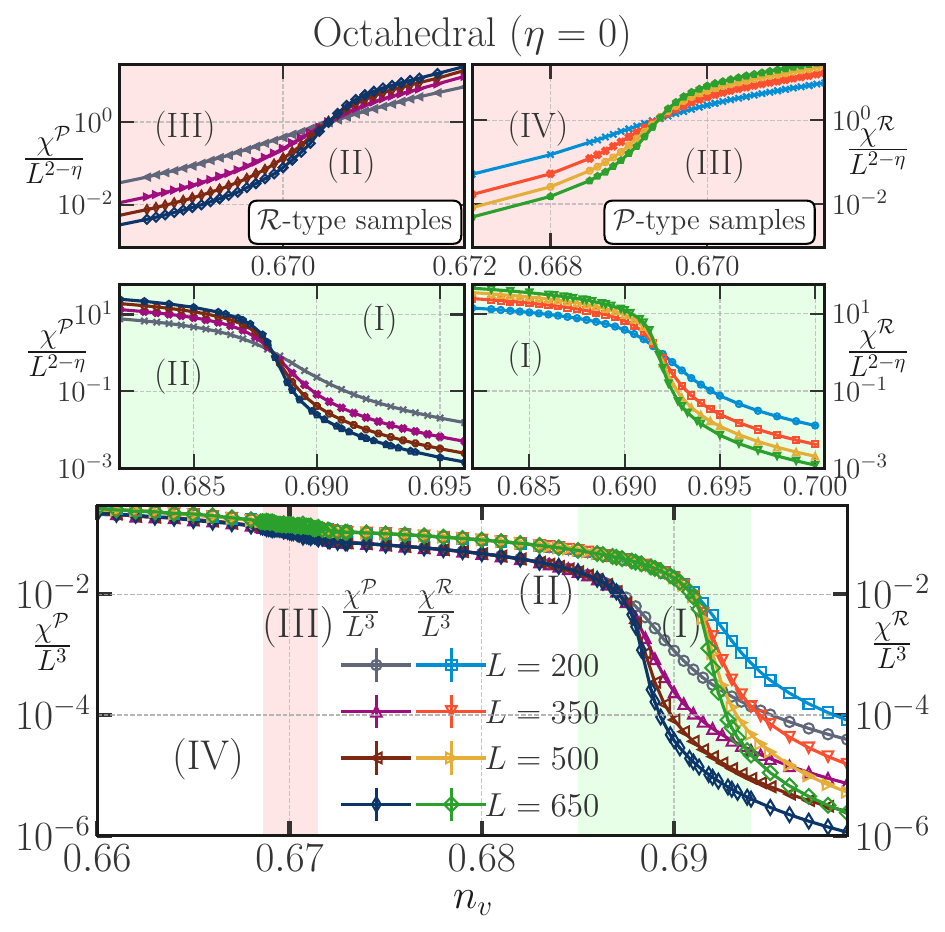}\\	
	\end{tabular}
	\caption{Bottom panels: The susceptibilities $\chi^{\calR}$ and $\chi^{\calP}$ scale as $L^3$ whenever there is a nonzero probability for Gallai-Edmonds clusters of the corresponding type to wrap around the torus in some direction(s). In phase III, the size of the largest $\calP$-type region in $\calP$-type samples (which constitute a fraction $1-f_{\calR} \approx 0.50(3)$ of all samples in phase III) is nearly twice as large as the corresponding size of the largest $\calP$-type region of a sample in phase II just to the right of the phase boundary between phases II and III. This more than compensates for the fact that $\calR$-type samples (which constitute a fraction $f_{\calR} \approx 0.50(3)$) in phase III have no macroscopic $\calP$-type region, thereby accounting for the slight but surpising-at-first-sight {\em rise} in the mean value of $\chi^{\calP}/L^3$ upon crossing into phase III from phase II; a similar slight rise in $\chi^{\calR}/L^3$, also visible in the data, is of course not particularly counter-intuitive. See Fig.~\ref{fig:hist_mtot3D}, Fig.~\ref{fig:hist_mmaxpmax3D} and Fig.~\ref{fig:hist_RmaxRRmaxP3D} for the relevant terminology and overview of the phase diagram. Middle panels: When scaled by $L^{2-\eta}$ (with the choice $\eta = 0$), data for $\chi^{\calR}$ and $\chi^{\calP}$ for various sizes $L$ display a sharp crossing at values of $n_v$ consistent within errors with $n_v^{c_1}$ (the onset of phase I) and $n_v^{c_2}$ (the phase boundary between phases I and II) respectively.  Top panels: When scaled by $L^{2-\eta}$ (again, with the choice $\eta = 0$),  curves of $\chi^{\calP}$ corresponding to various sizes $L$, computed within the sub-ensemble of $\calR$-type samples,  display a sharp crossing at a value of $n_v$ consistent within errors with $n_v^{c_3}$, the phase boundary between phase II and phase III. When similarly scaled,  curves for $\chi^{\calR}$ corresponding to various sizes $L$, computed within the sub-ensemble of $\calP$-type samples, display a sharp crossing at a value of $n_v$ consistent within errors with $n_v^{c_4}$, the phase boundary between phase III and phase IV.  See also Sec.~\ref{sec:ComputationalMethodsObservables} and Sec.~\ref{sec:ComputationalResults} for additional details. 	
		\label{fig:chi3D}}
\end{figure*}

		Although this scaling analysis appears to be fairly conclusive, we note that the scaling collapse appears to deteriorate rather rapidly on the high-dilution side of the transition. We attribute this to the presence of the crossover identified earlier, {\em i.e.}, the existence of single-wrapping clusters of large linear size just above the actual percolation transiton. 
		Since this crossover occurs so close to the actual percolation transition, it is interesting to ask if $\xi$ and $P_{\rm single}$, whose non-monotonic behavior directly reflects the presence of this crossover, also exhibit finite-size scaling in the vicinity of the percolation transition, and whether the scaling function for these quantities is non-monotonic. 
			 					 \begin{figure*}[ht]
	\begin{tabular}{cccc}
		\includegraphics[width=0.5\columnwidth]{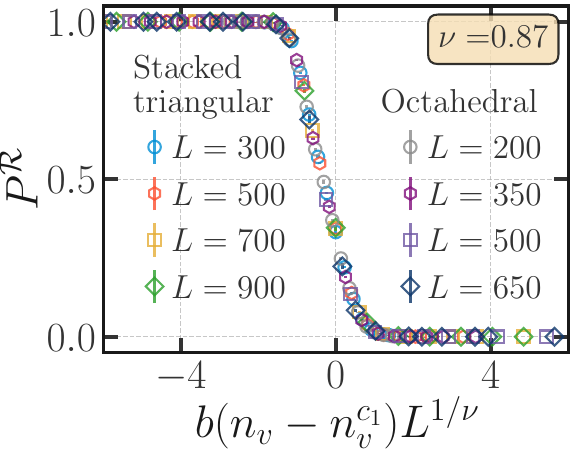}		
	&	\includegraphics[width=0.5\columnwidth]{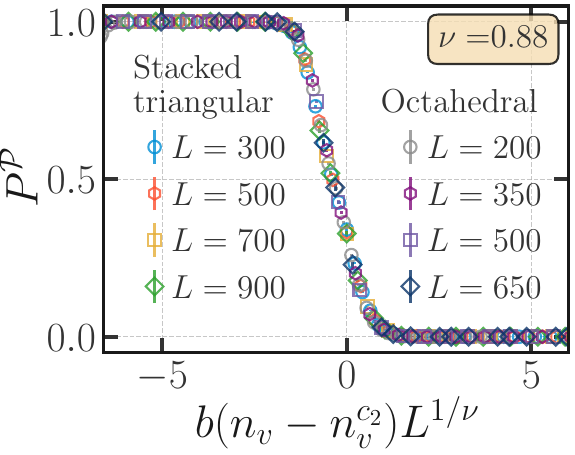}
		\includegraphics[width=0.5\columnwidth]{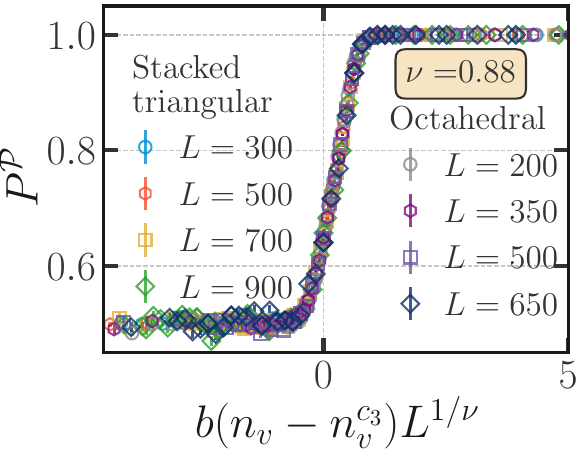}		
	&	\includegraphics[width=0.5\columnwidth]{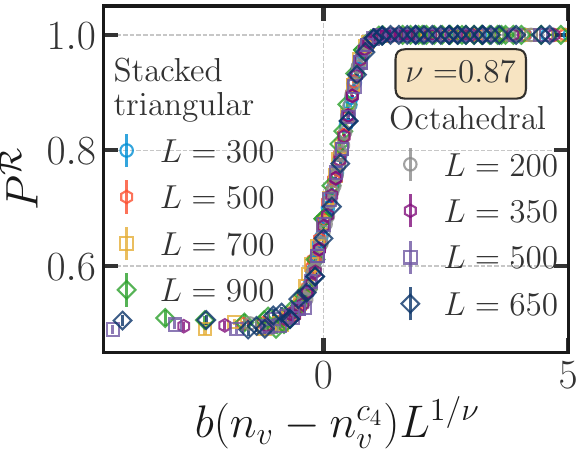}\\		
	\end{tabular}
	\caption{The $L$ and $n_v$ dependence of $P^{\calR}$ ($P^{\calP}$), the probability that a sample has at least one $\calR$-type ($\calP$-type) region that wraps around the torus in some way, displays finite-size scaling behavior in the vicinity of $n_v^{c_1}$ and $n_v^{c_4}$ ($n_v^{c_2}$ and $n_v^{c_3}$). The figure displays our evidence for this finite-size scaling collapse at these critical points. At each critical point, the displayed scaling collapse employs a value of $\nu$ chosen to yield reasonably high quality scaling collapse for that particular critical point. The  range of values that yield comparably high quality collapse at a particular critical point and the variation in these optimal values across different critical points, in conjuncton with the results of a similar finite-size scaling of other variables (see Figs.~\ref{fig:chiRchiPscaled3D} and \ref{fig:xiRxiPscaled3D}) at these phase transitions, sets the error bar in our common estimate of $\nu = 0.92(6)$ for the correlation length exponent at all these transitions. Here $b = 0.65$ ($b=1.0$) for the stacked-triangular lattice (corner-sharing octahedral lattice) is a lattice-dependent (non-universal) scale factor chosen to collapse all data for both lattices onto a single scaling curve.  See  Sec.~\ref{sec:ComputationalMethodsObservables} and Sec.~\ref{sec:ComputationalResults} for terminology and details. 	\label{fig:PRPPscaled3D}
	}
		
\end{figure*}

		With this motivation, we have  performed a scaling analysis of our data for $\xi/L$ and $P_{\rm single}$ in the vicinity of the percolation threshold $n^{\rm crit}_v$. The results of this analysis are shown in Fig.~\ref{fig:scaling_xiPsingle2D}.  Based on this analysis, we do find that curves for $\xi/L$ as well as $P_{\rm single}$ corresponding to different values of $L$ display finite-size scaling behavior and collapse onto a single scaling curve when plotted as a function of the scaling variable $(n_v - n_v^{\rm crit})L^{1/\nu}$, with values of $\nu$ that lie within the error estimate for $\nu$ obtained from our earlier analysis of the scaling of $P_{\rm cross}$ and $\chi$. The non-monotonic behavior of $\xi$ in the crossover region leads to a slight non-monotonicity (at the very edge of the scaling regime) of the scaling function defined by this data collapse.

In addition, we have studied the finite-size scaling behavior of $\langle m_{\rm max} \rangle$, the ensemble average of the mass of the largest Gallai-Edmonds cluster (either $\calR$-type or $\calP$-type region) in a sample, as well as the behavior of the ensemble average of the corresponding radius of gyration $\langle \Rmax \rangle$. We find that both these quantities display finite-size scaling behavior with values of exponents consistent within errors with the estimates obtained above. The results of this analysis are relegated to Fig.~\ref{fig:App:scaling_Rmaxmmax_2D}  in the Appendix.

	Thus, we have a fairly consistent picture of a critical point at $n_v^{\rm crit} \approx 0.48068(4)$ [$n_v^{\rm crit} \approx 0.41820(5)$] on the triangular [Shastry-Sutherland] lattice. This critical point separates a high-dilution phase in which all $\calR$-type and $\calP$-type regions are small from a low-dilution phase in which each sample has one infinite cluster, which can either be an $\calR$-type cluster with probability $f_{\calR} \approx 0.50(2)$ or an infinite $\calP$-type cluster with probability $1-f_{\calR}$. Interestingly, we find that $f_{\calR}$ is, within our statistical errors, independent of the dilution $n_v$ throughout the percolated phase.
	The percolation transition itself is characterized by the exponents $\nu \approx 1.45(10)$ and $\eta \approx 0.09(2)$ on both lattices. Curiously, this estimate of $\nu$ is quite close to but {\em not entirely consistent with} the exact value of the correlation length exponent $\nu_{\rm Bernoulli2D} = 4/3$ at the ordinary geometric percolation transition in two dimensions~\cite{Nijs_1979, Nienhuis_Riedel_Schick_1980, Pearson_1980}. Moreover, our value of $\eta$ is {\em very different} from the value of the corresponding anomalous exponent $\eta_{\rm Bernoulli2D} = 5/24$ at the geometric percolation transition in two dimensions. Thus, not only does the percolated phase display an unusual lack of self-averaging in the thermodynamic limit, the critical point appears to not be in the same universality class as usual geometric percolation in two dimensions.

\subsection{Results in $d=3$}
\label{subsec:Results3D}
We now turn to a study of the three-dimensional case. Our analysis in $d=3$ is informed by what we have learnt from the foregoing study of the two-dimensional case, namely, that the large-scale geometry of $\calR$-type and $\calP$-type regions need not exhibit self-averaging in the thermodynamic limit at low dilution. This experience now allows us to chart the most direct route towards an overall understanding of the different behaviors exhibited at successively lower values of $n_v$ in the three-dimensional case. 
	 \begin{figure*}
	\begin{tabular}{cccc}
		\includegraphics[width=0.5\columnwidth]{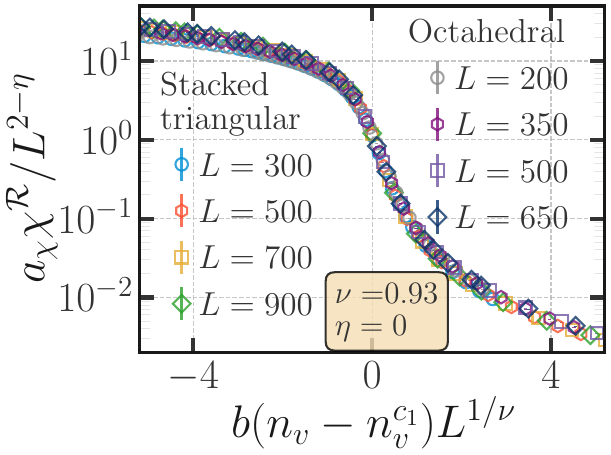}		
	&	\includegraphics[width=0.5\columnwidth]{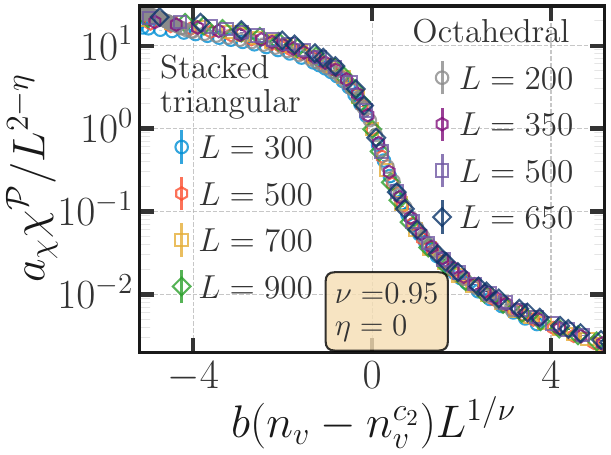}
		\includegraphics[width=0.5\columnwidth]{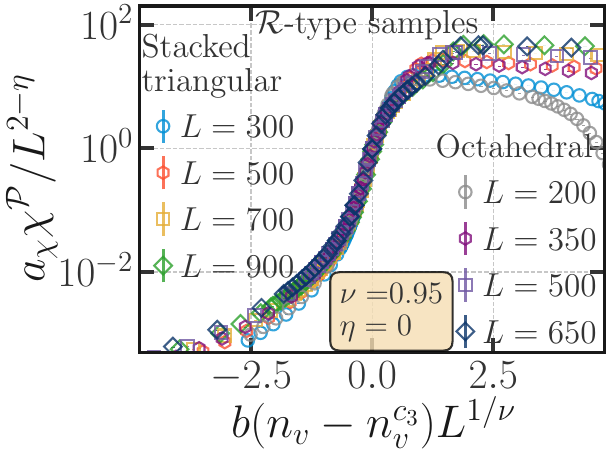}		
	&	\includegraphics[width=0.5\columnwidth]{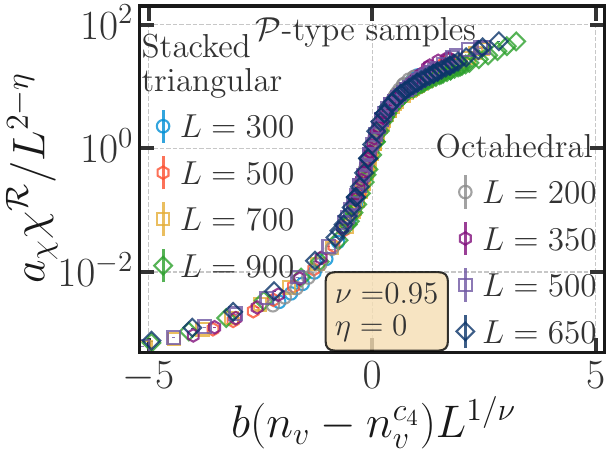}\\		
	\end{tabular}
	\caption{The $L$ and $n_v$ dependence of $\chi^{\calR}/L^{2-\eta}$ ($\chi^{\calP}/L^{2-\eta}$), the rescaled susceptibility associated with the correlation function $C^{\calR}(r-r')$ ($C^{\calP}(r-r')$), displays finite-size scaling behavior in the vicinity of $n_v^{c_1}$ ($n_v^{c_2}$) for all samples, and in the vicinity of $n_v^{c_4}$ ($n_v^{c_3}$) for $\calP$-type ($\calR$-type) samples. The figure displays evidence for this finite-size scaling collapse  at these critical points. At each critical point, the displayed scaling collapse employs a value of $\nu$ and $\eta$ chosen to yield reasonably high quality scaling collapse for that particular critical point. The  range of values that yield comparably high quality collapse at a particular critical point and the variation in these optimal values across different critical points, in conjuncton with the results of a similar finite-size scaling of other variables (see Figs.~\ref{fig:PRPPscaled3D} and \ref{fig:xiRxiPscaled3D}) at these phase transitions, sets the error bar in our common estimate of $\nu = 0.92(6)$ for the correlation length exponent and $\eta = 0.00(4)$ for the anomalous exponent at all these transitions. Here $b = 0.65$ ($b=1.0$) and $a_{\chi} = 6.0$ ($a_{\chi} = 1.0$ ) for the stacked-triangular lattice (corner-sharing octahedral lattice) are lattice-dependent (non-universal) scale factors chosen to collapse all data for both lattices onto a single scaling curve.   See Sec.~\ref{sec:ComputationalMethodsObservables} and Sec.~\ref{sec:ComputationalResults} for terminology and details.
		\label{fig:chiRchiPscaled3D}}
\end{figure*}

\subsubsection{Phase diagram in $d=3$}
\label{subsubsec:PhaseDiagram3D}
We begin by examining the $n_v$ dependence of $P^{\calR}$, the probability that a sample has at least one $\calR$-type region that wraps around the torus in some way, and the analogously defined probability $P^{\calP}$ that  a sample has at least one $\calP$-type region that wraps around the torus in some way.
Fig.~\ref{fig:POverview3D} displays our results for these probabilities, focusing on a relatively narrow range of $n_v$ that already displays the entire gamut of interesting behavior. From this data, it is clear that there are five distinct regimes, which we now proceed to describe: Reading from right to left, {\em i.e.}, from the largest to the smallest value of $n_v$ displayed in Fig.~\ref{fig:POverview3D}, we see that initially, both probabilities are small and tend rapidly to zero with increasing system size, implying that all $\calR$-type and $\calP$-type regions remain finite in the thermodynamic limit. As we lower the value of $n_v$, we observe a sharp threshold for $P^{\calR}$ at  $n_v^{c_1}= 0.6970(5)$ ($n_v^{c_1}=0.6917(5)$) on the stacked triangular (corner-sharing octahedral) lattice. Below this threshold, $P^{\calR} =1 $ within our statistical errors, while $P^{\calP}$ shows no change from the earlier behavior. 
As $n_v$ is decreased further, the system encounters a second equally sharp threshold at $n_v^{c_2}=0.6920(3)$ ($n_v^{c_2}=0.6882(3)$) on the stacked triangular (corner-sharing octahedral) lattice, visible in the behavior of $P^{\calP}$. Below this threshold, we have $P^{\calP} = P^{\calR} =1$  within our statistical errors. This behavior persists as we lower the dilution, until a third sharp threshold is reached at $n_v^{c_3}= 0.66975(10)$ ($n_v^{c_3}=0.6705(4)$) on the stacked triangular (corner-sharing octahedral) lattice. 
At this third threshold, $P^{\calP}$ drops down to a smaller but {\em nonzero} value of $P^{\calP} \approx 0.50(3)$, while $P^{\calR}$ remains unity. 
Finally, there is another equally sharp threshold at a slightly lower value of $n_v$, which we estimate to be $n_v^{c_4}=0.6680(2)$ ($n_v^{c_4}=0.6694(4)$) on the stacked triangular (corner-sharing octahedral) lattice. Below this threshold, {\em both } $P^{\calR}$ and $P^{\calP}$ are equal within our statistical errors -- $P^{\calR} =  P^{\calP} \approx 0.50(2)$.

Apart from the unusually rich spectrum of behavior seen in such a narrow range of $n_v$, another striking thing about this data is the fact that the corresponding thresholds for the stacked triangular and the corner-sharing octahedral lattice are so close to each other on the $n_v$ axis. While we do not fully understand the reason for this close correspondence, and our initial choice to study these two lattice geometries certainly did not anticipate this feature, we note that it may be related to the fact that the thresholds for the ordinary site percolation transition on these lattices are also very close to each other (we have checked this by obtaining an estimate for the percolation threshold of the corner-sharing octahedaral lattice and comparing it with the published result for the stacked triangular lattice~\cite{Schrenk_Araujo_Herman_2013}).


In any case, at each of these thresholds, our data for the $n_v$ dependence of the relevant wrapping probabilities displays sharp crossings of the curves corresponding to different sizes $L$. This is the typical behavior expected of a dimensionless quantity (such as a wrapping probability)  at a scale-invariant critical point. We may therefore conclude that these thresholds represent sharp thermodynamic phase transitions, and each of these regimes is thus a distinct phase in the thermodynamic limit.
Thus, we have the following picture of the phase diagram in three dimensions: Regime I corresponds to a thermodynamic phase in which $\calR$-type regions percolate with probability one, but $\calP$-type regions remain finite. Regime II corresponds to a thermodynamic phase in which both $\calR$-type and $\calP$-type regions percolate with probability one.

Regime III corresponds to a thermodynamic phase in which $\calR$-type regions percolate with probability one and $\calP$-type regions percolate with probability $1-f_{\calR} \approx 0.50(3)$; one therefore has two kinds of samples, $\calR$-type samples, which occur with probability $f_{\calR}$, in which there is a single percolating region of type $\calR$, and $\calP$-type samples, which occur with probability $1-f_{\calR}$, in which there are two percolating regions, one of type $\calR$ and another of type $\calP$. 
Finally, regime IV is a phase that is identical to the percolated phase in two dimensions. In this phase, any given sample has a single percolating region which is of type $\calR$ with probability $f_{\calR} \approx 0.50(2)$, and of type $\calP$-with probability $1-f_{\calR}$. 
Further evidence in favor of this interpretation comes for instance from a study of the $n_v$ dependence $n_{\rm cross}^{\calR}$ and $n_{\rm cross}^{\calP}$, respectively the number of $\calR$-type and $\calP$-type clusters that wrap around the torus in three independent directions simultaneously. The data for these quantities is shown in Fig.~\ref{fig:ncrossOverview3D}.  

Note that in summarizing the behavior thus, we have intentionally conflated two somewhat distinct characteristics of a sample: On the one hand, the question is whether a sample has at least one $\calP$-type region that wraps around the three dimensional torus in some way, with the corresponding probability being defined as $P^{\calP}$. The data in Fig.~\ref{fig:POverview3D} answers this question. On the other hand, the question is whether a sample belongs to the left peak or the right peak in the histogram of $\mtotR$, with the corresponding probabilities defined to be $1-f_{\calR}$ and $f_{\calR}$ respectively, exactly as in the discussion of the two-dimensional case in Sec.~\ref{subsubsec:ThermodynamicDensities2D}. 
						 \begin{figure*}[ht]
	\begin{tabular}{cccc}
		\includegraphics[width=0.5\columnwidth]{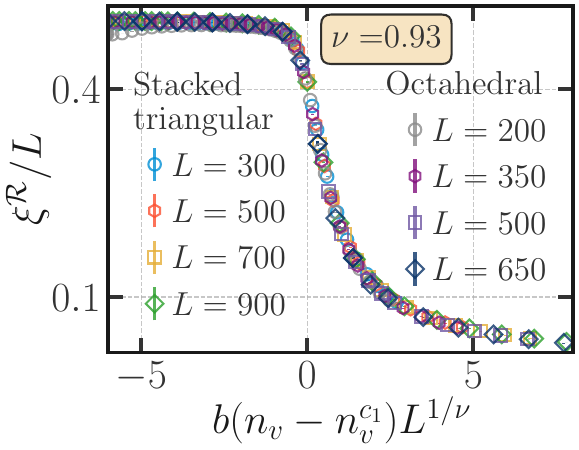}		
	&	\includegraphics[width=0.5\columnwidth]{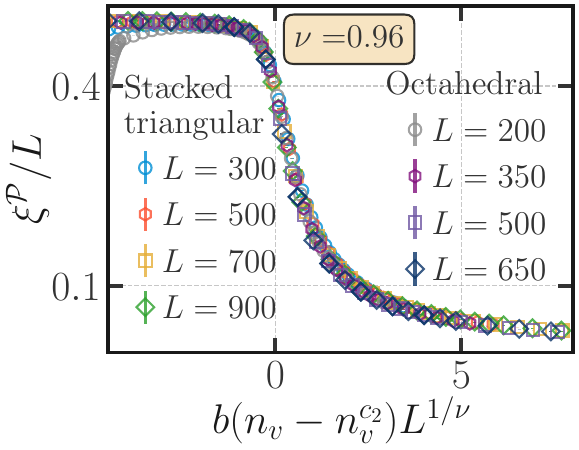}
		\includegraphics[width=0.5\columnwidth]{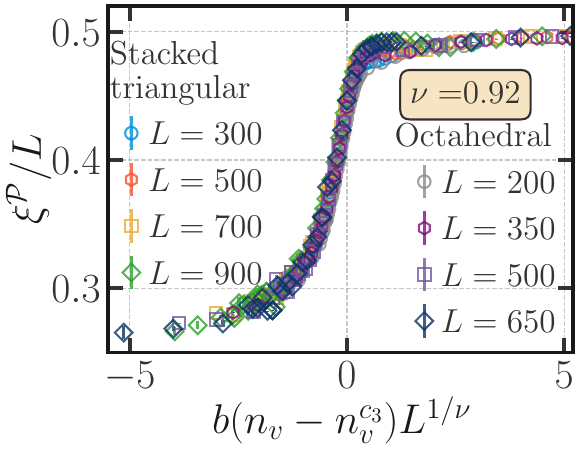}		
	&	\includegraphics[width=0.5\columnwidth]{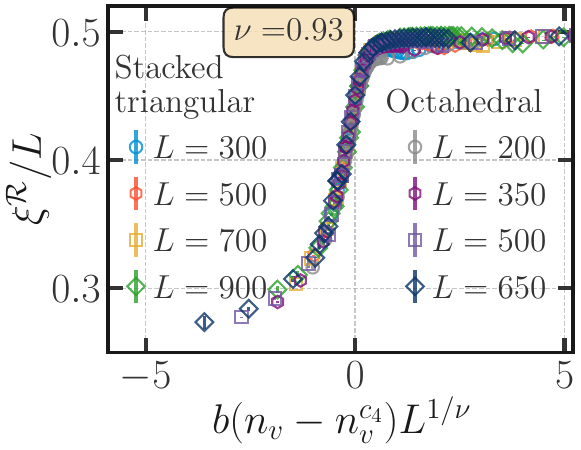}\\		
	\end{tabular}
	\caption{The $L$ and $n_v$ dependence of $\xi^{\calR}/L$ ($\xi^{\calP}/L$), the rescaled correlation length [see Eq.~\ref{eq:xi}] associated with the correlation function $C^{\calR}(r-r')$ ($C^{\calP}(r-r')$), displays finite-size scaling behavior in the vicinity of $n_v^{c_1}$ ($n_v^{c_2}$) and $n_v^{c_4}$ ($n_v^{c_3}$). The figure displays evidence for this finite-size scaling collapse at these critical points. At each critical point, the displayed scaling collapse employs a value of $\nu$ chosen to yield reasonably high quality scaling collapse for that particular critical point. The range of values that yield comparably high quality collapse at a particular critical point and the variation in these optimal values across different critical points, in conjuncton with the results of a similar finite-size scaling of other variables (see Figs.~\ref{fig:PRPPscaled3D} and \ref{fig:chiRchiPscaled3D}) at these phase transitions, sets the error bar in our common estimate of $\nu = 0.92(6)$ for the correlation length exponent at all these transitions. Here $b = 0.65$ ($b=1.0$) for the stacked-triangular lattice (corner-sharing octahedral lattice) is a lattice-dependent (non-universal) scale factor chosen to collapse all data for both lattices onto a single scaling curve.  See  Sec.~\ref{sec:ComputationalMethodsObservables} and Sec.~\ref{sec:ComputationalResults} for terminology and details.
		\label{fig:xiRxiPscaled3D}}
\end{figure*}

However, this second question {\em assumes}, by analogy to our earlier results in two dimensions, that the distribution of $\mtotR$ in phase III and phase IV is bimodal, with two macroscopically separated and sharply defined peaks, and that this bimodality is perfectly correlated with the existence of percolating clusters of type $\calR$ or $\calP$. 
Consequently, our summary above of the behavior in phases III and IV implicitly assumes the validity of this conflation of the two notions, and further assumes the equivalence $P^{\calP} \equiv 1-f_{\calR}$ in these two low-dilution phases. 

As we discuss next, this does turn out to be the case: We find that $P^{\calP} = 1-f_{\calR}$ within our statistical errors throughout phase III and phase IV so long as $n_v$ is not too close to $n_v^{c_3}$, the phase boundary between phase III and phase II; indeed, in both phases III and IV, as long as $n_v$ is not close to $n_v^{c_3}$, we find that no sample from the right peak of the histogram of $\mtotR$ has a percolating $\calP$-type region. Additionally, in phase IV, we also find that $P^{\calR} = f_{\calR}$ within our statistical errors so long as $n_v$ is not too close to the $n_v^{c_4}$, the phase boundary between  phase IV and phase III; in other words, in phase IV, we find that no sample from the left peak of the histogram of $\mtotR$ has a percolating $\calR$-type region.

\subsubsection{Violations of thermodynamic self-averaging in $d=3$}
\label{subsubsec:SelfAveragingViolations3D}
This overall picture of the phase diagram implies interesting violations of thermodynamic self-averaging in phases III and IV, but not in phase I and II. With this in mind, we have also studied the distributions of various thermodynamic densities for various representative values of $n_v$. In phases I and II, we indeed find that all the thermodynamic densities have single-peaked distributions, with the width of the peak narrowing with increasing $L$, as one would expect if conventional thermodynamic self-averaging holds. In contrast to this, the behavior of these distributions in phase IV shows clear violations of self-averaging,
identical in nature to the behavior seen in the percolated phase in two dimensions.

In other words, we find that $m_{\rm tot}^{\calR}$ in phase IV has a bimodal distribution, with two well-separated peaks, each of which narrows and becomes more sharply defined with increasing $L$, allowing samples to be unambiguously labeled $\calR$-type or $\calP$-type exactly as in the two dimensions, with $f_{\calR}$, the fraction of $\calR$-type samples, again taking on the $n_v$ independent value $f_{\calR} \approx 0.50(2)$ throughout the phase (except in the immediate vicinity of the phase boundary at $n_v^{c_4}$). Other physical quantities such as $\RmaxR$, $\RmaxP$, $\mmaxR$, $\mmaxP$, $n_{\calR}$, and $n_{\calP}$ all  behave very differently in $\calR$-type and $\calP$-type samples. This dichotomy of behavior is also exactly analogous to corresponding behavior in the percolated phase in two dimensions. However, the overall density $w$ of monomers does not show this anomalous behavior in the thermodynamic limit. Since all of this is completely analogous to the behavior seen in the percolated phase in two dimensions (and analyzed at length in Sec.~\ref{subsec:Results2D}), we relegate the corresponding three-dimensional data and its analysis to the Appendix (see Figs.~\ref{fig:App:hist_mtotIV_3D}, \ref{fig:App:hist_mmaxpmaxIV_3D}~\ref{fig:App:hist_RmaxPRmaxRIV_3D}, and ~\ref{fig:App:hist_nRnPIV_3D}).
  	\begin{figure*}
\begin{tabular}{ccc}
a)\includegraphics[width=0.66\columnwidth]{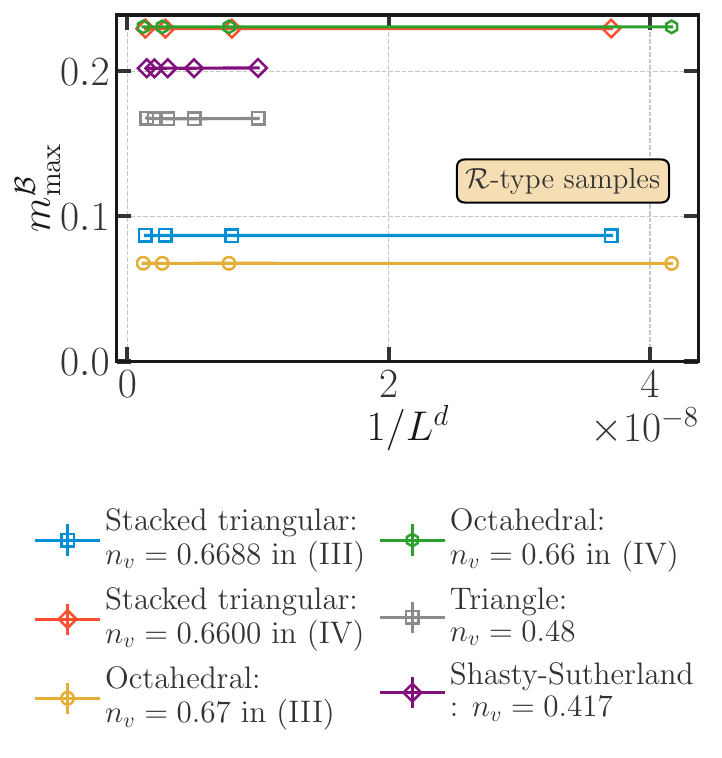}
& b)\includegraphics[width=0.66\columnwidth]{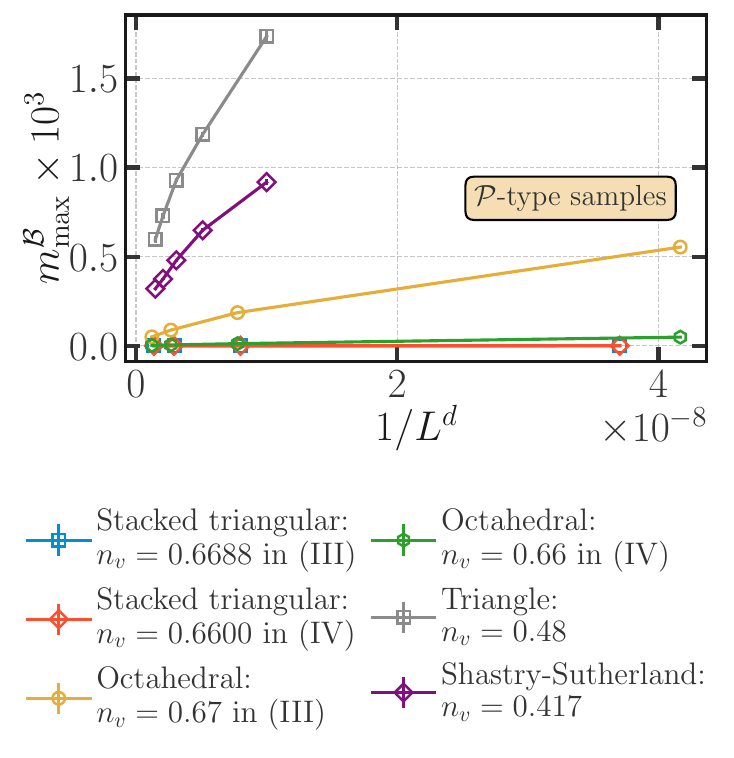}
& c) \includegraphics[width=0.66\columnwidth]{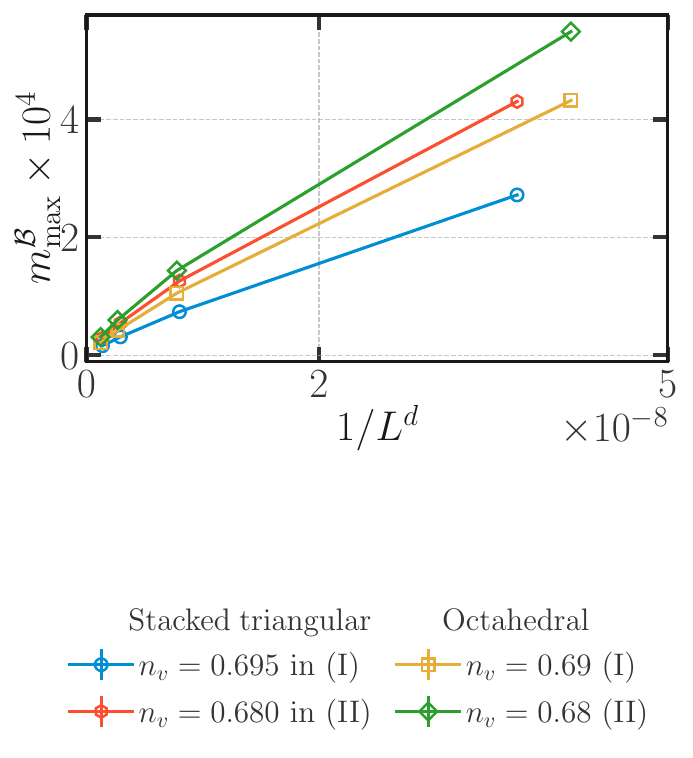}\\
\end{tabular}
\caption{(c) $m^{\calB}_{\rm max}$, the mass of the largest blossom in the largest $\calR$-type region of a sample scaled by the mass of the largest Gallai-Edmonds region of that sample, goes to zero in the thermodynamic limit in phases I and II in three dimensions, although the largest $\calR$-type region is always macroscopic in size in these phases. (b) This is also true in ${\calP}$-type samples in phase III in three dimensions, although even such $\calP$-type samples always a percolating $\calR$-type region in this phase. Since $\calP$-type samples have no percolating $\calR$-type region in phase IV, this is also trivially true in $\calP$-type samples in  phase IV in three dimensions. (a) In sharp contrast to this behavior, $m^{\calB}_{\rm max}$ tends to a nonzero value in the thermodynamic limit in the percolated phase in two dimensions, and in $\calR$-type samples in phases III and IV in three dimensions.  See Fig.~\ref{fig:hist_mtot3D}, Fig.~\ref{fig:hist_mmaxpmax3D} and Fig.~\ref{fig:hist_RmaxRRmaxP3D}, as well as Sec.~\ref{sec:ComputationalMethodsObservables} and Sec.~\ref{sec:ComputationalResults} for terminology and details.
\label{fig:BlossomMorphology}}
\end{figure*}

Phase III on the other hand does merit a separate discussion here, since it displays a different type of violation of self-averaging, which has no analog in the two dimensional phase diagram studied earlier. Before we focus on these differences, it is useful to explicitly list aspects of the behavior that are entirely analogous to the behavior in phase IV and the two-dimensional percolated phase:  First, we again find that  $m_{\rm tot}^{\calR}$ has a bimodal distribution, with two well-separated peaks, each of which narrows and becomes more sharply defined with increasing $L$. This is illustrated in Fig.~\ref{fig:hist_mtot3D}.  This separation between the peaks indicates a macroscopic difference in the morphology of the corresponding samples, {\em i.e.}, a difference that survives in the thermodynamic limit (since $m_{\rm tot}^{\calR}$, as defined by us, is an intensive density variable).  Second, we also find that the weight $f_{\calR}$ in the right peak is independent of $n_v$: $f_{\calR} \approx 0.50(3)$.  Additionally, the overall density $w$ of monomers does not show this anomalous behavior in the thermodynamic limit, and has a histogram that features a single sharply defined peak.

However, this is where the similarity stops. 
The features that are unique to phase III become clear when we separate the samples by asking which of the two peaks each sample belongs to. Such a separation is unambiguous because the peaks are well-separated even at the smallest size we study. Following the same convention as in our earlier study of the bimodal phase in two dimensions, samples belonging to the right peak with larger $m_{\rm tot}^{\calR}$ are labeled $\calR$-type, and the other samples, all of which belong to the peak with smaller $m_{\rm tot}^{\calR}$, are labeled $\calP$-type. With this separation in place, we now examine the statistics of other observables within each of these two groups of samples. When analyzed in this manner, the data paints a distinct picture which we now sketch. 

Consider first the histograms of $m^{\calR}_{\rm max}$ and $m^{\calP}_{\rm max}$ in both groups of samples.  From Fig.~\ref{fig:hist_mmaxpmax3D}, which displays these histograms, we see that the histograms of $m^{\calP}_{\rm max}$ in $\calP$-type samples have a single peak at a relatively large $L$-independent value, with this peak becoming sharper with increasing $L$, while the corresponding histograms in $\calR$-type samples have a single peak at a low $L$-dependent value that continually shifts to even lower values with increasing $L$.  While this behavior is identical to that seen in phase IV and in the two-dimensional percolated phase, $m^{\calR}_{\rm max}$ behaves very differently: we see that histograms of $m^{\calR}_{\rm max}$ in both $\calR$-type and $\calP$-type samples display a single sharp peak at an $L$-independent value, with the peak becoming sharper with increasing $L$. The only difference between the two types of samples is that the peak for $\calP$-type samples is located at a significantly smaller value of $m^{\calR}_{\rm max}$ compared to the corresponding peak for $\calR$-type samples.

Similar behavior is also seen in the histograms of the scaled radii of gyrations $\RmaxR/L$ and $\RmaxP/L$. From Fig.~\ref{fig:hist_RmaxRRmaxP3D}, which displays these histograms, we see that the histograms of $\RmaxP/L$ in $\calP$-type samples have a single peak at a relatively large $L$-independent value, with this peak becoming sharper with increasing $L$, while the corresponding histograms in $\calR$-type samples have a single peak at a low $L$-dependent value that continually shifts to even lower values with increasing $L$. In contrast, the histogram of $\RmaxR/L$ has a single sharply defined peak at an $L$-independent value for both $\calR$-type and $\calP$-type samples, and the only difference between the two types of samples is that this peak is at significantly smaller values of $\RmaxR/L$ in $\calP$-type samples relative to $\calR$-type samples.

It is thus clear that the violations of thermodynamic self-averaging in phase III are very different from those seen in phase IV and in the two-dimensional percolated phase. Specifically, the largest $\calR$-type region is macroscopic in size in this phase both in $\calR$-type and in $\calP$-type samples. The only difference between the two sub-ensembles is that its size is {\em much smaller} in $\calP$-type samples compared to the corresponding size in $\calR$-type samples. On the other hand, the largest $\calP$-type regions display the same dichotomy of behavior (between the two sub-ensembles) that is a feature of phase IV and the percolated phase in two dimensions. In the Appendix, we also present some additional evidence (Fig.~\ref{fig:App:hist_nRnPIII_3D}), related to the dichotomy in behavior of the densities $n_{\calR}$ and $n_{\calP}$ in $\calR$-type and $\calP$-type samples, which fills in this picture with additional detail.
  		\begin{figure*}
\begin{tabular}{cccc}
a)\includegraphics[width=0.5\columnwidth]{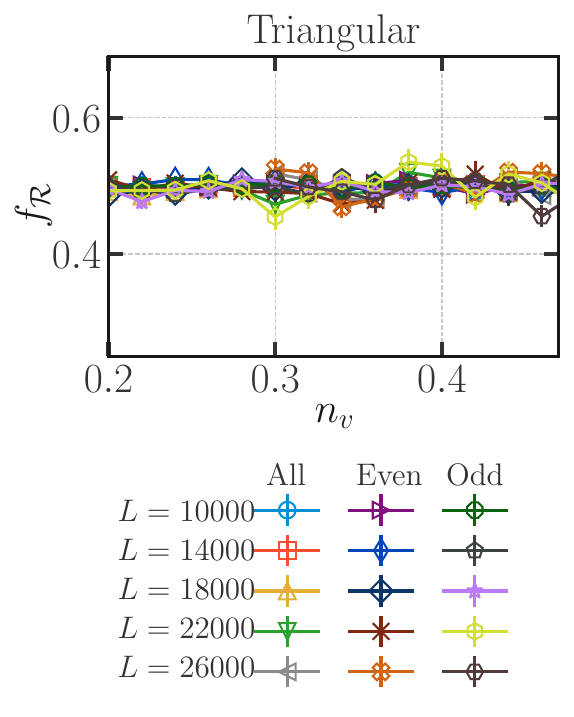}
& b)\includegraphics[width=0.5\columnwidth]{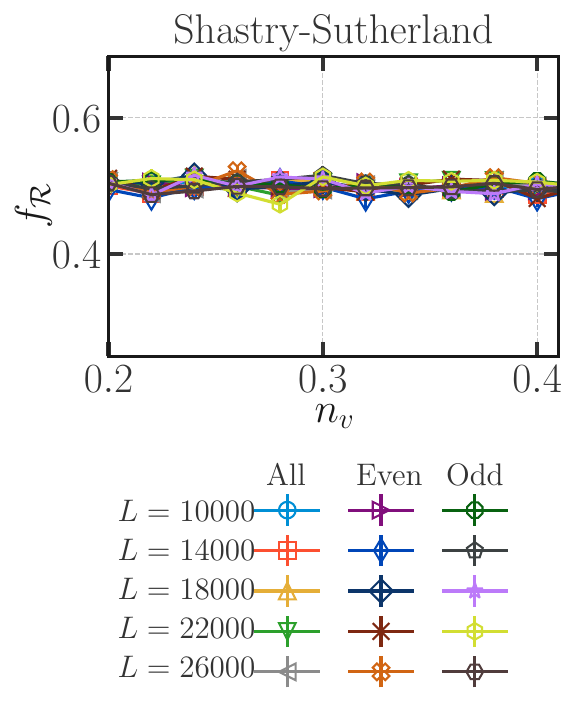}
&c)\includegraphics[width=0.5\columnwidth]{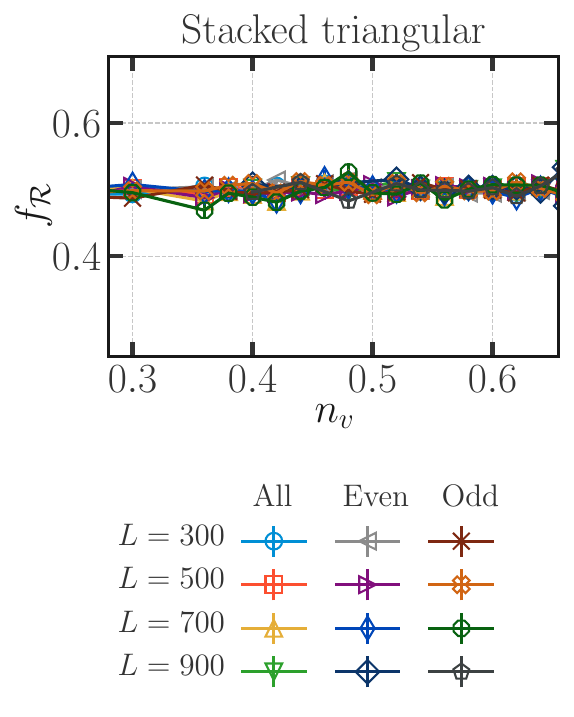}
& d)\includegraphics[width=0.5\columnwidth]{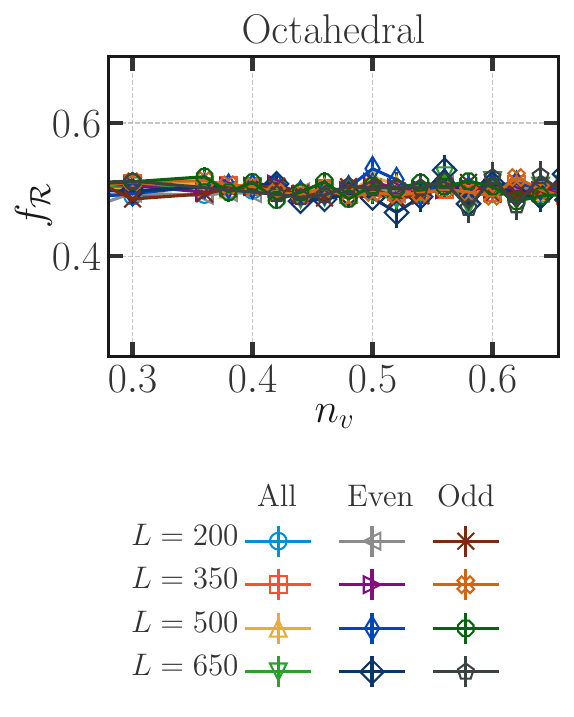}\\
\end{tabular}
\caption{In phases III and IV in three dimensions, and in the percolated phase in two dimension, the probability $f_{\calR}$ for a sample to belong to the right peak in the histogram of $\mtotR$ is pinned to $f_{\calR} \approx 0.50(3)$ throughout these phases (see Figs.~\ref{fig:mtotR2D}, \ref{fig:hist_mtot3D} and \ref{fig:App:hist_mtotIV_3D} respectively for the analysis of these histograms at representative points in these regimes). This is also seen to be {\em separately true} in sub-ensembles of samples, obtained by splitting the samples into two groups based on the parity $m(\calG)$, the number of sites in the largest geometric cluster $\calG$ of the lattice. Thus the interesting lack of self-averaging exhibited by the low-dilution percolated phases is {\em not} a simple parity effect having to do with the parity of $m(\calG)$. See Sec.~\ref{sec:ComputationalMethodsObservables} and Sec.~\ref{sec:ComputationalResults} for terminology and details.
\label{fig:EmergentSymmetry}}
\end{figure*}

This analysis of the conditional probability distributions confirms the basic picture painted by our data for the wrapping probabilities and numbers of crossing clusters: In phase I, the Gallai-Edmonds decomposition of $\calG$ has a single giant component, which is of type $\calR$. In phase II, there are two giant components, one of type $\calR$ and the other of type $\calP$. In phase III, each sample has a giant component of type $\calR$, but only a fraction $1-f_{\calR} \approx 0.50(3)$ of the samples have a second giant component, which is of type $\calP$. Finally, in phase IV, each sample has a single giant component, which is of type $\calR$ in a fraction $f_{\calR} \approx 0.50(2)$ of the samples, and type $\calP$ in all the other samples, exactly analogous to the behavior in the percolated phase in the two dimensional case. 

\subsubsection{Other probes of the percolated phases in $d=3$}
\label{subsubsec:OtherProbes3D}

We now examine the behavior of other quantities whose $L$ dependence can corroborate this picture of the four different percolated phases. We begin by studying the behavior of the correlation lengths $\xi^{\calR}$ and $\xi^{\calP}$ corresponding to the correlation functions $C^{\calR}(r,r')$ and $C^{\calP}(r,r')$ respectively (see Eq.~\ref{eq:xi} and the definitions immediately preceding it). The corresponding results are shown in Fig.~\ref{fig:xi3D}. We see that our data for $\xi^{\calR}/L$ displays a sharp crossing point at a value of $n_v$ consistent with the onset of phase I (as inferred previously from the behavior of $P^{\calR}$ and $n_{\rm cross}^{\calR}$). In an entirely analogous manner, $\xi^{\calP}/L$ displays a sharp crossing at a value of $n_v$ consistent with the location of the phase boundary between phase I and phase II (as inferred previously from the behavior of $P^{\calP}$ and $n_{\rm cross}^{\calP}$). 

As $n_v$ is lowered further, $\xi^{\calP}/L$ displays another sharp crossing at a value of $n_v$ consistent with the location of the phase boundary between phases II and III, while $\xi^{\calR}/L$ shows no signature of this phase boundary. We have checked that $\xi^{\calP}/L $ has a bimodal distribution in phase III, with the value of $\xi^{\calP}$ scaling as $L$ only in $\calP$-type samples, while remaining finite in $\calR$-type samples. Thus, it is $\calP$-type samples that are responsible for the crossing point seen in $\xi^{\calP}/L$ at the phase boundary between phases II and III. 

Finally, $\xi^{\calR}/L $ shows a similar crossing point at a smaller value of $n_v$ consistent with the location of the phase boundary between phases III and IV; $\xi^{\calP}/L$ is however insensitive to this phase transition. In phase IV, we have checked that both $\xi^{\calP}$
 and $\xi^{\calR}$ have a bimodal distribution, with the former scaling as $L$ only in $\calP$-type samples and the latter scaling as $L$ only in $\calR$-type samples.

Next, we consider the susceptibilities $\chi^{\calR}$
and $\chi^{\calP}$ defined in Eq.~\ref{eq:chi}. The corresponding results are shown in Fig.~\ref{fig:chi3D}. We see that the susceptibility shows clear signatures of macroscopic Gallai-Edmonds regions forming at the corresponding percolation transitions. Indeed, as one reduces $n_v$ to cross into phase I, $\chi^{\calR}$ starts scaling as $L^3$, while $\chi^{\calP}$ starts scaling as $L^3$ as soon as $n_v$ is reduced further to cross into phase II. In phase III, $\chi^{\calP}$ restricted to $\calR$-type samples ({\em i.e.}, with the ensemble average performed only over $\calR$-type samples) stops scaling as $L^3$, since only $\calP$-type samples have a macroscopically large $\calP$-type region. Finally, in phase IV, $\chi^{\calP}$ and $\chi^{\calR}$ scale as $L^3$ in $\calP$-type and $\calR$-type samples respectively, but not vice-versa. Further, when scaled by $L^{2-\eta}$ (with the choice $\eta = 0$), we see that curves for these susceptibilities corresponding to various sizes $L$ show sharp crossings at the corresponding critical points. Thus, $\chi^{\calR}/L^{2-\eta}$
and $\chi^{\calP}/L^{2-\eta}$ have a sharp crossing point respectively at values of $n_v$ consistent with $n_v^{c_1}$ and $n_v^{c_2}$. Similarly, $\chi^{\calR}/L^{2-\eta}$ ($\chi^{\calP}/L^{2-\eta}$), restricted to the sub-ensemble of $\calP$-type ($\calR$-type) samples, displays a sharp crossing point at a value of $n_v$ consistent with $n_v^{c_4}$ ($n_v^{c_3}$). 

In addition, by way of further confirmation of this basic picture of the four percolated phases, we have also analyzed the behavior of other quantities such as $R_{\rm max}^{\calR}$, $R_{\rm max}^{\calP}$, $m_{\rm max}^{\calR}$, and $m_{\rm max}^{\calP}$, and found that these quantities also paint the same consistent picture. 

\subsubsection{Scaling at critical points in $d=3$}
\label{subsubsec:Scaling3D}
As far as the three-dimensional phase diagram is concerned, we may summarize the story thus far as follows: Our data (at various sizes $L$) for $P^{\calR}$, the probability that there is an $\calR$-type region that wraps around the torus in some way, and $P^{\calP}$, the probability that there is a $\calP$-type region that wraps around the torus in some way, yield the following estimates for a sequence of four percolation transitions at successively lower values of $n_v$ on the stacked-triangular lattice (corner-sharing octahedral lattice): $n_v^{c_1}= 0.6970(5)$ ($n_v^{c_1}=0.6917(5)$), $n_v^{c_2}=0.6920(3)$ ($n_v^{c_2}=0.6882(3)$), $n_v^{c_3}= 0.66975(10)$ ($n_v^{c_3}=0.6705(4)$), and $n_v^{c_4}=0.6680(2)$ ($n_v^{c_4}=0.6694(4)$).


Given that the phase boundaries between the different percolated phases are characterized by such sharply-defined thresholds for the relevant observables, it is interesting to ask if various properties obey universal critical scaling in the vicinity of these transitions? If the answer is yes, it is also interesting to ask if the corresponding exponents are different from each other, and compare them to the exponents that are known to characterize the geometric Bernoulli percolation transition of randomly diluted lattices in three dimensions. These questions motivate our detailed finite-size scaling analysis of various other observables in the vicinity of each of these thresholds.

Essentially the same estimates for the location of these critical points are also obtained from the completely analogous crossing points displayed by $\xi^{\calR}/L$ and $\xi^{\calP}/L$, where $\xi^{\calR}$ and $\xi^{\calP}$ are the correlation lengths defined in Eq.~\ref{eq:xi},  which respectively measure the typical sizes of $\calR$-type and $\calP$-type regions in a sample.
 We also find that $\chi^{\calR}/L^{2-\eta}$ displays sharp crossings at values of $n_v$ consistent within errors with $n_v^{c_1}$ and $n_v^{c_4}$ (with the ensemble average restricted to $\calP$-type samples in the latter case) when $\eta$ is chosen to be within a fairly narrow band around $\eta = 0$. Likewise, we find that $\chi^{\calP}/L^{2-\eta}$ also displays comparably sharp crossings at values of $n_v$ consistent with $n_v^{c_2}$ and $n_v^{c_3}$ (with the ensemble average restricted to $\calR$-type samples in the latter case) when $\eta$ is again chosen in a narrow band around $\eta = 0$. This yields the common estimate $\eta = 0.00(4)$ at all four critical points.

In addition, we find that the standard finite-size scaling ansatz works well for the dimensionless quantities $P^{\calR}$, $\chi^{\calR}/L^{2-\eta}$ (with $\eta = 0.00(4)$), and $\xi^{\calR}/L$ in the vicinity of the critical points at $n_v^{c_1}$ and $n_v^{c_4}$ (with the  ensemble average restricted to $\calP$-type samples when calculating $\chi^{\calR}/L^{2-\eta}$ in the latter critical region). In other words, data at various sizes $L$ for these quantities in the vicinity of these critical points collapses onto a scaling curve when plotted against the scaling variable $(n_v - n_v^{c})L^{1/\nu}$ (with  the superscript $c =c_1$ or $c=c_4$) for an appropriate choice of $\nu$ within a reasonably well-determined and narrow band of values around $\nu = 0.87$ for $P^{\calR}$, around $\nu = 0.94$ for $\chi^{\calR}/L^{2-\eta}$, and around $\nu = 0.93$ for $\xi^{\calR}/L$.

The same is true of the dimensionless quantities $P^{\calP}$, $\chi^{\calP}/L^{2-\eta}$ (with $\eta = 0.00(4)$), and $\xi^{\calP}/L$ in the vicinity of the critical points at $n_v^{c_2}$ and $n_v^{c_3}$ (with the  ensemble average restricted to $\calR$-type samples when calculating $\chi^{\calP}/L^{2-\eta}$ in the latter critical region). Again, the corresponding band of values of $\nu$ lie in a reasonably narrow band in each case, centered around  $\nu = 0.88$ for $P^{\calP}$, around $\nu = 0.95$ for $\chi^{\calP}/L^{2-\eta}$, and around $\nu = 0.94$ for $\xi^{\calP}/L$.

 Further, we find it is possible to rescale the scaling variable $(n_v - n_v^{c})L^{1/\nu}$ (with $n_v^{c}$ chosen to be the appropriate critical value) by a {\em single overall constant} $b \approx 0.65$ on the stacked-triangular lattice in such a way that the resulting scaling functions for $P^{\calR}$, $P^{\calP}$, $\xi^{\calR}/L$, and $\xi^{\calP}/L$ in the octahedral case coincide with the corresponding function on the stacked triangular lattice (viewed as a function of $(n_v - n_v^{c})bL^{1/\nu}$. For the finite-size scaling collapse of   $\chi^{\calR}/L^{2-\eta}$, and $\chi^{\calP}/L^{2-\eta}$, we find that this is again true if in addition to this rescaling the $x$-axis by $b$,, we also multiply these quantities on the stacked-triangular lattice by an  appropriately chosen scale factor $a_{\chi} \approx 6.0$. 
 
 These scaling collapses are shown for the wrapping probabilities $P^{\calR}$ and $P^{\calP}$ in Fig.~\ref{fig:PRPPscaled3D}, for $\chi^{\calR}/L^{2-\eta}$, and $\chi^{\calP}/L^{2-\eta}$ in Fig.~\ref{fig:chiRchiPscaled3D}, and for the scaled correlation lengths $\xi^{\calR}/L$ and $\xi^{\calP}/L$ in Fig.~\ref{fig:xiRxiPscaled3D}. 
  
As is clear from this description of our scaling analysis, the corresponding best-fit values of the  correlation length exponent $\nu$ are slightly different for different quantities, and for different transition points. This is also true for the exponent $\eta$. But these differences are within the uncertainties of our determination of these exponents in the following sense: For each quantity at each critical point on a given lattice, we find roughly comparable data collapse over a range of values of the corresponding exponents around their best-fit values, and these ranges overlap across these various cases. 

From this scaling analysis, we therefore conclude that all the four critical points display universal scaling, and, moreover,  belong to the same universality class. For the common values of exponents at these transitions, we obtain the following consolidated estimates: $\nu \approx 0.92(6)$ and $\eta = 0.00(4)$. Interestingly, our estimate for $\nu$ is consistent within errors with the best-known estimates for the corresponding exponent at the usual geometric percolation transition in three dimensions, {\em i.e.}, $\nu_{\rm Bernoulli3D} \approx 0.876$. Our estimate for $\eta$ is also very close to the known estimate  $\eta_{\rm Bernoulli3D} \approx -0.046$~\cite{Wang_Zhou_Zhang_Garoni_2013} for this exponent at the geometric percolation transition. The slight discrepancy in $\eta$ may be due to a small drift (with increasing $L$) in our estimated value of $\eta$ or an underestimation of our errors. 

Thus, although the percolated phases themselves are extremely unusual, all the percolation transitions
 that separate these phases appear to fall in the same universality class as geometric percolation (except for this slight discrepancy in the value of $\eta$). This
  is quite different from the two dimensional case studied earlier. In this latter case, our estimate for $\eta$ at the two-dimensional Gallai-Edmonds percolation transition is quite different from the known value of $\eta_{\rm Bernoulli2D}=5/24$ at the geometric percolation transition in two dimensions. Moreover, our estimate for $\nu$ in this two-dimensional case is also slightly different, and our estimated error bars seem to exclude the known value of $\nu_{\rm Bernoulli2D} = 4/3$ at the geometric percolation transition. In this respect, the critical behavior at these Gallai-Edmonds percolation transitions bears some resemblance with the previously studied Dulmage-Mendelsohn percolation transitions of site-diluted bipartite lattices: In that bipartite setting too, the behavior in two dimensions is very different from geometric percolation, while the three-dimensional Dulmage-Mendelsohn percolation transition has exponents that match, within errors, with the known values for geometric percolation in three dimensions.

\subsection{Blossoms in the percolating cluster}
\label{subsec:BlossomMorphology}
We have thus far focused on the total mass $m(\calR_i)$ and the total imbalance (monomer number) $\calI(\calR_i)$ of $\calR$-type regions, in particular, of the largest $\calR$-type region in a sample, and used observables constructed from these quantities to study the evolution of the large-scale geometry as a function of the dilution. To probe the internal structure of $\calR$-type regions in the sample, we now go beyond this and study the size of the blossoms that make up an $\calR$-type region.   This is a particularly interesting quantity since we expect the character of the collective Majorana mode wavefunctions hosted inside the largest $\calR$-type region to depend sensitively on the size of the blossoms. This is because each blossom, if isolated from the rest of the sample, hosts a single Majorana zero mode, and the collective Majorana modes inside an $\calR$-type region represent particular linear combinations of these precursor modes that survive the mixing between these modes.

Specifically, we study the $L$ dependence of the mass density $m_{\rm max}^{\calB}$, defined in Eq.~\ref{eq:MaxBlossomDensity} as the ratio of the mass of the largest blossom in the largest $\calR$-type region of a sample  to the mass of the largest Gallai-Edmonds region of the sample (which can either be the $\calR$-type region that contains this blossom, or a larger $\calP$-type region). Clearly, this is most interesting in the various percolated phases, and that is what we focus on here. 
Before we proceed to our results for $m_{\rm max}^{\calB}$, it is useful to clarify a few things about the way we have defined this extremal quantity: In phase I, the largest Gallai-Edmonds region is always of type $\calR$ since $\calR$-type regions percolate and $\calP$-type regions do not. Interestingly, we find that the same is true even in phase II, although each sample has a percolating $\calP$-type region in addition to a percolating $\calR$-type region. In phase III on the other hand, as is clear from the histograms displayed in Fig.~\ref{fig:hist_mmaxpmax3D}, the largest Gallai-Edmonds region in $\calR$-type samples is of type $\calR$, while the largest such region in $\calP$-type samples is of type $\calP$. Naturally, this is also true in the percolated phase in two dimensions, and in phase IV in three dimensions, since there is just one percolating region in each sample (which is of type $\calR$ or $\calP$ depending on the type of sample).

Turning to our results for $m_{\rm max}^{\calB}$, we find that the $L$ dependence of $m_{\rm max}^{\calB}$ is a useful probe of the internal structure of the largest $\calR$-type region in all the percolated phases, confirming expectations in some cases, and yielding somewhat surprising information in other cases: For instance,  from the data analyzed in Fig.~\ref{fig:BlossomMorphology}, we see that $m_{\rm max}^{\calB}$ tends to zero in the thermodynamic limit in phases I and II, and in $\calP$-type samples in phase III; this is in spite of the fact that there is a percolating $\calR$-type region present in all these cases. On the other hand, $m_{\rm max}^{\calB}$ does tend to a nonzero value in the large $L$ limit in $\calR$-type samples in phase III. This is also seen quite clearly in Fig.~\ref{fig:BlossomMorphology}. In phase IV, and in the percolated phase in two dimensions, we find that  $m_{\rm max}^{\calB}$ tends to a nonzero value in the thermodynamic limit  only in $\calR$-type samples and not in $\calP$-type samples. This is also seen quite clearly in Fig.~\ref{fig:BlossomMorphology}, and is of course entirely consistent with the fact that there are no percolating $\calR$-type regions in $\calP$-type samples in these phases.

\subsection{Does the parity of $m(\calG)$ play a role?}
\label{subsec:DueDiligence}
	Having established that the low-dilution percolated phase in two dimensions, as well as the percolated phases III and IV in three dimensions, are all characterized by an unusual lack of self-averaging, we now proceed to examine the most ``obvious'' possible explanation for this behavior. We have in mind here the parity of $m(\calG)$,  the total number of sites in the largest geometric component $\calG$ of the diluted lattice. Of course, this line of thought is simply motivated by the elementary fact that it is impossible to find a perfect matching for a lattice with an odd number of sites. 

Although this rationale might appear naive, it is still well worth taking this possibility seriously and testing for it, since no other gross property of the sample comes to mind as a possible source of the macroscopic differences between $\calR$-type and $\calP$-type samples in these low-dilution phases. With this motivation, we split our ensemble of samples into two sub-ensembles at each dilution, the first sub-ensemble comprised of all samples in which $m(\calG)$ is even, and the other consisting of all samples in which $m(\calG)$ is odd. 
With this bifurcation in place, we separately analyze the random geometry of $\calR$-type and $\calP$-type regions in each sub-ensemble.  

If the unusual and interesting lack of self-averaging noted earlier is caused by just the parity of the largest geometric cluster, we would expect that the even-parity sub-ensemble would be dominated by $\calP$-type samples in which $\calP$-type regions show a percolation transition while $\calR$-type regions remain finite in size. Likewise, we would expect the odd-parity sub-ensemble to be dominated by $\calR$-type samples in which $\calR$-type regions show a percolation transition and $\calP$-type regions remain bounded in size.

We find that such a bifurcated analysis of our data reveals no distinction between the two sub-ensembles in any of the quantities that we have studied. To illustrate this point, we focus here on $f_{\calR}$, the fraction of all samples that belong to the right peak in the histogram of $\mtotR$, and compare this to $f_{\calR}$ computed by measuring the corresponding fraction separately in each sub-ensemble. 
In Fig.~\ref{fig:EmergentSymmetry}, we display our results for $f_{\calR}$ in the low-dilution regime in each of these sub-ensembles, and compare them with our results for the same quantity obtained by averaging over all samples without regard to the parity of $m(\calG)$. We see from this figure that {\em there is no statistically significant distinction in the three sets of results}. This illustrates the fact that the parity of $m(\calG)$ plays no role in determining the $\calR$-type or $\calP$-type character of a sample, and is therefore not responsible for the surprising violations of thermodynamic self-averaging seen at low dilution. Apart from unequivocally ruling out any role for the parity of $m(\calG)$, Fig.~\ref{fig:EmergentSymmetry} also highlights another rather intriguing aspect of the low-dilution phase, namely the fact that $f_{\calR}$ is independent of $n_v$  and appears to be pinned to a value of $1/2$ in all cases (within our statistical errors).

Should this be viewed as a signature of some emergent statistical symmetry between $\calR$-type and $\calP$-type regions?  Since we have been unable to come up with any theoretical description that incorporates such a symmetry, this is just an interesting and natural speculation at this point, which we hope will motivate some follow-up work aimed at a better understanding of this aspect of our results. In this context, what is particularly intriguing about any such scenario is that this emergent symmetry would be required to be a property of an entire phase, rather than being contingent on fine-tuning the dilution to a particular value such as a critical point. While several examples of such enlarged symmetries are known in other contexts at particular finely-tuned values of some coupling constant, examples where this happens throughout some phase are harder to come by. In Sec.~\ref{sec:DiscussionandOutlook}, we mention one concrete and perhaps more tractable version of this speculation in a more familiar setting.

\subsection{Gallai-Edmonds Chaos}
\label{subsec:Chaos}
Our observation of the lack of self-averaging in the thermodynamic limit of the percolated phases immediately leads to a natural and obvious question: Is there a sense in which the geometry of $\calR$-type and $\calP$-type regions is ``chaotic'' in this low-dilution regime? To formulate this more precisely, one can for instance set the vacancies in motion, allowing them to execute a random walk on the lattice with the constraint that the largest geometric cluster $\calG$ in the initial condition does not break up into multiple connected components or acquire additional sites. With this stringent restriction in place, one can now formulate precise questions about the dynamics of the $\calR$- and $\calP$-type regions of this ``time-invariant'' $\calG$, and ask if this dynamics is chaotic in some precise sense. 
\begin{figure} 
	\centering
	\includegraphics[width=\columnwidth]{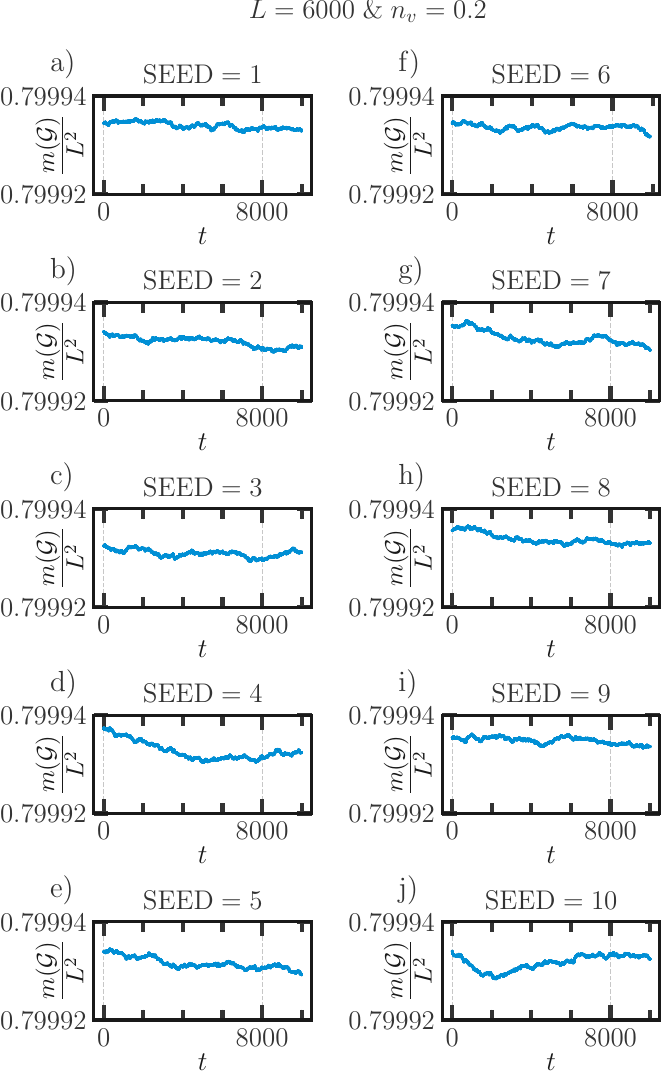}
	\caption{
Time series of $m_{{\mathcal G}}(t) = m({{\mathcal G}}(t))/L^2$ starting from ten different randomly chosen initial conditions of the site-diluted triangular lattice, {\em i.e.}, starting from arbitrarily-chosen maximum-density dimer packings of randomly-diluted samples labeled by the \textsc{seed} variable. See Sec.~\ref{subsec:Chaos} for details.} 
	\label{fig:TimeSeriesmG} 
\end{figure}

To explore global (as opposed to spatiotemporal) aspects, it is enough to study the following simple but representative model of the dynamics, with a less restrictive constraint on the behavior of $\calG$: A small fraction $5 \times 10^{-5}$ of randomly chosen vacancies exchange position with a randomly chosen occupied neighbor at each time step,  with mild restrictions ensuring that ${\mathcal G}(t)$ only fluctuates slightly with time $t$ (but no demand that the sites in $\calG$ remain unchanged). Such a model for the dynamics is in any case perhaps more relevant for the kind of question that is posed by the possibility of slow annealing of defects in the triangular vortex lattice.

In this minimal model, at each time step, a small fraction $5 \times 10^{-5}$ of randomly-chosen vacancies are swapped with an occupied nearest-neighbor using the following protocol: A vacancy is chosen randomly from the list of vacancies in the original diluted lattice. If it has three or more occupied nearest-neighbors that belong to the largest connected component ${\mathcal G}$ of the diluted lattice, it exchanges position with one of these neighbors, chosen randomly but subject to the proviso that this neighboring vertex itself have two or more occupied nearest-neighbors that belong to ${\mathcal G}$. 
	\begin{figure} 
	\centering
	\includegraphics[width=\columnwidth]{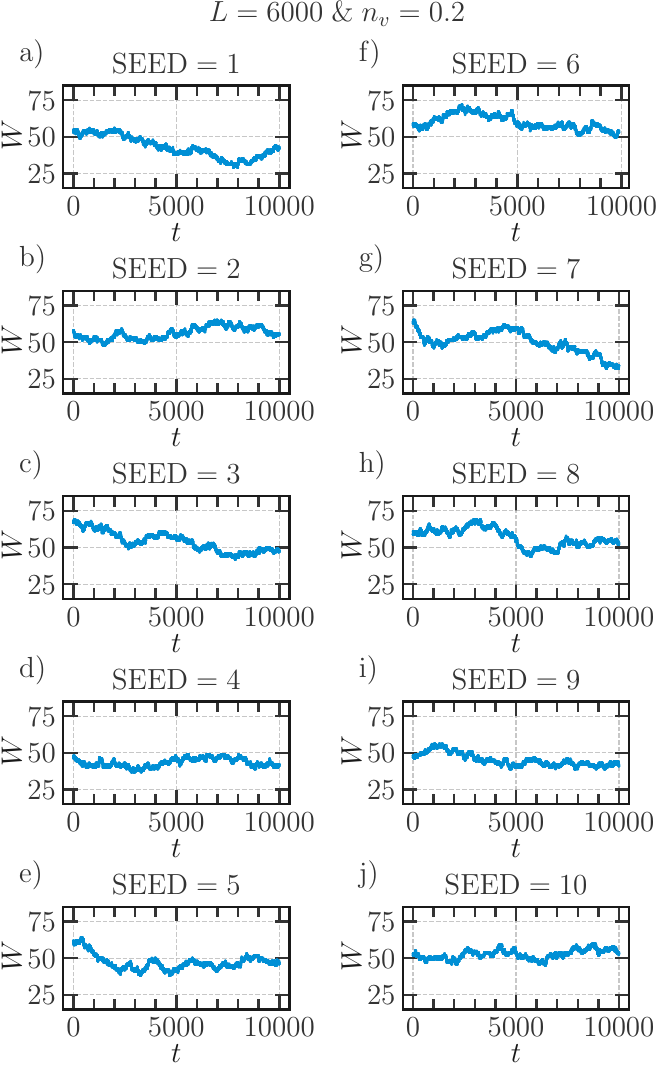}
	\caption{
Time series of $W(t)$ starting from ten different randomly chosen initial conditions of the site-diluted triangular lattice, {\em i.e.}, starting from arbitrarily-chosen maximum-density dimer packings of randomly-diluted samples labeled by the \textsc{seed} variable. See Sec.~\ref{subsec:Chaos} for details.} 
	\label{fig:TimeSeriesW} 
\end{figure}

As mentioned above, ${\mathcal G}$, the largest connected component of the diluted lattice, can change after each time-step, since this dynamics can lead to vertices being added to or removed from ${\mathcal G}$. The restrictions incorporated into our model dynamics have only one role, which is to severely limit these changes in ${\mathcal G}$ and the resulting fluctuations in $m({\mathcal G})$, the number of vertices that belong to ${\mathcal G}$. That this is indeed the case for this model dynamics is clear from Fig.~\ref{fig:TimeSeriesmG}. And the purpose of imposing these limits is of course to highlight the fact that even very small changes in the vacancy configuration lead to dramatic changes to the large-scale geometry of maximum-density dimer packings. 

In Fig.~\ref{fig:TimeSeriesW} we also monitor the time series of $W(t)$, the total number of monomers hosted by $\calG$. We see that this has much larger fluctuations (compared to those seen in $m(\calG(t))$).
This is consistent with the fact that very small changes to the vacancy configuration lead to very large changes in the structure of maximum matchings. However, $W(t)$ does not exhibit any lack of self-averaging or bimodal behavior, which would have corresponded to $W(t)$ switching between the vicinity of two very distinct values without taking on any of the values in between.
\begin{figure} 
	\centering
	\includegraphics[width=\columnwidth]{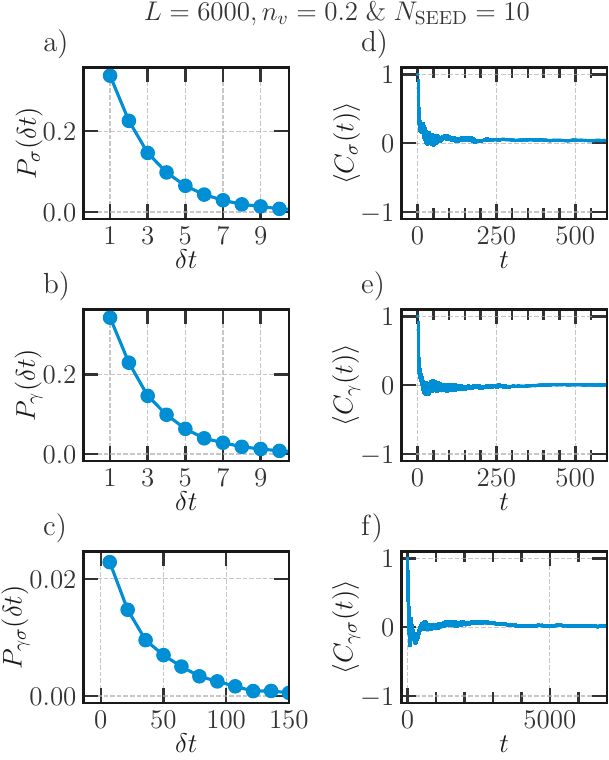}
	\caption{a), b), and c) The normalized histograms $\langle P_q(\delta t) \rangle$ (for $q=\sigma$, $\gamma$, and $\zeta=\gamma \sigma$ respectively) all fall off monotonically with increasing $\delta t$ and go to zero for $\delta t \gg \tau_q$, where $\tau_{\sigma} \approx 3$, $\tau_{\gamma} \approx 3$, and $\tau_{\gamma \sigma} \approx 30$. d), e), and f) The corresponding autocorrelation functions $\langle C_q(t) \rangle $ all decay to zero when $t \gtrsim 150\tau_q$. Angular brackets around $P_q$ and $C_q$ denote an average over just ten initial conditions at $t=0$, corresponding to arbitrarily-chosen maximum-density dimer packings of ten different randomly-diluted samples of the site-diluted triangular lattice. Essentially the same (but more noisy) behavior is seen at the level of the individual time series that start from these ten initial conditions. See Sec.~\ref{subsec:Chaos} for details.}
	\label{fig:TimeSeries} 
\end{figure}

We use $\sigma(t)$ and $\gamma(t)$ to encode the even ($\sigma=-1$) or odd ($\sigma=1$) parity of $m({\mathcal G}(t))$ and the ${\mathcal R}$-type ($\gamma = 1$) or ${\mathcal P}$-type ($\gamma = -1$) character of a sample, and $\zeta(t) = \sigma(t) \gamma(t)$ to monitor their correlations.  All three variables change sign frequently at characteristic rates $\tau_q^{-1}$ (for $q= \sigma$, $\gamma$, $\zeta$), as seen in the normalized histograms $P_q(\delta t)$  of time intervals $\delta t$ between jumps (Figure~\ref{fig:TimeSeries}). The corresponding integrated autocorrelation functions $C_q(t)  = \sum_{0 \leq j \leq t}  q(0)q(j)/(t+1)$ all decay to zero when $t \gg \tau_q$ (Figure 5).
 Thus, although $\gamma(t)$ is completely determined by the sample geometry, we cannot predict it from $\sigma(t)$. Nor are we able to predict it from any other macroscopic property of the diluted lattice. In this fairly precise sense, the large-scale geometry of maximum-density dimer packings of the site-diluted triangular lattice exhibits chaos for $n_v < n_v^{\rm crit}$.
Naturally, there is much more to study along these lines, including, for instance, spatiotemporal aspects of the chaotic rearrangement of the boundaries of $\calR$-type and $\calP$-type regions in response to very localized changes in the vacancy concentration. We have not attempted that here.

	\section{Discussion}
	\label{sec:DiscussionandOutlook}	
Perhaps the most interesting (and certainly the most unusual) aspect of our results is the identification of extended regimes of vacancy concentration in which usual thermodynamic self-averaging fails for various thermodynamic densities. In fact, not only does self-averaging fail, but identically-prepared random samples fall into two populations that are in two different thermodynamic phases, since monomers are deconfined in ${\mathcal R}$-type samples but localized in ${\mathcal P}$-type samples. As emphasized earlier in the Introduction, this aspect of the Gallai-Edmonds percolation phenomena identified here appears to be without a known parallel.

What {\em observable} consequences is this expected to have? This is arguably the most crucial question to be thrown up by our study. Below, Sec.~\ref{subsec:PhysicalImplications} addresses this question and discusses physical implications of our results.
Our finding of a sharply-defined percolation transition also leads to several other natural and interesting questions about this Gallai-Edmonds percolation phenomenon itself, as well as questions about how this transition affects the wavefunctions of the collective Majorana modes.
These are also discussed briefly in Sec.~\ref{subsec:Wavefunctions} and Sec.~\ref{subsec:GallaiEdmondsPercolation}.

\subsection{Physical implications}
\label{subsec:PhysicalImplications}
Does the observed lack of self-averaging (alluded to above) in the thermodynamic limit  have an observable physical consequence? The answer is perhaps not immediately obvious, since we have already seen that there are quantities such as the overall thermodynamic density $w$ of monomers in maximum matchings of the disordered graph (equivalently, collective Majorana zero modes of the associated Majorana network Hamiltonian) which remain self-averaging and sharply-defined in spite of the violations of self-averaging seen in other properties; this is clear from the fact that the histogram of $w$ has a single well-defined peak that gets narrower with increasing $L$.  

Nevertheless, we now argue that macroscopic differences between $\calR$-type and $\calP$-type samples at small values of dilution are expected to indeed have an observable consequence in the pinned disordered vortex lattice state of topological superconductors~\cite{Laumann_Ludwig_Huse_Trebst_2012,Kraus_Stern_2011,Read_Green_2000}.
In this pinned vortex lattice state, each vortex hosts a Majorana mode in its core, and the quantum-mechanical mixing of these modes on neighboring vortices can be modeled by a Majorana network Hamiltonian, defined as in Eq.~\ref{eq:HMajorana}. The vacancy disorder we consider has a natural interpretation in terms of missing vortices, {\em i.e.}, vacancy defects in the pinned vortex lattice.

So long as the bond disorder (in the values of $a_{r r'}$) places the system in a phase with delocalized zero energy wavefunctions, one expects that the localization length $\xi_{\Delta G}$ associated with $\Delta G$, the basis-independent zero energy on-shell Green function defined in Eq.~\ref{eq:DeltaG}, will be infinite at small nonzero dilution $n_v$ if $\calR$-type regions percolate, and finite if all $\calR$-type regions are finite in size. This is because the thermal metal phase identified in Ref.~\cite{Laumann_Ludwig_Huse_Trebst_2012,Kraus_Stern_2011} is {\em stabilized} by disorder. By analogy to the Green function description of electrical conductance~\cite{Baranger_Stone_1989}, we expect that this Green function $\Delta G$ controls the contribution of these Majorana modes to the thermal conductance $\kappa_T$. 

This in turn leads us to expect that the percolation transitions studied here has a direct signature in the thermal conductance $\kappa_T$, since a sample with a giant system-spanning  $\calR$-type region will be a thermal metal of these Majorana modes, while a sample in which all $\calR$-type regions are small will be a thermal insulator.  It also leads us to expect that $\kappa_T$ will carry direct signatures of the violations of self-averaging seen deep in the percolated phase at small values of dilution.


Within the framework of the triangular vortex lattice model introduced in Ref.~\cite{Laumann_Ludwig_Huse_Trebst_2012}, one can make this expectation somewhat more precise: In this model, the underlying triangular lattice is undisturbed in terms of its connectivity, but the nearest neighbor mixing amplitudes $a_{rr'}$ in Eq.~\ref{eq:HMajorana} have sign disorder and take on values $\pm J$ with probabilities $p$ and $1-p$ respectively. For $p$ near $p=1/2$, which corresponds to maximal sign disorder, a thermal metal phase was found~\cite{Laumann_Ludwig_Huse_Trebst_2012}. 
Our results lead us to expect that this thermal metal phase will show violations of thermodynamic self-averaging in its heat conductance as soon as there is a small nonzero density $n_v$ of quenched vacancies (missing vortices), while larger values of $n_v$ (above the percolation threshold of maximum matchings) will destroy this effect since the system of Majorana excitations will now be a thermal insulator in all the samples.

This conclusion follows immediately from the fact that small $n_v$ places the sample deep in the percolated phase that violates thermodynamic self-averaging. However, as alluded to above, it is contingent on the expectation that the zero energy wavefunctions within a large percolating $\calR$-type region of an $\calR$-type sample remain extended for a small nonzero value of $n_v$ if $p$ lies in the metallic phase of the model considered in Ref.~\cite{Laumann_Ludwig_Huse_Trebst_2012}. To put it another way: For the Majorana modes of $\calR$-type samples to {\em not} have a thermal conductance that is macroscopically different from that of $\calP$-type samples at small nonzero $n_v$ and $p$ near $p=1/2$, the vacancy disorder in this regime would have to induce a finite and rather small localization length for the collective Majorana modes inside the system-spanning $\calR$-type regions, in spite of the fact that $\xi_{\Delta G}$ is infinite at $n_v = 0$. This is rather unlikely since the thermal metal phase is actually stabilized by disorder in the calculations of Ref.~\cite{Laumann_Ludwig_Huse_Trebst_2012,Kraus_Stern_2011}.

As we have noted earlier in Sec.~\ref{subsec:VacancyInducedMoments}, the morphology of $\calR$-type and $\calP$-type regions is also expected to play a key role in determining the nature of the vacancy-induced local moment instability~\cite{Ansari_Damle_2024} of gapped short-range resonating valence bond states of frustrated magnets. To recap: The presence of a nonzero density $w$ of monomers in maximum matchings of the diluted lattice implies that vacancy disorder leads at intermediate energies (well below the gap of the pure system) to a nonzero density of emergent local moments whose physics determines the ultimate fate of the system and the character of its ground state. 

The character of this ground state is expected to be strongly influenced by the manner in which these emergent moments are distributed in the sample, since the interactions between them are expected to be short-ranged, and therefore confined to within individual $\calR$-type regions. This leads us to expect that the morphology and size of $\calR$-type regions hosting these emergent moments plays a key role in this physics.
Indeed, one has a heuristic picture in terms of a spin-liquid background confined to $\calP$-type regions coexisting with a possibly glassy locally ordered state of the emergent local moments inside the $\calR$-type regions. 

{\em Independent of the ultimate fate of this system at $T=0$}, {\em i.e.}, independent of the precise nature of the disordered ground state, the susceptibility at low but nonzero temperature is anyway expected to exhibit macroscopically significant differences between ${\mathcal R}$-type and ${\mathcal P}$-type samples when $n_v$ is small, {\em i.e.}, with weak vacancy disorder. This is of course because one is always in the percolated phase (say on the triangular lattice) of the maximum matchings at such small $n_v$. This percolated phase exhibits dramatic violations of thermodynamic self-averaging, with $\calR$-type samples (that constitute a fraction $f_{\calR} \approx 0.50(2)$ of all samples) having small $\calP$-type regions and one infinite $\calR$-type region and $\calP$-type samples (that constitute $1-f_{\calR}$ of all samples) having small $\calR$-type regions and one infinite $\calP$-type region.

In addition, the ground state of $\calR$-type samples could potentially be quite different from that of $\calP$-type samples.  In this scenario, $\calR$-type samples would have long-range spin-glass order that breaks spin-rotation symmetry, while $\calP$-type samples would break no symmetries. Clearly it would be interesting to study this possibility in more detail in future work.

\subsection{Wavefunctions of collective Majorana modes}
\label{subsec:Wavefunctions}
The most basic question regarding collective Majorana mode wavefunctions concerns their localization properties {\em within} a single $\calR$-type region. Since each $\calR$-type region $\calR_i$ hosts $\calI_i$ linearly independent such modes, the basis independent way of asking this question is to study the localization length $\xi_{\Delta G}$ of the on-shell zero-energy Green function $\Delta G$ defined in Eq.~\ref{eq:DeltaG}. Is this localization length small compared to the typical size of $\calR$-type regions in phases in which $\calR$-type regions have a nonzero probability of percolating in the thermodynamic limit? 
This question needs to be formulated separately for $\calR$-type and $\calP$-type samples in the  low-dilution phase in two dimensions, and phase III and phase IV in three dimensions. 
A related question is the following: If $\xi_{\Delta G}$ diverges when $\calR$-type regions percolate, is this divergence governed by an independent exponent, or is it controlled by the same correlation length exponent $\nu$ that controls the divergence of the mean size of the $\calR$-type regions themselves? Clearly, the answer to this question will depend on the nature of the disorder distribution for the mixing amplitudes $a_{ r r'}$. 
For the disorder ensemble studied in Ref.~\cite{Laumann_Ludwig_Huse_Trebst_2012,Kraus_Stern_2011} and alluded to earlier, {\em i.e.}, with $a_{rr'}$ taken to be $\pm J$ with probability $p$ and $1-p$ respectively on the triangular lattice,  there is a regime with extended states at zero energy in the vicinity of $p=1/2$, while the zero energy ``impurity band'' states are localized in the vicinity of $p=0$ and $p=1$. It would clearly be interesting to extend this to include vacancy disorder and study the $n_v$ dependence of the localization length $\xi_{\Delta G}$ for both $\calR$-type and $\calP$-type samples at small $n_v$, as well as in the vicinity of the percolation transition. 

Another interesting question has to do with the manner in which these collective Majorana modes arise from the mixing of ``primordial'' zero modes localized on each blossom: as we have already reviewed in Sec.~\ref{subsec:CollectiveMajoranaModes}, each blossom $\calB_i$, if isolated from the o-type sites to which it is connected by links of the lattice, hosts a single Majorana zero mode. These primordial zero modes on different blossoms within a single $\calR$-type region mix via their coupling to the o-type sites of the region, to leave behind $\calI_i$ Majorana modes that survive this mixing. 
Thus, the wavefunctions of these collective Majorana zero modes are expected to depend quite sensitively on the typical size of individual blossoms. Since the largest blossom $\calB_{\rm max}$ scales in mass as $m(\calR_{\rm max})$ in $\calR$-type samples in phases III and IV in three dimensions and the percolated phase in two dimensions, but not in $\calP$-type samples in these phases, and not in phases I and II in three dimensions (although $\calR$-type regions do percolate with unit probability in these latter phases), this raises the following interesting question: Does the zero-energy Green function of the Majorana network show signatures of the phase transition from phase II to phase III in three dimensions?

In the triangular lattice case, another natural question has to do with the fact that one can view the triangular lattice as a square lattice with one set of diagonal bonds added to increase the coordination number to six. These diagonal bonds connect two sites on the same sublattice of the square lattice and break the bipartite symmetry of the parent square lattice. In Ref.~\cite{Bhola_Biswas_Islam_Damle_2022}, it was argued that {\em odd} $\calR_A$ and $\calR_B$ regions (with an odd number of monomers) of a bipartite lattice were guaranteed to host at least one perturbatively stable (to additional non-bipartite hopping) collective Majorana mode of the corresponding non-bipartite Majorana network obtained by perturbing the original bipartite problem with small non-bipartite mixing amplitudes. This leads to the natural question: What is the connection between the topologically protected collective Majorana zero modes  of the site-diluted triangular lattice, and such perturbatively stable modes obtained by starting with the parent site-diluted square lattice and introducing a small mixing amplitude along one diagonal of each surviving square plaquette?

\subsection{Gallai-Edmonds percolation}
\label{subsec:GallaiEdmondsPercolation}
	Our results also lead very naturally to a number of interesting questions about the Gallai-Edmonds percolation phenomena themselves.
One such question that arises immediately has to do with the relation between the unusual low-dilution phase identified here on the triangular lattice, and the large-scale structure of the Dulmage-Mendelsohn decomposition of the site-diluted square lattice in the corresponding dilution range. In the square lattice case, $\calR_A$ and $\calR_B$ type regions that host the monomers of maximum matchings and topologically protected zero modes of bipartite hopping problems grow in spatial extent as $n_v$ is reduced, but never percolate except in the limit $n_v \rightarrow 0$~\cite{Bhola_Biswas_Islam_Damle_2022}. 

This unusual {\em incipient percolation} phenomenon also occurs on the site-diluted honeycomb lattice, and to that extent, appears to be universal~\cite{Bhola_Biswas_Islam_Damle_2022}. Importantly, $\calP$-type regions appear to play no role in it at all, in the sense that their size remains small in this limit (of course, the entire pure sample is $\calP$-type {\em at} $n_v =0$, so  this statement is only true when the thermodynamic limit is taken keeping $n_v$ fixed at a small but {\em nonzero} value). The question then is: How does this regime, with incipient percolation of $\calR_A$ and $\calR_B$ regions and microscopically small $\calP$-type regions, go over to the unusual percolated phase found here on the triangular lattice when one randomly starts adding in some of the additional diagonal links needed to convert the square lattice into a triangular lattice? If the fraction of diagonal links added is denoted by $x$, then is there a critical value $x_c$ at which there is a percolation transition as a function of $x$? (with $n_v$ held fixed at a small enough value in this low-dilution regime).
	
	Another equally immediate question has to do with the nature of the Gallai-Edmonds critical point in two dimensions. Based on the comparison of our results on the triangular and Shastry-Sutherland lattices, one may conclude that the large-scale behavior in the critical region is certainly universal. As mentioned earlier,  our estimate of the corresponding correlation length exponent $\nu =1.45(10) $ of the Gallai-Edmonds percolation transition is quite close to but not entirely consistent with the exact value of the correlation length exponent $\nu = 4/3$ at the ordinary geometric percolation transition in two dimensions ~\cite{Nijs_1979, Nienhuis_Riedel_Schick_1980, Pearson_1980}, since this exact result lies just outside our error bars which have been estimated quite conservatively. In addition, our value of $\eta = 0.09 (2)$ is certainly very different from the value of the corresponding anomalous exponent $\eta = 5/24$ at the geometric percolation transition in two dimensions.	This raises the questions: Is there a different continuum field theory description of the Gallai-Edmonds percolation transition in two dimensions? Does the Gallai-Edmonds critical point exhibit conformal invariance like the usual percolation transition in two dimensions?

In contrast, our estimate of the exponent $\nu$ at all the successive transitions in $d=3$ appears to be consistent with the corresponding result for ordinary geometric percolation, and our analysis cannot really distinguish between $\eta = 0$ and a small nonzero value for $\eta$ at any of these transitions. This is also true of the estimates for the analogous exponents in the bipartite Dulmage-Mendelsohn percolation transition studied earlier~\cite{Bhola_Biswas_Islam_Damle_2022}. Thus, it appears that our results for these exponents in three dimensions are essentially consistent with what is known from numerical studies of ordinary geometric percolation in three dimensions. However, as we have emphasized earlier, the properties of the percolated phases themselves in the non-bipartite case are very different from the usual behavior in geometric percolation. Any attempt at a coarse-grained field-theoretical description would therefore need to reconcile these two findings.

Another question has to do with bond dilution. Are there such percolation phenomena of $\calR$-type and $\calP$-type regions driven by bond dilution instead of site dilution, and do they share the universal properties identified here? The answer to this is not clear even in the bipartite case and deserves further study in that case, as well as in the general non-bipartite case studied here. Additionally, since the phase diagram as a function of site dilution in three dimensions is so different from the corresponding phase diagram in two dimensions, it is also interesting to ask if these dramatic differences seen in three dimensions arise as a result of the higher dimensionality, or are already present in {\em non-planar} two dimensional lattices. Again, the answer to this is not clear even in the bipartite case studied earlier.

Further, in the general case of non-bipartite lattices, there is another curious fact that adds to the interest in this question: On a certain class of lattices, of which the kagome lattice is the most prominent two-dimensional example, site dilution alone does not lead to a nonzero density of monomers in maximum matchings of the diluted lattice for any nonzero dilution fraction $n_v$~\cite{Ansari_Damle_2024,Bhola_Damle_ClawFree2024,Faudree_1997,Sumner_1974}. Indeed, on such lattices, each geometrically connected component of the site-diluted lattice hosts one monomer (no monomers) in any maximum matching if the size of this component (number of sites) is odd (even). 

As a result, with site dilution, there is no Gallai-Edmonds percolation distinct from the ordinary geometric percolation transition on such lattices.  Given this, it would be indeed very interesting to ask: Does {\em bond dilution} on such lattices drive Gallai-Edmonds percolation transitions  that are in the same universality class as the Gallai-Edmonds transitions identified here in the site-diluted case? And do the low-dilution phases share the unusual properties identified here?

Finally, building on the comparison made earlier (in the Introduction) between Gallai-Edmonds percolation and Fortuin-Kasteleyn percolation in spin $S=1$ Ising models, one can ask: Is there a modification of the standard spin $S=1$ Ising model with a phase in which $S=0$ clusters percolate half the time and $S= \pm 1$ clusters percolate half the time? If yes, this phase would exhibit the kind of emergent symmetry that we have speculatively referred to in Sec.~\ref{subsec:BlossomMorphology}.

These are some of the interesting questions that are brought to the fore by our study, and we hope to return to them in future work.	

	\section{Acknowledgements}
	\label{sec:Acknowledgements}
	We thank  Mustansir Barma, Deepak Dhar, Subhajit Goswami, and Piyush Srivastava for useful discussions. We are also grateful to the departmental system administrators K. Ghadiali and A. Salve, who have ensured the efficient utilization of departmental computational resources by their prompt actions and timely interventions.
	The work of RB was supported by a graduate fellowship of the DAE, India at the Tata Institute of Fundamental Research (TIFR) and contributed to his Ph.D thesis at the TIFR. KD was supported at the TIFR by DAE, India and in part by a J.C. Bose Fellowship  (JCB/2020/000047) of SERB, DST India, and by the Infosys-Chandrasekharan Random Geometry Center (TIFR). All computations were performed using departmental computational resources of the Department of Theoretical Physics, TIFR and additional resources supported by a J.C. Bose Fellowship grant (JCB/2020/000047).

	\bibliography{3000wordsNoURL}  	

\onecolumngrid
\newpage
\appendix


\section{Additional details in two dimensions}
\label{subsec:App:AdditionalDetails2D}
In this appendix we present some additional results that shed further light on the percolated phase and the phase transition in two dimensions. This is done via a sequence of figures that display the relevant data with captions that should be read in conjunction with the corresponding discussion in Sec.~\ref{subsec:Results2D}.

		\begin{figure}[h]
		\begin{tabular}{cc}
			a)\includegraphics[width=0.5\columnwidth]{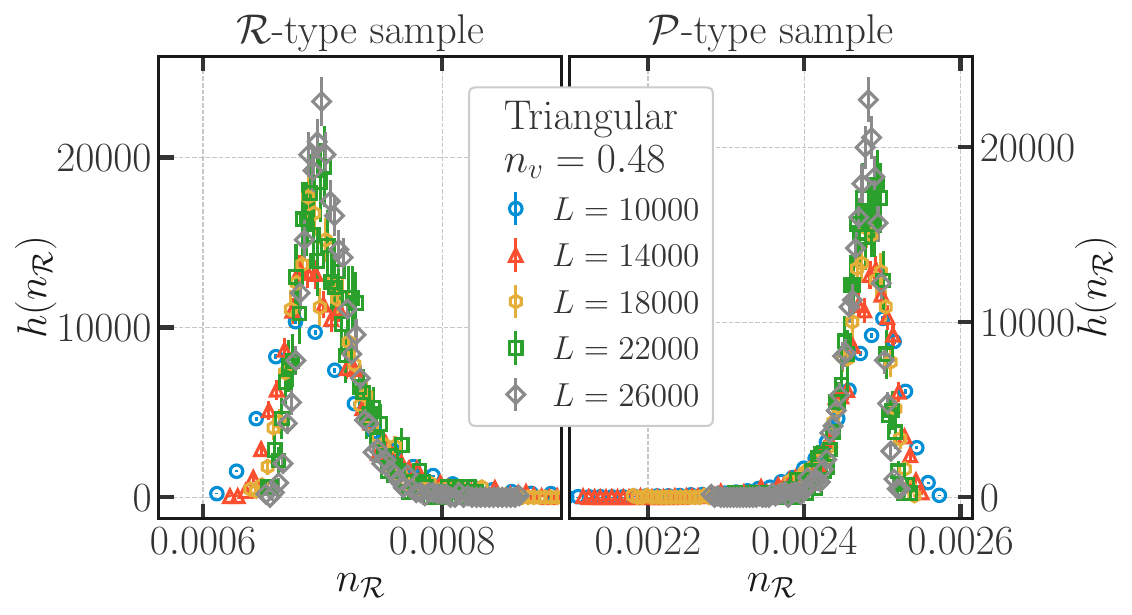}&
			b)\includegraphics[width=0.5\columnwidth]{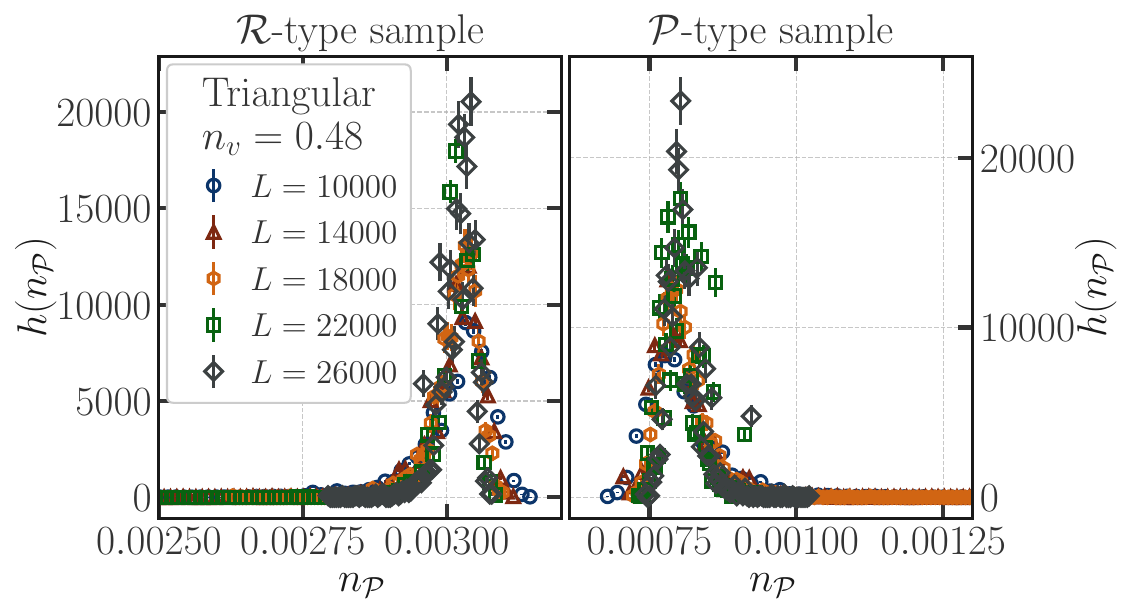}\\
			c)\includegraphics[width=0.5\columnwidth]{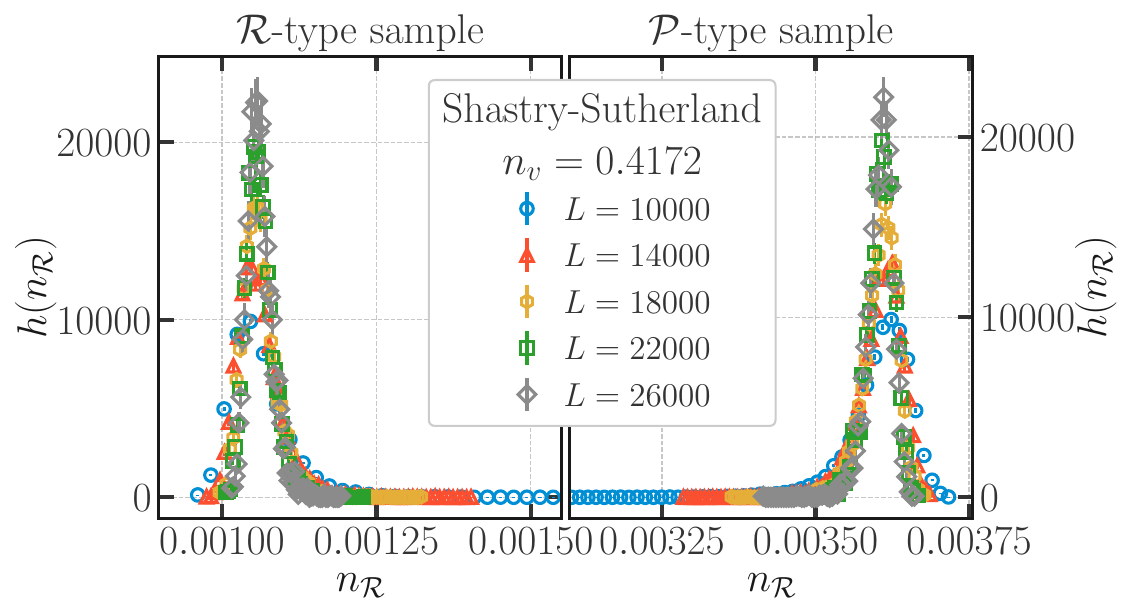}&
			d)\includegraphics[width=0.5\columnwidth]{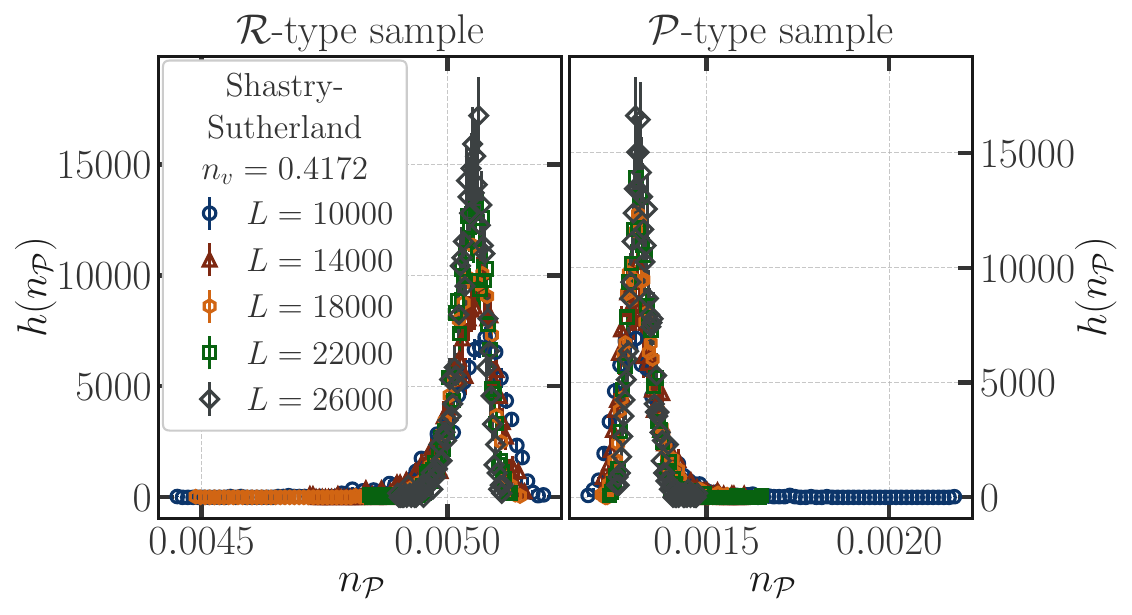}\\
		\end{tabular}
		\caption{
		(a) and (c) The histogram $h_{\calR}(\nR)$ of $\nR$, the number of distinct $\calR$-type regions scaled by the mass $m(\calG)$ of the largest geometric cluster, has a single sharply defined  peak at an $L$-independent value in both $\calR$-type and $\calP$-type samples, but the positions of these two peaks are macroscopically distinct: $\calR$-type samples have a significantly smaller density $\nR$ relative to $\calP$-type samples. (b) and (d) The histogram of $\nP$, the number of distinct $\calP$-type regions in $\calG$ scaled by the mass $m(\calG)$ of $\calG$, shows similar behavior, with the role of $\calR$-type and $\calP$-type samples being interchanged.	
		 This provides additional confirmation of the overall picture that emerges from the results presented in Sec.~{\protect{\ref{subsec:Results2D}}}.}
		\label{fig:App:hist_nR_nP_2D}
	\end{figure}

	\begin{figure*}
		\includegraphics[width=0.45\columnwidth]{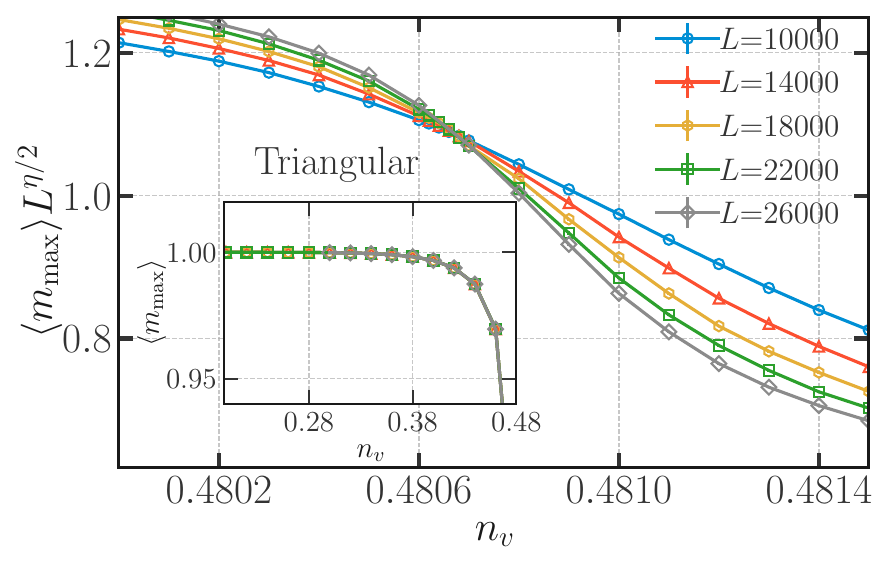}
		\includegraphics[width=0.45\columnwidth]{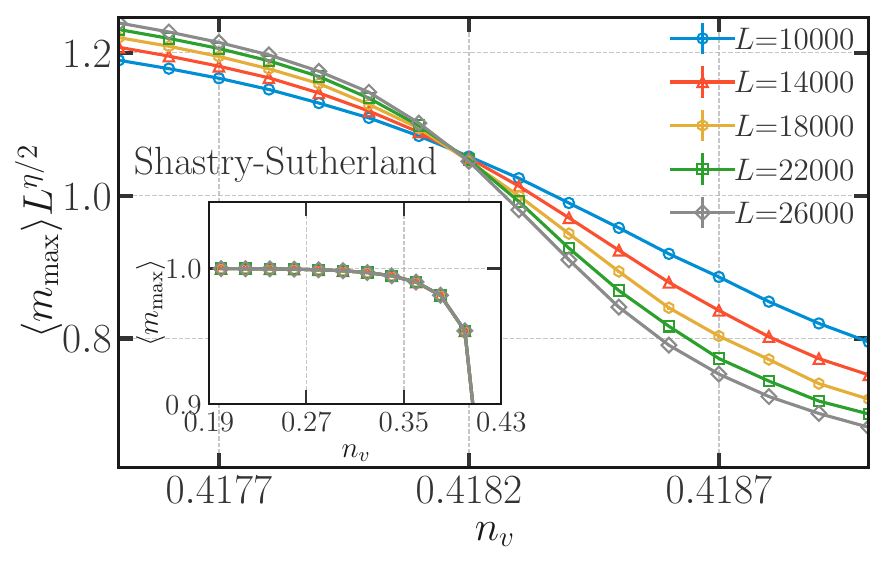}	
		\caption{$\langle m_{\rm max} \rangle$, the mean  mass of the largest Gallai-Edmonds region inside $\calG$ (the largest geometric cluster) in units of $m(\calG)$, the mass of $\calG$, saturates to an $L$ independent value in the low-dilution phase.  When scaled by $L^{-\eta/2}$, it shows a clear crossing at $n_v^{\rm crit}$.  This provides additional confirmation of the overall picture that emerges from the results presented in Sec.~{\protect{\ref{subsec:Results2D}}}.}
		\label{fig:App:mmax_2D}
	\end{figure*}

	\begin{figure*}
		\includegraphics[width=0.45\columnwidth]{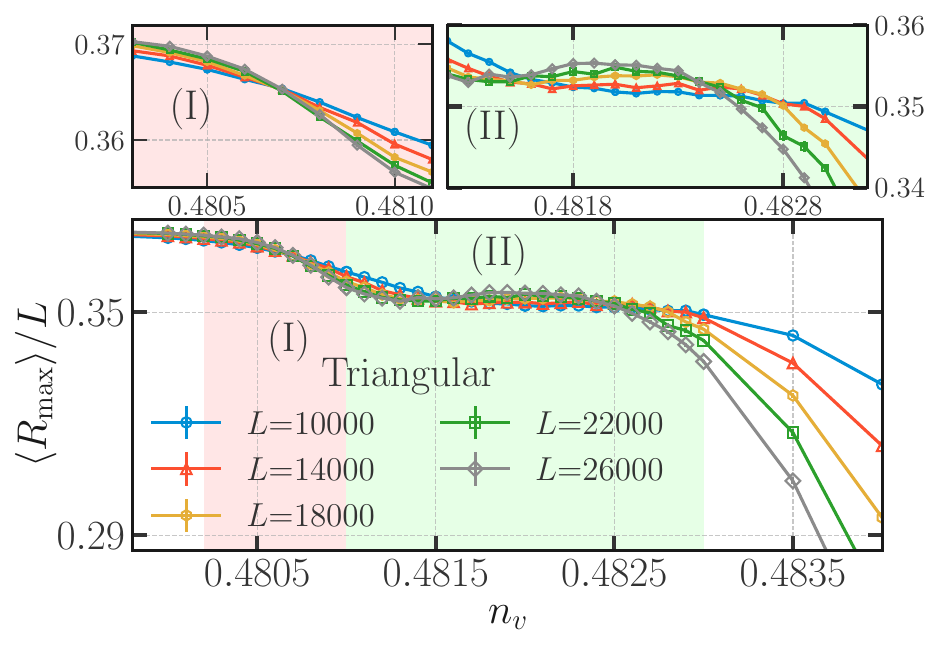}
		\includegraphics[width=0.45\columnwidth]{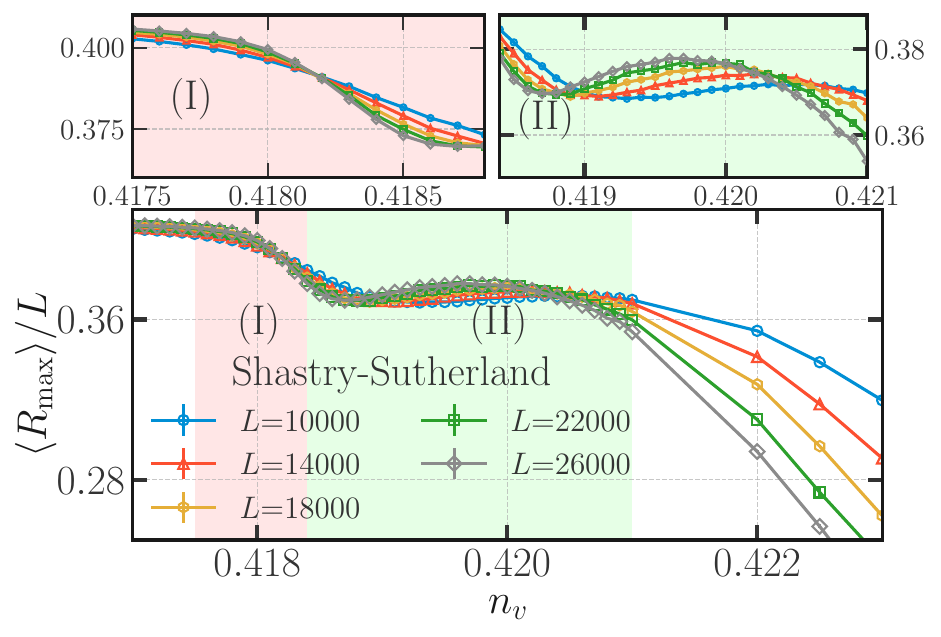}	
		\caption{Data for the dimensionless ratio $\langle \Rmax \rangle/L$, where $\langle \Rmax \rangle$ is the average radius of gyration of the largest Gallai-Edmonds region in a sample, shows a clear crossing at the critical dilution. In addition, it shows non-monotonic behavior very similar to that of $\xi/L$ on the high-dilution side of the crossing.  This provides additional confirmation of the overall picture that emerges from the results presented in Sec.~{\protect{\ref{subsec:Results2D}}}.}
		\label{fig:App:Rmax_2D}
	\end{figure*}
	
\begin{figure*}
\begin{tabular}{cc}
		a)	\includegraphics[width=0.45\columnwidth]{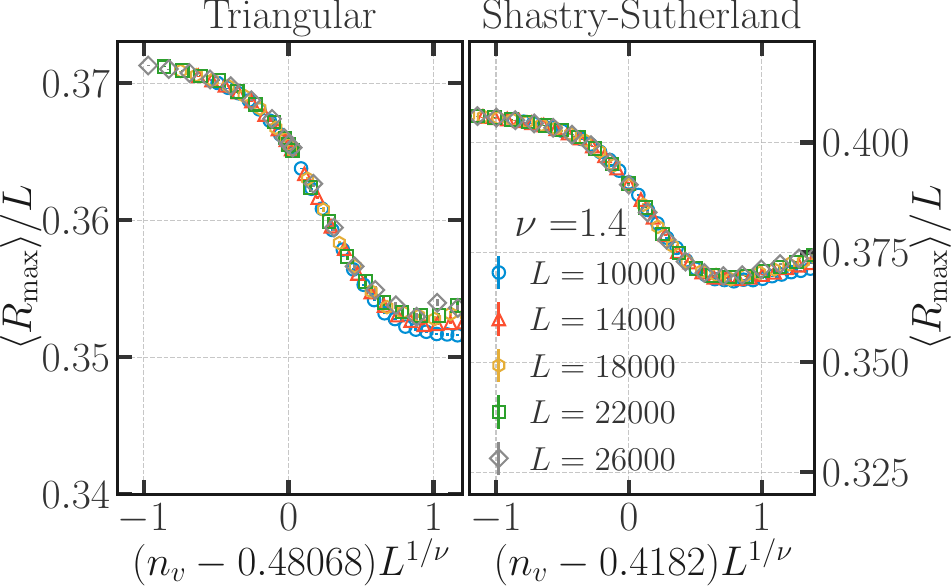}
		& b) \includegraphics[width=0.45\columnwidth]{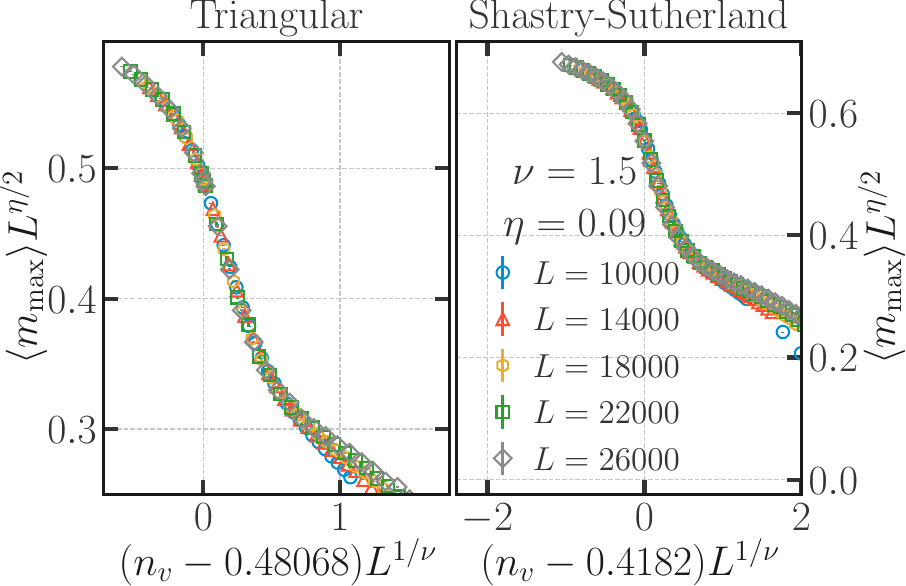}	\\
				\end{tabular}{cc}
		\caption{ (a) Data for the dimensionless ratio $\Rmax/L$, where $\Rmax$ is the radius of gyration of the largest Gallai-Edmonds region in $\calG$, the largest connected component of the diluted $L \times L$ lattice, collapses onto a single scaling curve when plotted against the scaling variable $(n_v-n_v^{\rm crit})L^{1/\nu}$ for a suitable choice of $n_v^{\rm crit}$ and $\nu$ consistent with the position of the critical point and the value of $\nu$ inferred from the scaling of other observables.  (b) Data for $m_{\rm max}$, the mass of the largest Gallai-Edmonds region  in $\calG$, the largest connected component of the diluted $L \times L$ lattice, when scaled by $L^{-\eta/2}$ and plotted versus $(n_v - n_v^{\rm crit})L^{1/\nu}$, collapses onto a single scaling curve for suitable choices of $n_{v}^{\rm crit}$, $\nu$ and $\eta$ consistent with the position of the critical point and values of these exponents inferred from the scaling of other observables.  This provides additional confirmation of the overall picture that emerges from the results presented in Sec.~{\protect{\ref{subsec:Results2D}}}.}
		\label{fig:App:scaling_Rmaxmmax_2D}

\end{figure*}

\newpage
\section{Additional details in three dimensions}
\label{subsec:App:AdditionalDetails3D}
We now present some additional results on the intricate phase diagram found in three dimensions. This is again done via a sequence of figures that display the relevant data with captions that should be read in conjunction with the corresponding discussion in the main text, {\em i.e.}, Sec.~\ref{subsec:Results3D} in this case.
 	\begin{figure*}
		\begin{tabular}{cc}
			a)\includegraphics[width=0.5\columnwidth]{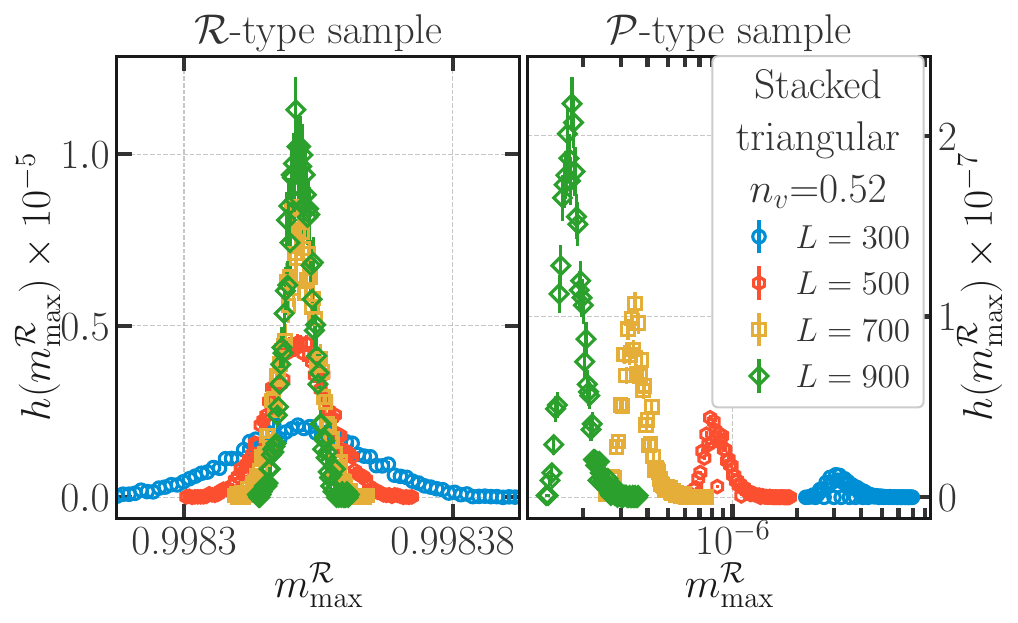} &	
			b)\includegraphics[width=0.5\columnwidth]{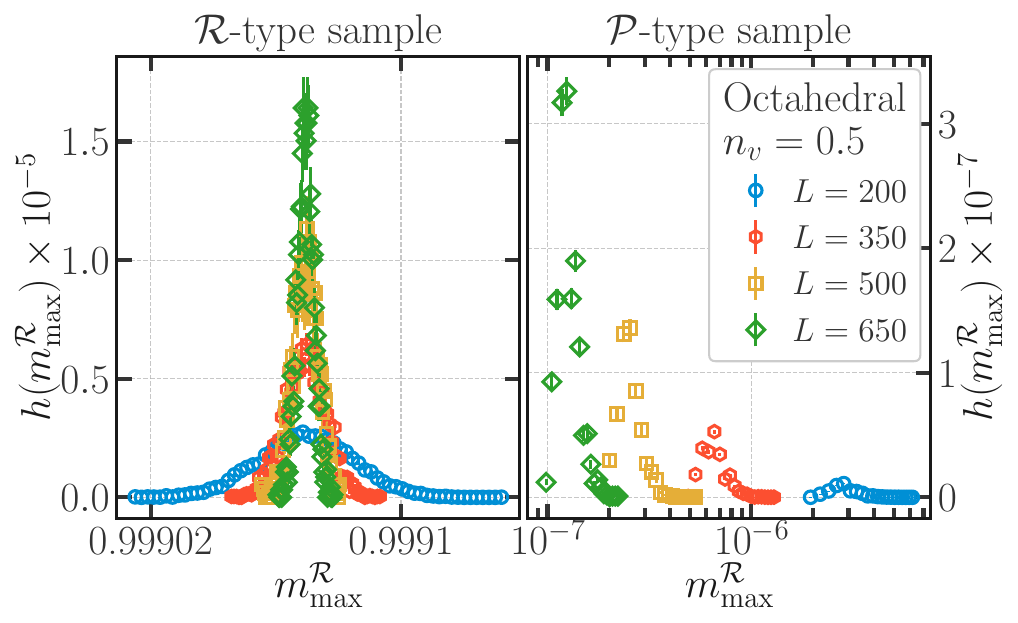}\\
			c)\includegraphics[width=0.5\columnwidth]{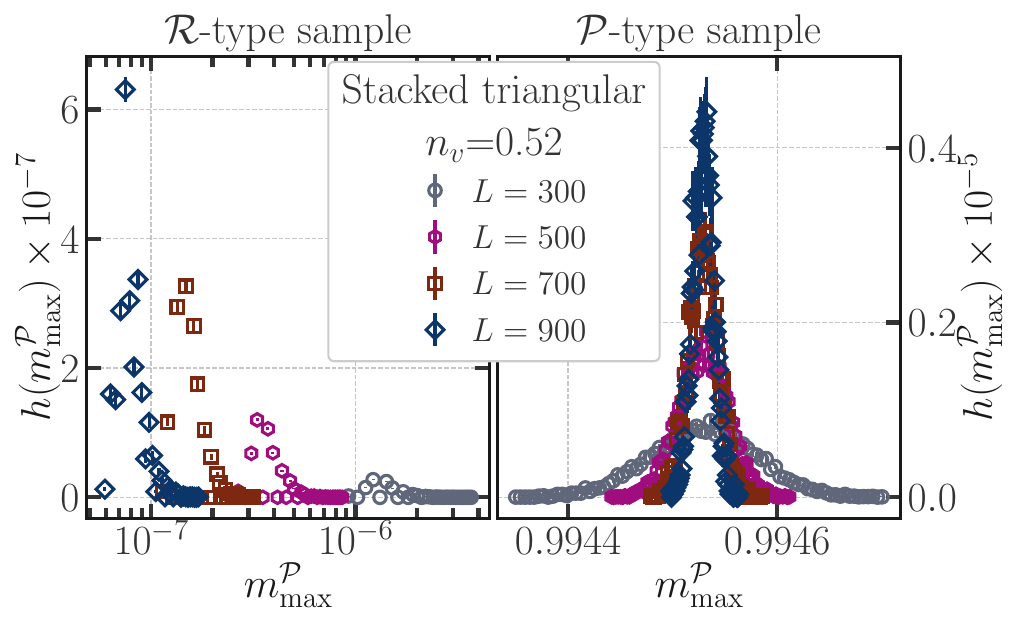} &	
			d)\includegraphics[width=0.5\columnwidth]{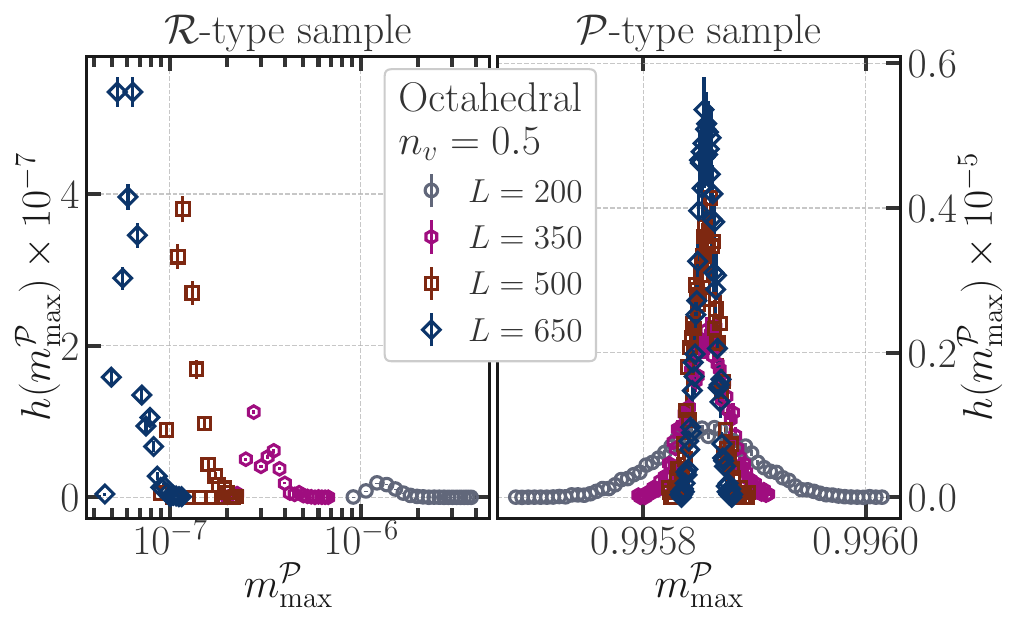}\\
		\end{tabular}
		\caption{As mentioned earlier in Sec.~\ref{subsec:Results3D}, deep in phase IV, the largest $\calR$-type region is macroscopic in size in $\calR$-type samples but not in $\calP$-type samples (where it has a small $L$-independent size in the large $L$ limit). Similarly, the largest $\calP$-type region is macroscopic in size in $\calP$-type samples but not in $\calR$-type samples (where it has a small $L$-independent size in the large $L$ limit). This is reflected in the sharply-defined peaks at $L$-independent positions in the histogram of the corresponding mass densities $\mmaxR$ and $\mmaxP$ respectively in $\calR$-type and $\calP$-type samples. On the other hand, the histogram of $\mmaxR$ ($\mmaxP$) in $\calP$-type ($\calR$-type) samples has a peak that shifts in position continually to lower and lower values as $L$ increases, consistent with the fact that the largest $\calR$-type ($\calP$-type) region of $\calP$-type ($\calR$-type) samples is finite in size in the thermodynamic limit.
			\label{fig:App:hist_mmaxpmaxIV_3D}}	
	\end{figure*}				
\begin{figure*}
\begin{tabular}{cc}
		a)	\includegraphics[width=0.45\columnwidth]{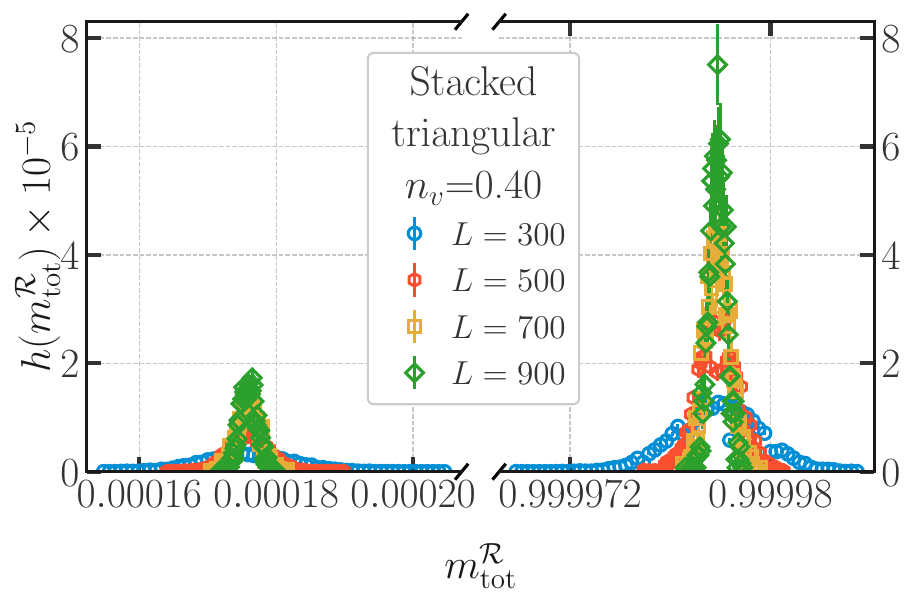}
		& b) \includegraphics[width=0.45\columnwidth]{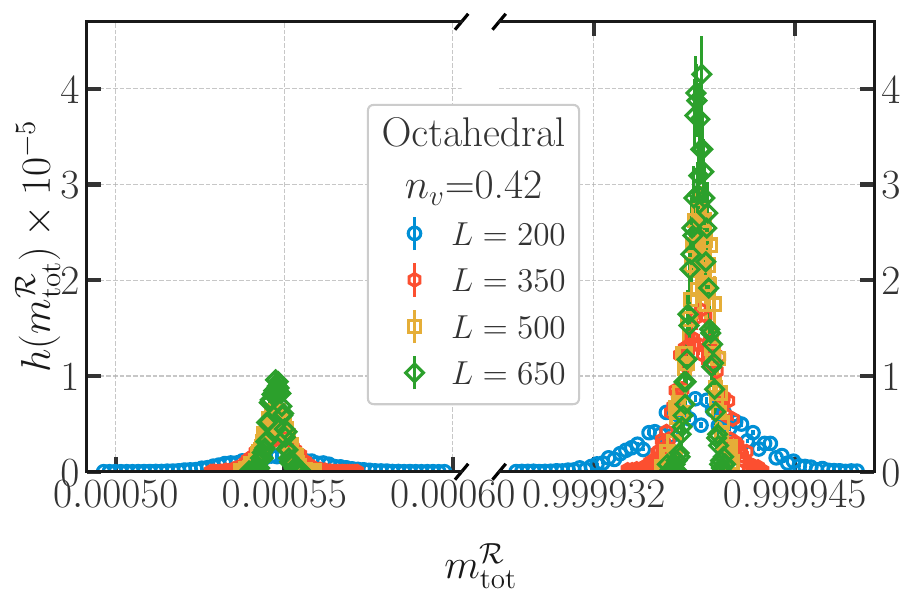}	\\
				\end{tabular}{cc}
		\caption{ As mentioned earlier in Sec.~\ref{subsec:Results3D},  the total mass density in $\calR$-type regions has a bimodal histogram deep in phase IV in three dimensions, and thus behaves in the same way as in the percolated phase in two dimensions. This is illustrated by the data shown here. Based on this data and analogous data at other values of $n_v$ $f_{\calR}$, the integrated weight in the peak to the right is estimated to be $f_{\calR} \approx 0.50(2)$ independent of $n_v$ (Note that the scale on the $n_v$ axis has been changed on the right to enable us to zoom in on this narrower peak; contrary to superficial appearances, the weight in the two peaks is indeed equal within statistical errors).}
		\label{fig:App:hist_mtotIV_3D}
\end{figure*}
	
 	\begin{figure*}
		\begin{tabular}{cc}
			a)\includegraphics[width=0.5\columnwidth]{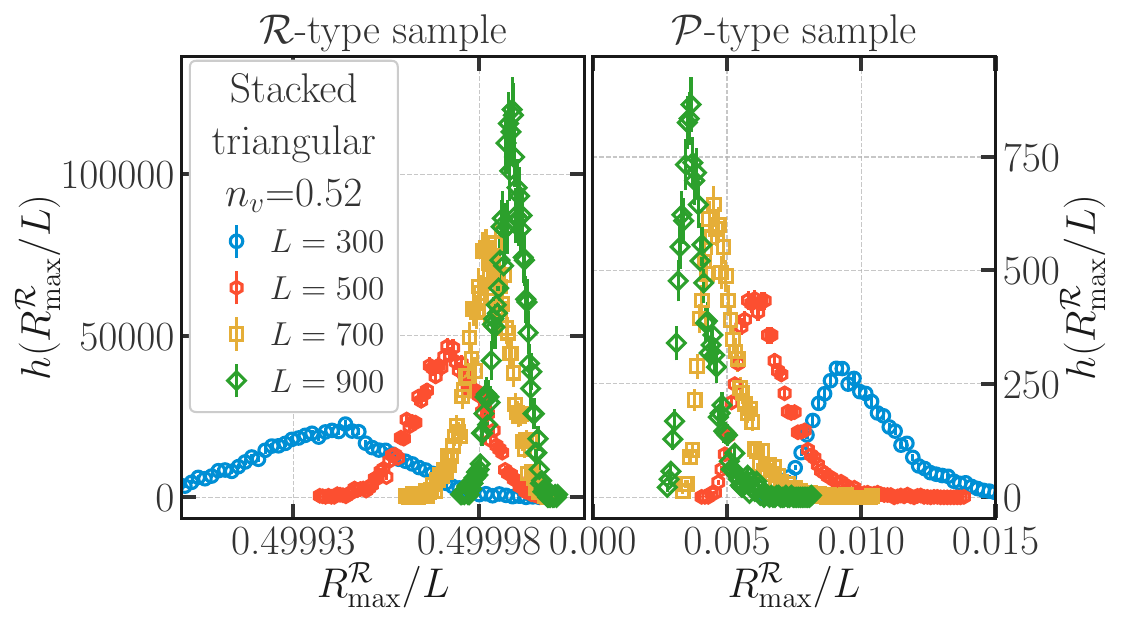} &	
			b)\includegraphics[width=0.5\columnwidth]{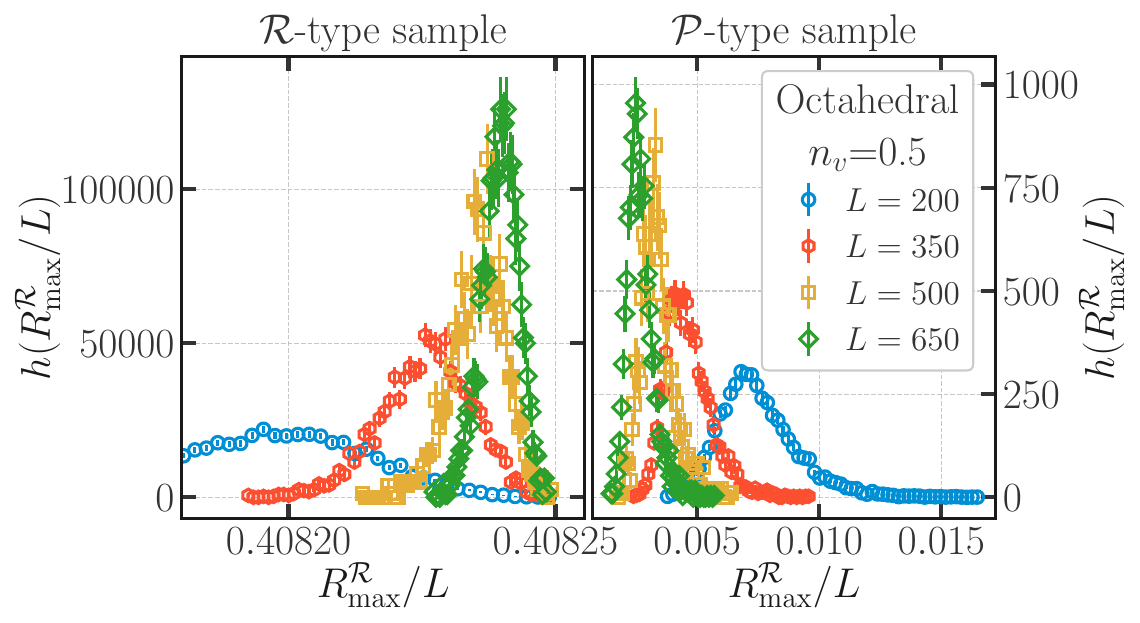}\\
			c)\includegraphics[width=0.5\columnwidth]{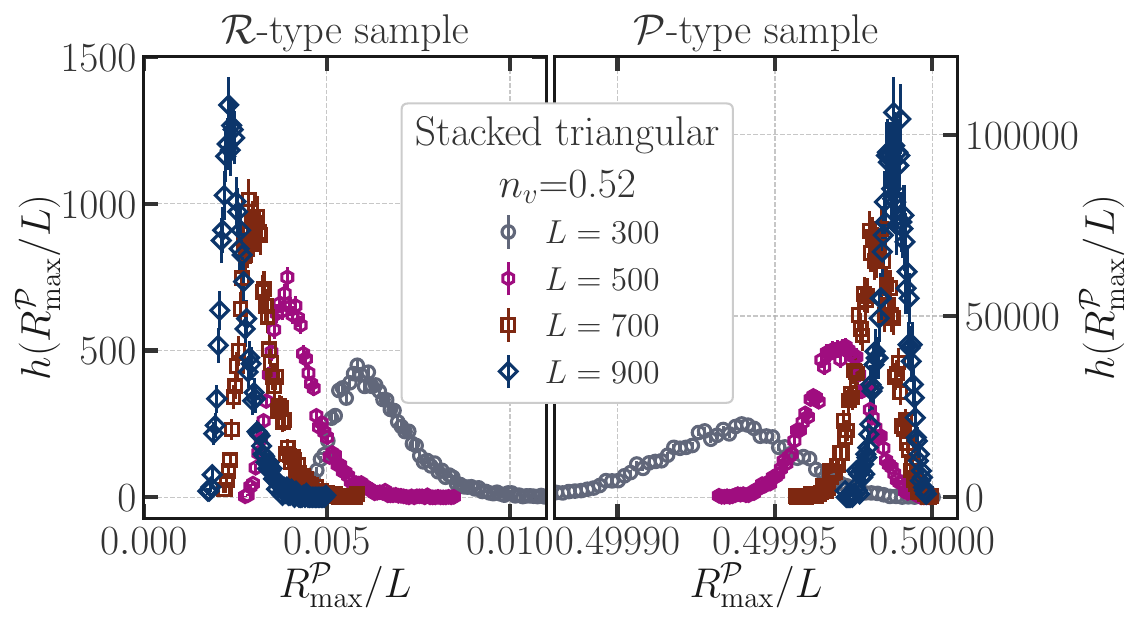} &	
			d)\includegraphics[width=0.5\columnwidth]{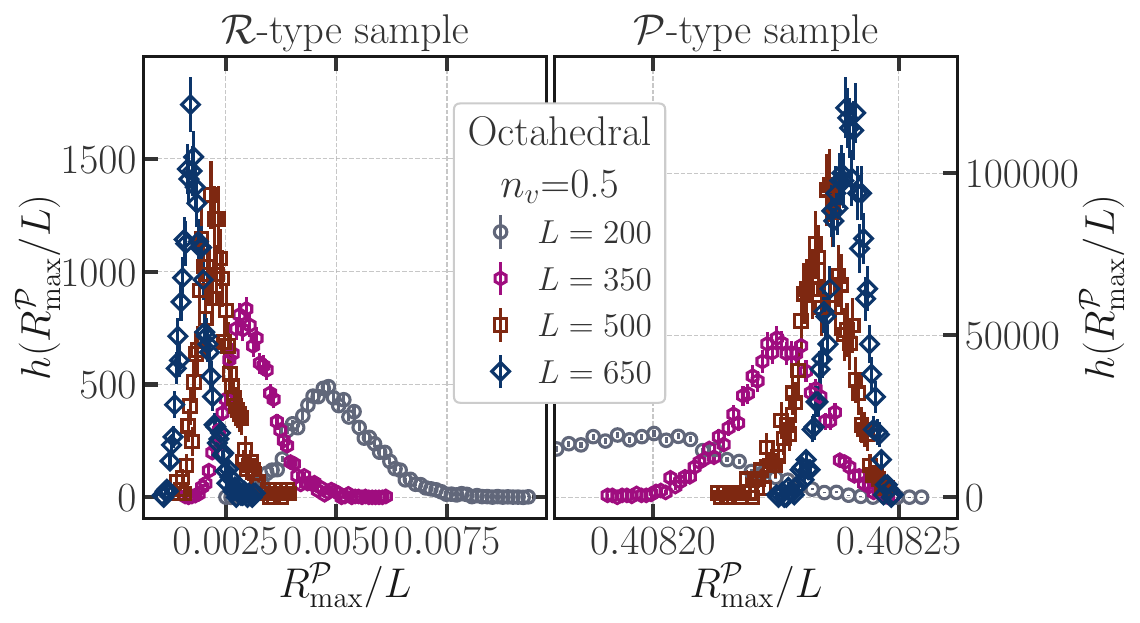}\\
		\end{tabular}
		\caption{As mentioned earlier in Sec.~\ref{subsec:Results3D}, deep in phase IV, the radius of gyration of the largest $\calR$-type region scales with $L$ in $\calR$-type samples but not in $\calP$-type samples (where it has a small $L$-independent value in the large $L$ limit). Similarly, the radius of gyration of the largest $\calP$-type region is scales with $L$ in $\calP$-type samples but not in $\calR$-type samples (where it has a small $L$-independent value in the large $L$ limit). This is reflected in the sharply-defined peaks at $L$-independent positions in the histogram of $\RmaxR/L$ and $\RmaxP/L$ respectively in $\calR$-type and $\calP$-type samples. On the other hand, the histogram of $\RmaxR/L$ ($\RmaxP/L$) in $\calP$-type ($\calR$-type) samples has a peak that shifts in position continually to lower and lower values as $L$ increases, consistent with the fact that the radius of gyration of the largest $\calR$-type ($\calP$-type) region of $\calP$-type ($\calR$-type) samples is finite and $L$-independent in the thermodynamic limit.
			\label{fig:App:hist_RmaxPRmaxRIV_3D}}	
	\end{figure*}				
	
 	\begin{figure*}
		\begin{tabular}{cc}
			a)\includegraphics[width=0.5\columnwidth]{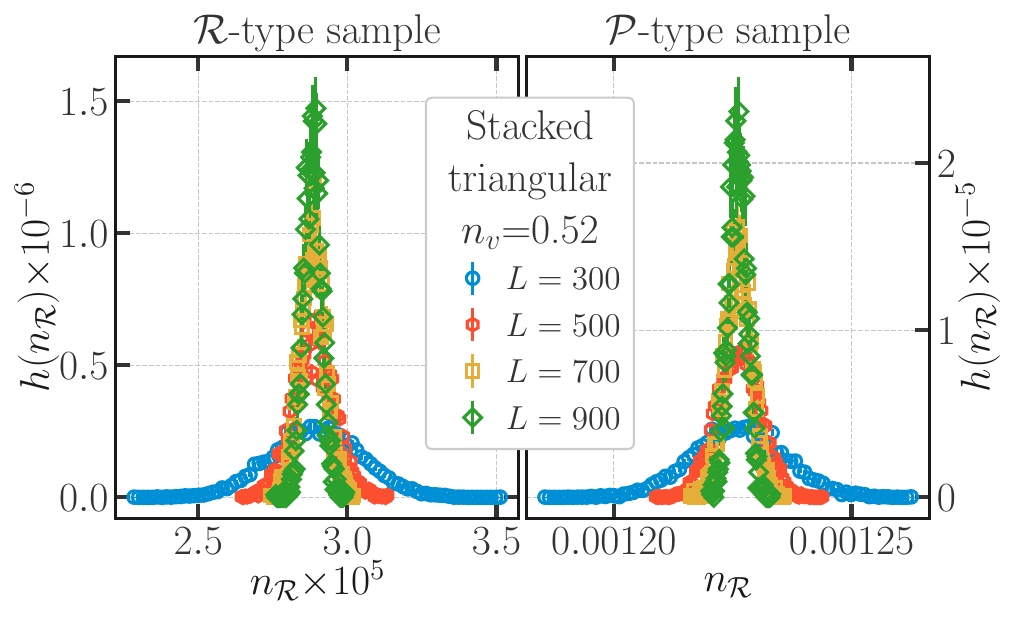} &	
			b)\includegraphics[width=0.5\columnwidth]{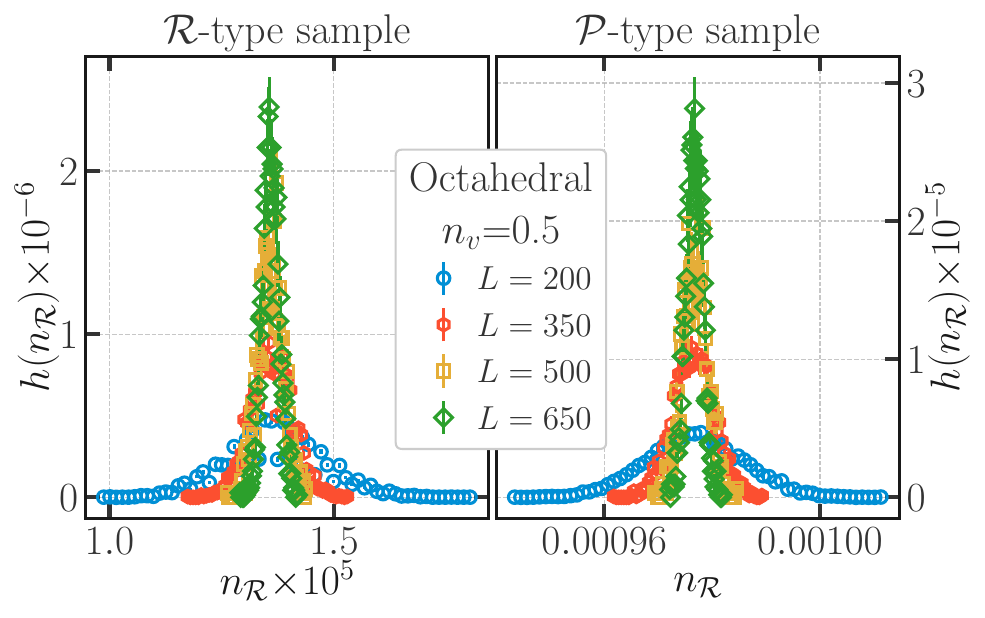}\\
			c)\includegraphics[width=0.5\columnwidth]{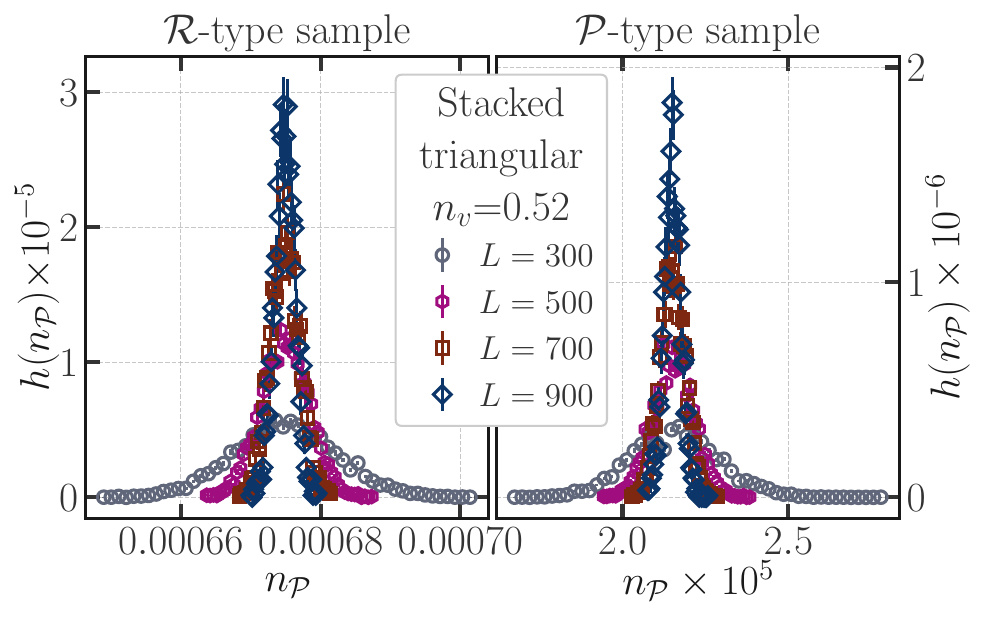} &	
			d)\includegraphics[width=0.5\columnwidth]{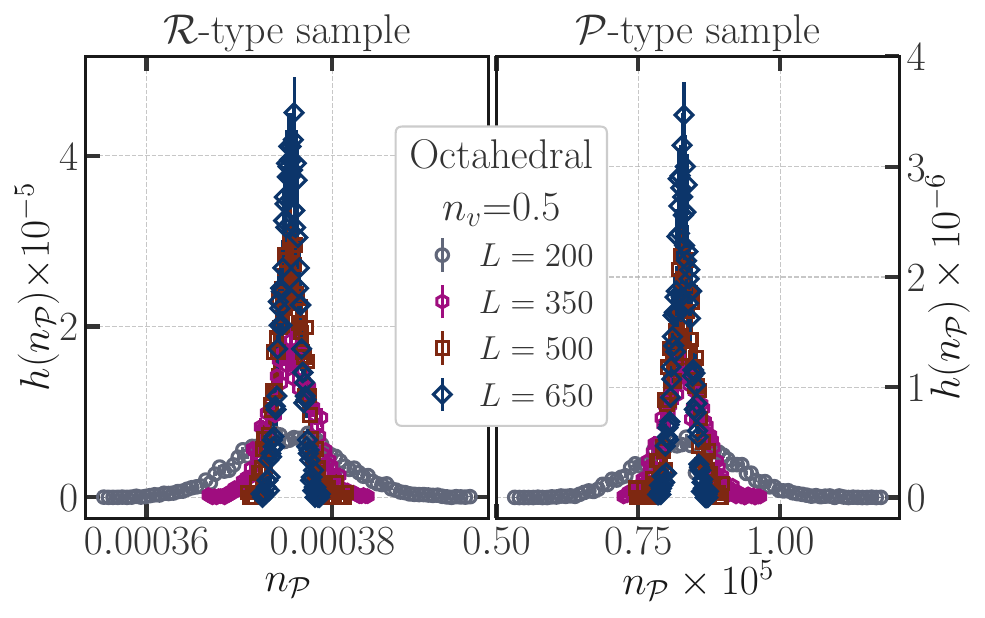}\\
		\end{tabular}
		\caption{(a) and (b) In phase IV in three dimensions, the histogram $h_{\calR}(\nR)$ of $\nR$, the number of distinct $\calR$-type regions scaled by the mass $m(\calG)$ of the largest geometric cluster, has a single sharply defined  peak at an $L$-independent value in both $\calR$-type and $\calP$-type samples, but the positions of these two peaks are macroscopically distinct: $\calR$-type samples have a significantly smaller density $\nR$ relative to $\calP$-type samples. (c) and (d) The histogram of $\nP$, the number of distinct $\calP$-type regions in $\calG$ scaled by the mass $m(\calG)$ of $\calG$, shows similar behavior, with the role of $\calR$-type and $\calP$-type samples being interchanged.	
		 This provides additional confirmation of the fact that the violations of thermodynamic self-averaging in phase IV in three dimensions (discussed in Sec.~\ref{subsec:Results3D}) are entirely analogous to those of the two-dimensional percolated phase (discussed in Sec.~{\protect{\ref{subsec:Results2D}}}).
			\label{fig:App:hist_nRnPIV_3D}}	
	\end{figure*}				

 	\begin{figure*}
		\begin{tabular}{cc}
			a)\includegraphics[width=0.5\columnwidth]{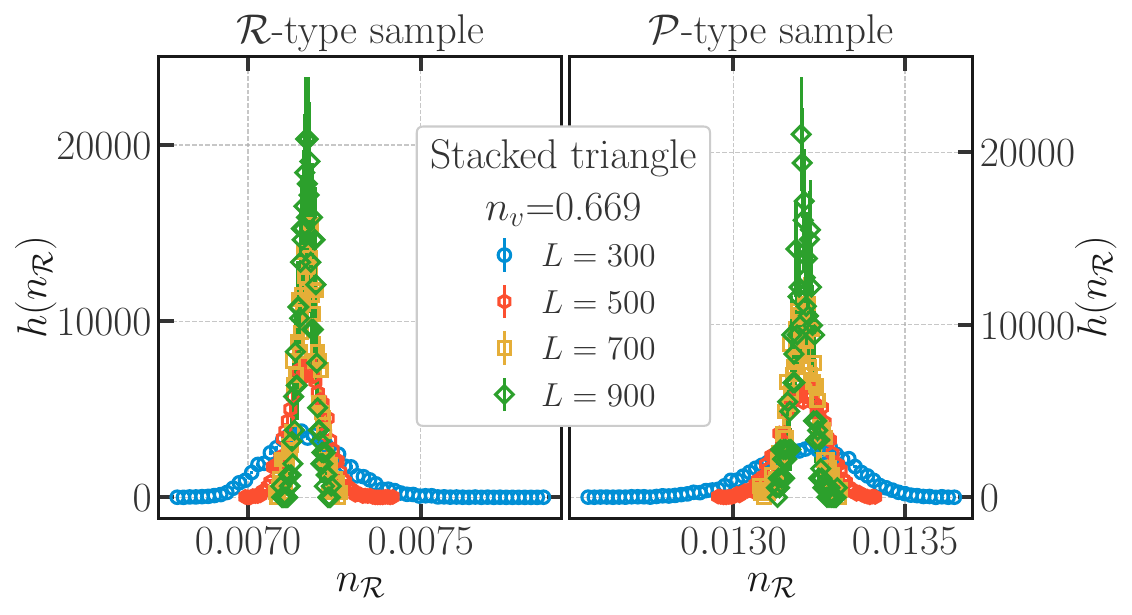} &	
			b)\includegraphics[width=0.5\columnwidth]{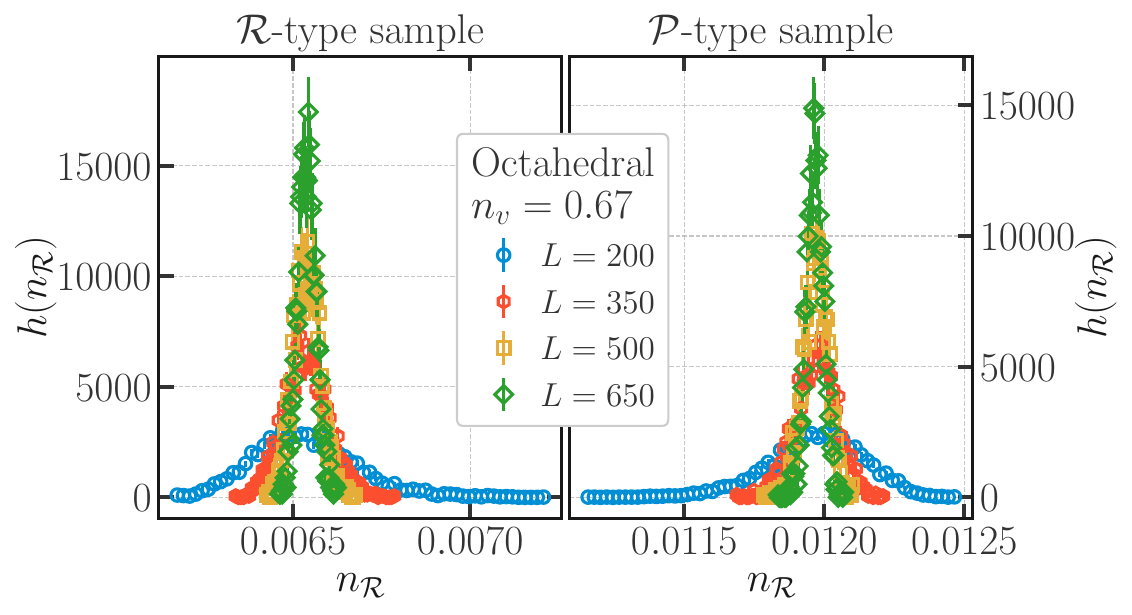}\\
			c)\includegraphics[width=0.5\columnwidth]{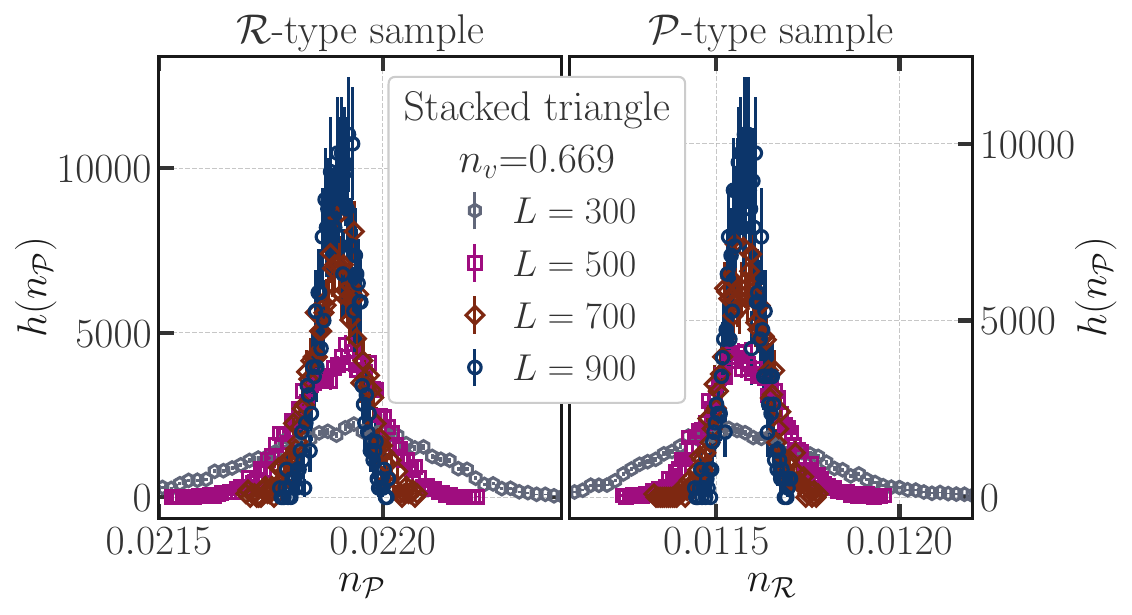} &	
			d)\includegraphics[width=0.5\columnwidth]{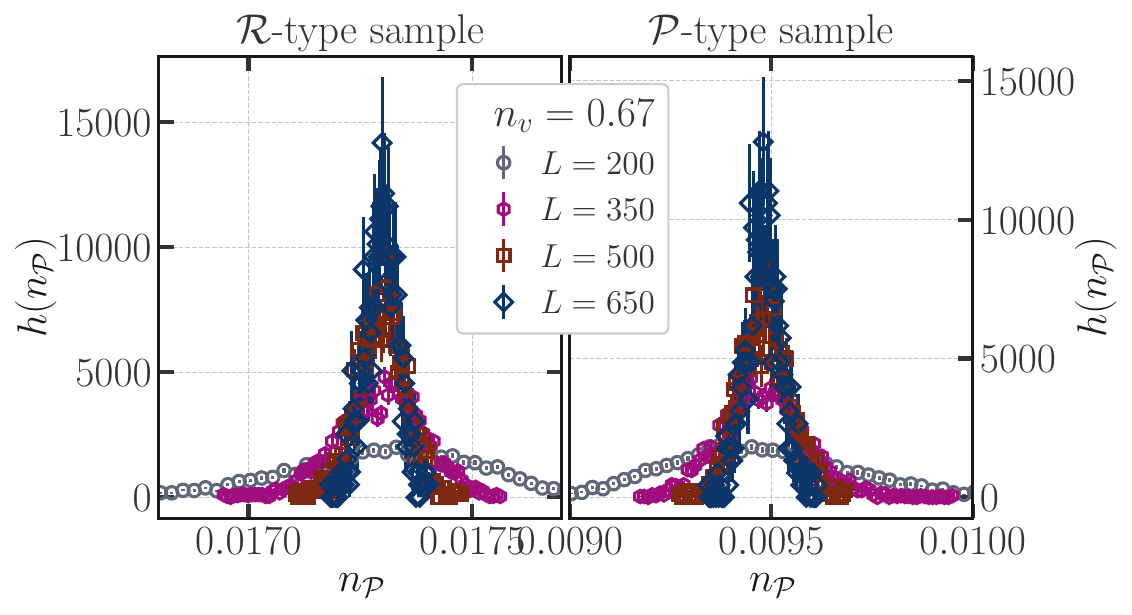}\\
		\end{tabular}
		\caption{(a) and (b) In phase III in three dimensions, the histogram $h_{\calR}(\nR)$ of $\nR$, the number of distinct $\calR$-type regions scaled by the mass $m(\calG)$ of the largest geometric cluster, has a single sharply defined  peak at an $L$-independent value in both $\calR$-type and $\calP$-type samples, but the positions of these two peaks are macroscopically distinct: $\calR$-type samples have a significantly smaller density $\nR$ relative to $\calP$-type samples. (c) and (d) The histogram of $\nP$, the number of distinct $\calP$-type regions in $\calG$ scaled by the mass $m(\calG)$ of $\calG$, shows similar behavior, with the role of $\calR$-type and $\calP$-type samples being interchanged. This behavior is entirely analogous to that displayed in the previous figure in phase IV. See the discussion in Sec.~\ref{subsec:Results3D} for further details.
			\label{fig:App:hist_nRnPIII_3D}}	
	\end{figure*}	

	 \begin{figure*}
	\begin{tabular}{cccc}
		\includegraphics[width=0.25\columnwidth]{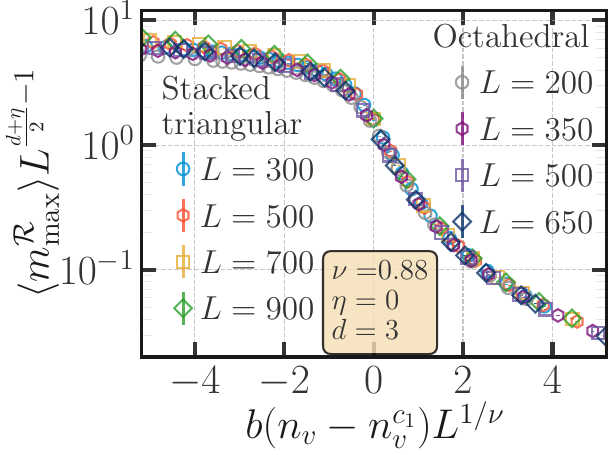}		
	&	\includegraphics[width=0.25\columnwidth]{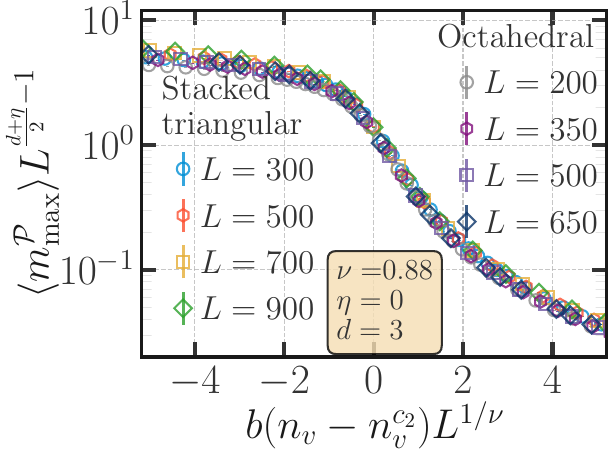}
		\includegraphics[width=0.25\columnwidth]{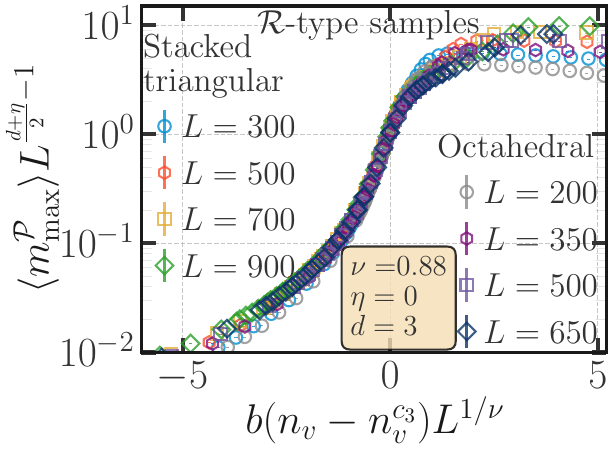}		
	&	\includegraphics[width=0.25\columnwidth]{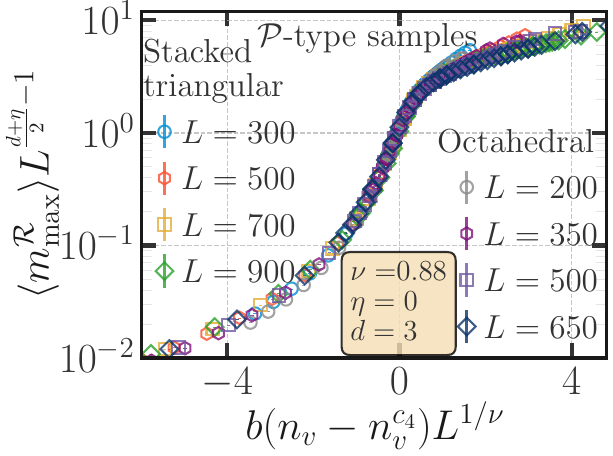}\\		
	\end{tabular}
	\caption{
	The $L$ and $n_v$ dependence of $\langle\mmaxR \rangle/L^{-\eta/2}$ ($\langle\mmaxP \rangle/L^{-\eta/2}$), the rescaled mean mass [see Eq.~\ref{eq:chi}] of the largest $\calR$-type ($\calP$-type) region of a sample, displays finite-size scaling behavior in the vicinity of $n_v^{c_1}$ ($n_v^{c_2}$) for all samples, and in the vicinity of $n_v^{c_4}$ ($n_v^{c_3}$) for $\calP$-type ($\calR$-type) samples. The figure displays evidence for this finite-size scaling collapse  at these critical points. At each critical point, the displayed scaling collapse employs a value of $\nu$ and $\eta$ chosen to yield reasonably high quality scaling collapse for that particular critical point. The  range of values that yield comparably high quality collapse at a particular critical point and the variation in these optimal values across different critical points, in conjuncton with the results of a similar finite-size scaling of other variables at these phase transitions, sets the error bar in our common estimate of $\nu = 0.92(6)$ for the correlation length exponent and $\eta = 0.00(4)$ for the anomalous exponent at all these transitions. Here $b = 0.65$ ($b=1.0$) for the stacked-triangular lattice (corner-sharing octahedral lattice) is a lattice-dependent (non-universal) scale factor chosen to collapse all data for both lattices onto a single scaling curve. It turns out that the non-universal rescaling parameter $a_{\mmax}$ that is needed in principle for rescaling the $y$-axis in the stacked-triangular case relative to the corner-sharing octahedral data (analogous to $a_{\chi}$ used in Fig.~\ref{fig:chiRchiPscaled3D}) is close enough to unity that omitting it makes no visible difference in this case.  See Sec.~\ref{sec:ComputationalMethodsObservables} and Sec.~\ref{sec:ComputationalResults} for terminology and details.
		\label{fig:mmaxRmmaxPscaled3D}}
\end{figure*}
	
	\begin{figure*}
		\begin{tabular}{cccc}
		\includegraphics[width=0.25\columnwidth]{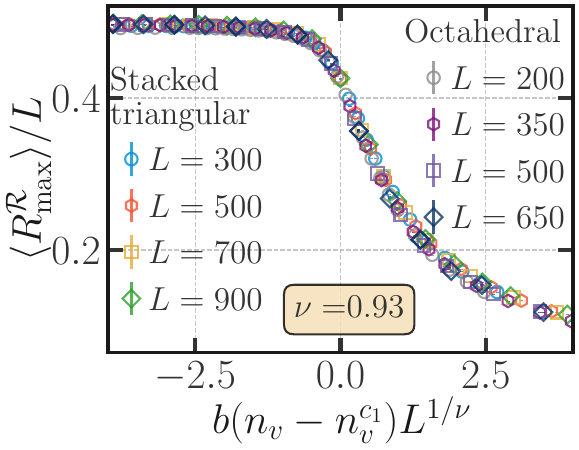}		
	&	\includegraphics[width=0.25\columnwidth]{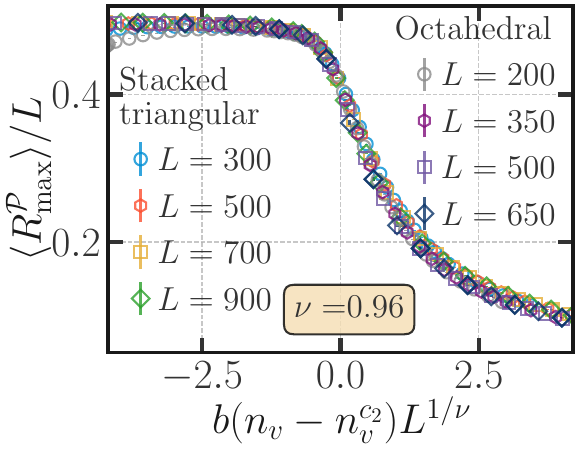}
		\includegraphics[width=0.25\columnwidth]{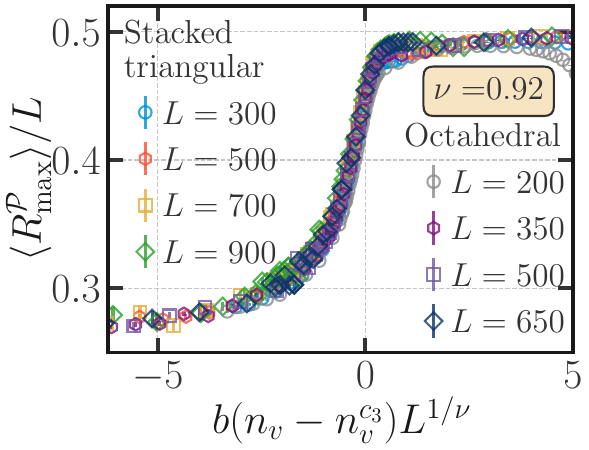}		
	&	\includegraphics[width=0.25\columnwidth]{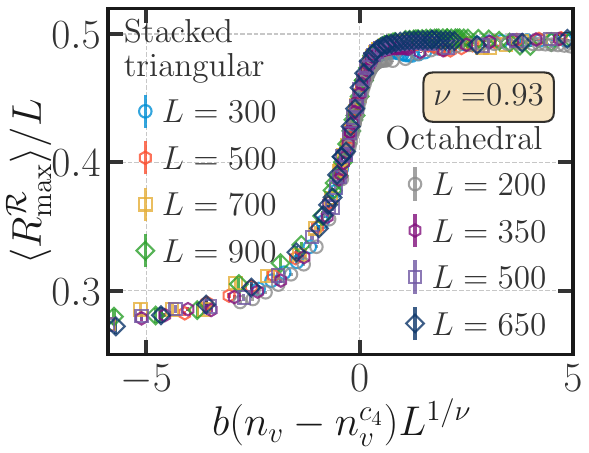}\\		
	\end{tabular}
	\caption{The $L$ and $n_v$ dependence of $\langle \RmaxR \rangle/L$ ($\langle \RmaxP \rangle/L$), the rescaled radius of gyration  of the largest $\calR$-type region ($\calP$-type region) in $\calG$, the largest connected component of the diluted $L \times L \times L$ lattice, displays finite-size scaling behavior in the vicinity of $n_v^{c_1}$ ($n_v^{c_2}$) and $n_v^{c_4}$ ($n_v^{c_3}$). The figure displays evidence for this finite-size scaling collapse, using a value of $\nu$ chosen for each data set to provide high quality scaling collapse for that data set. This analysis, in conjuncton with the finite-size scaling of other variables yields a common estimate of $\nu = 0.92(6)$ for the correlation length exponent at all these transitions. Here $b = 0.65$ ($b=1.0$) for the stacked-triangular lattice (corner-sharing octahedral lattice) is a lattice-dependent (non-universal) scale factor chosen to collapse all data for both lattices onto a single scaling curve. See  Sec.~\ref{sec:ComputationalMethodsObservables} and Sec.~\ref{sec:ComputationalResults} for terminology and details. 
		\label{fig:RmaxRRmaxPscaled3D}}
\end{figure*}

\end{document}